\documentclass[iop]{emulateapj}
\let\pwiflocal=\iffalse \let\pwifjournal=\iffalse

\usepackage{amsmath,amssymb}
\usepackage[breaklinks,colorlinks,urlcolor=blue,citecolor=blue,linkcolor=blue]{hyperref}

\usepackage[T1]{fontenc}
\pwifjournal\else
  \usepackage{microtype}
\fi

\pwifjournal\else
  \makeatletter
  \renewcommand\plotone[1]{%
    \centering \leavevmode \setlength{\plot@width}{0.95\linewidth}
    \includegraphics[width={\eps@scaling\plot@width}]{#1}%
  }%
  \makeatother
\fi

\makeatletter

\newcommand\@simpfx{http://simbad.u-strasbg.fr/simbad/sim-id?Ident=}

\newcommand\MakeObj[4][\@empty]{% [shortname]{ident}{url-escaped}{formalname}
  \pwifjournal%
    \expandafter\newcommand\csname pkgwobj@c@#2\endcsname[1]{\protect\object[#4]{##1}}%
  \else%
    \expandafter\newcommand\csname pkgwobj@c@#2\endcsname[1]{\href{\@simpfx #3}{##1}}%
  \fi%
  \expandafter\newcommand\csname pkgwobj@f#2\endcsname{#4}%
  \ifx\@empty#1%
    \expandafter\newcommand\csname pkgwobj@s#2\endcsname{#4}%
  \else%
    \expandafter\newcommand\csname pkgwobj@s#2\endcsname{#1}%
  \fi}%

\newcommand\MakeTrunc[2]{% {ident}{truncname}
  \expandafter\newcommand\csname pkgwobj@t#1\endcsname{#2}}%

\newcommand{\obj}[1]{%
  \expandafter\ifx\csname pkgwobj@c@#1\endcsname\relax%
    \textbf{[unknown object!]}%
  \else%
    \csname pkgwobj@c@#1\endcsname{\csname pkgwobj@s#1\endcsname}%
  \fi}
\newcommand{\objf}[1]{%
  \expandafter\ifx\csname pkgwobj@c@#1\endcsname\relax%
    \textbf{[unknown object!]}%
  \else%
    \csname pkgwobj@c@#1\endcsname{\csname pkgwobj@f#1\endcsname}%
  \fi}
\newcommand{\objt}[1]{%
  \expandafter\ifx\csname pkgwobj@c@#1\endcsname\relax%
    \textbf{[unknown object!]}%
  \else%
    \csname pkgwobj@c@#1\endcsname{\csname pkgwobj@t#1\endcsname}%
  \fi}

\makeatother

\pwifjournal\else
  \usepackage{etoolbox}
  \makeatletter
  \patchcmd{\NAT@citex}
    {\@citea\NAT@hyper@{%
       \NAT@nmfmt{\NAT@nm}%
       \hyper@natlinkbreak{\NAT@aysep\NAT@spacechar}{\@citeb\@extra@b@citeb}%
       \NAT@date}}
    {\@citea\NAT@nmfmt{\NAT@nm}%
     \NAT@aysep\NAT@spacechar\NAT@hyper@{\NAT@date}}{}{}
  \patchcmd{\NAT@citex}
    {\@citea\NAT@hyper@{%
       \NAT@nmfmt{\NAT@nm}%
       \hyper@natlinkbreak{\NAT@spacechar\NAT@@open\if*#1*\else#1\NAT@spacechar\fi}%
         {\@citeb\@extra@b@citeb}%
       \NAT@date}}
    {\@citea\NAT@nmfmt{\NAT@nm}%
     \NAT@spacechar\NAT@@open\if*#1*\else#1\NAT@spacechar\fi\NAT@hyper@{\NAT@date}}
    {}{}
  \makeatother
\fi

\MakeObj{n33370}{NLTT\%2033370}{NLTT~33370\,AB}
\MakeTrunc{n33370}{NLTT~33370}
\MakeObj{tvlm}{TVLM\%20513-46546}{TVLM~513--46546}

\newcommand\aafd{erg~s$^{-1}$~cm$^{-2}$~\AA$^{-1}$} % "AAngstrom flux density"
\newcommand\apx{\ensuremath{\sim}}
\newcommand\cgsflux{erg~s$^{-1}$~cm$^{-2}$}
\newcommand\cgslum{erg~s$^{-1}$}
\newcommand\chandra{\textit{Chandra}}
\newcommand\citeeg[1]{\citep[\eg,][]{#1}}
\newcommand\cps{ct~s$^{-1}$}
\newcommand\dd{\ensuremath{\text{d}}}
\newcommand\eg{\textit{e.g.}}

\newcommand\ewha{\ensuremath{\text{EW}(\ha)}}
\newcommand\fig[1]{Figure~\ref{f.#1}}
\newcommand\ha{{\ensuremath{\text{H}\alpha}}}

\newcommand\ie{\textit{i.e.}}
\newcommand\kms{km~s$^{-1}$}
\newcommand\ls{L$_\odot$}
\newcommand\mj{M$_\text{J}$}

\newcommand\nltt{\obj{n33370}}

\newcommand\percc{cm$^{-3}$}
\newcommand\rcs{\ensuremath{\chi_r^2}}
\newcommand\rj{R$_\text{J}$}
\newcommand\sect[1]{Section~\ref{s.#1}}
\newcommand\sherpa{\textsf{Sherpa}}
\newcommand\sti{Stokes~$I$}
\newcommand\stiv{Stokes~$I$ and~$V$}
\newcommand\stv{Stokes~$V$}
\newcommand\swift{\textit{Swift}}

\newcommand\tbl[1]{Table~\ref{t.#1}}
\newcommand\teff{\ensuremath{T_\text{eff}}}

\newcommand\ujy{$\mu$Jy}

\newcommand\vsi{\ensuremath{v \sin i}}

\newcommand\Lb{\ensuremath{L_\text{bol}}}
\newcommand\Lh{\ensuremath{L_\ha}}

\newcommand\sLr{\ensuremath{L_{\nu,\text{R}}}}
\newcommand\Lu{\ensuremath{L_\text{UVW1}}}
\newcommand\Lx{\ensuremath{L_\text{X}}}
\newcommand\Lxf{\ensuremath{L_{\text{X},f}}}
\newcommand\Lxq{\ensuremath{L_{\text{X},q}}}
\newcommand\lb{\ensuremath{[\Lb]}}
\newcommand\lh{\ensuremath{[\Lh]}}

\newcommand\slr{\ensuremath{[\sLr]}}
\newcommand\lu{\ensuremath{[\Lu]}}
\newcommand\lx{\ensuremath{[\Lx]}}
\newcommand\lxf{\ensuremath{[\Lxf]}}
\newcommand\lxq{\ensuremath{[\Lxq]}}
\newcommand\lhlb{\ensuremath{[\Lh/\Lb]}}

\newcommand\lulb{\ensuremath{[\Lu/\Lb]}}

\newcommand\lxflb{\ensuremath{[\Lxf/\Lb]}}
\newcommand\lxqlb{\ensuremath{[\Lxq/\Lb]}}
\newcommand\slrlb{\ensuremath{[\sLr/\Lb]}}

\begin{document}

\title{Simultaneous Multiwavelength Observations of Magnetic Activity in
  \\ Ultracool Dwarfs. IV. The Active, Young Binary \\ NLTT~33370~AB (=
  2MASS~J13142039$+$1320011)}
\author{
  P.~K.~G. Williams\altaffilmark{1},
  E. Berger\altaffilmark{1},
  J.~Irwin\altaffilmark{1},
  Z.~K.~Berta-Thompson\altaffilmark{2},
  D. Charbonneau\altaffilmark{1}
}
\email{pwilliams@cfa.harvard.edu}
\altaffiltext{1}{Harvard-Smithsonian Center for Astrophysics, 60 Garden Street,
  Cambridge, MA 02138, USA}
\altaffiltext{2}{MIT Kavli Institute, 77 Massachusetts Avenue,
  Cambridge, MA 02139, USA}

\slugcomment{Draft: \today}
\shorttitle{Multiwavelength Observations of Ultracool Dwarfs IV}
\shortauthors{Williams \textit{et al.}}

\begin{abstract}
  We present multi-epoch simultaneous radio, optical, \ha, UV, and X-ray
  observations of the active, young, low-mass binary NLTT~33370\,AB (blended
  spectral type M7e). This system is remarkable for its extreme levels of
  magnetic activity: it is the most radio-luminous ultracool dwarf (UCD)
  known, and here we show that it is also one of the most X-ray luminous UCDs
  known. We detect the system in all bands and find a complex phenomenology of
  both flaring and periodic variability. Analysis of the optical light curve
  reveals the simultaneous presence of two periodicities, $3.7859\pm0.0001$
  and $3.7130\pm0.0002$~hr. While these differ by only \apx2\%, studies of
  differential rotation in the UCD regime suggest that it cannot be
  responsible for the two signals. The system's radio emission consists of at
  least three components: rapid 100\% polarized flares, bright emission
  modulating periodically in phase with the optical emission, and an
  additional periodic component that appears only in the 2013 observational
  campaign. We interpret the last of these as a gyrosynchrotron feature
  associated with large-scale magnetic fields and a cool, equatorial plasma
  torus. However, the persistent rapid flares at all rotational phases imply
  that small-scale magnetic loops are also present and reconnect nearly
  continuously. We present an SED of the blended system spanning more than 9
  orders of magnitude in wavelength. The significant magnetism present in
  NLTT~33370\,AB will affect its fundamental parameters, with the components'
  radii and temperatures potentially altered by \apx$+20$\% and \apx$-10$\%,
  respectively. Finally, we suggest spatially resolved observations that could
  clarify many aspects of this system's nature.
\end{abstract}


\section{Introduction}
\label{s.intro}

Low-mass stars are the most common stars in the Universe \citep{rg97},
dominating the solar neighborhood in particular \citep{kgc+12}. The ultracool
dwarfs (UCDs) are those very low-mass stars and brown dwarfs (BDs) with
spectral types of M7 and later \citep{krl+99,mdb+99}, and they comprise some
of our closest neighbors \citeeg{l13,l14}. Despite this prevalence and
proximity, there are significant gaps in our understanding of the nature of
these objects, including their formation and multiplicity \citep{l12},
internal structure \citep{bcah02}, radiative output \citep{dli09b,x.dli14},
atmospheric chemistry \citep{mack13}, magnetism \citep{wcb14,cwb14}, and
rotational evolution \citep{ib08,rm12,gb13}.

Besides their intrinsic interest, these topics have received increased
attention because low-mass stars are appealing host candidates for exoplanets
\citep{sks+07,tbm+07}, in particular Earth-like planets in the habitable zone
\citep{bdu+13,k13b}. The inferred properties of such planets depend
sensitively on those of their host stars, making accurate knowledge of the
fundamental properties of the hosts a topic of paramount importance.
Simultaneously, searches for the coolest UCDs are penetrating into regions of
parameter space occupied by giant exoplanets amenable to direct imaging
\citeeg{clz05,bldc10,pkdrm10,bmkm11,dgm+12,dk13,lmd+13,l14}, driving interest
in understanding the physical properties of these analogous populations.

The presence of magnetic activity in the UCD regime has important implications
for both the fundamental physics of these objects and their role as exoplanet
hosts and analogs. Magnetic fields alter their internal structures, affecting
estimates of radii, temperatures, and mass by \apx5--30\% \citep{sksd12},
likely playing a role in explaining the frequent measurement of UCD radii that
are significantly larger than those predicted by models
\citeeg{lm07,rmj+08,mbi+11,mm13}. High flare rates and strong stellar winds
will dramatically impact the location or existence of habitable zones around
low-mass stars, prompting a significant amount of work investigating the
magnetic \citep{mjh06,krl+07,l13b,lvj+13,ckd+11,cdg+14} and radiative
\citep{skm+05,l07,lde07,t09,psw13,lff14,sb14} interactions between stars and
planets.

The generation of magnetic fields in UCDs, however, is not clearly understood.
UCDs are fully convective \citep{cb00} and thus lack a tachocline, the
shearing interface between stellar radiative and convective zones that is
understood to play a vital role in the generation of magnetic fields in
Sun-like stars \citep{o03}. Early theoretical work supported the idea that
these objects would therefore not be able to generate large-scale magnetic
fields via dynamo action \citep{ddyr93}. The detection of radio bursts from
the young BD \object{LP 944-20}, however, demonstrated that large-scale fields
could in fact be present \citep{bbb+01}, and subsequent observations have
established this using a wide range of techniques
\citep{b02,b06b,bp05,had+06,rb06,rb07,rb10,mdp+10,rw12}. More recent
simulations have also demonstrated the generation of large-scale magnetic
fields in fully-convective systems \citep{ck06,dsb06,b08,bbb+10}.

Observational tracers of UCD magnetism also present puzzles. The radio
detection of \object{LP 944-20}, for instance, was a surprise, because it
exceeded predictions based on the empirical G\"udel-Benz relation
\citep[GBR;][]{gb93,bg94} between the radio and X-ray luminosities of
magnetically active stellar systems by more than four orders of magnitude
\citep{bbb+01}. Subsequent observations have revealed that \apx5--10\% of UCDs
are similar outliers, and the origin of this divergence remains unclear
\citep{sab+12,wcb14}. Other observational tracers of magnetic activity, on the
other hand, fade away dramatically in UCDs. Chromospheric activity as traced
by \ha\ emission is generally ``saturated'' at luminosities of $\Lh/\Lb \apx
10^{-3.5}$ in dMe flare stars, but this ratio decreases rapidly in the UCD
regime \citep{gmr+00,whw+04,bbf+10}. X-ray emission follows a similar pattern,
with typical luminosities decreasing from a saturation value of $\Lx/\Lb \apx
10^{-3}$ \citep{v84,pmm+03} to virtually undetectable levels in L~dwarfs
\citep{smf+06,bbg+08,bbf+10,wcb14}; \object{Kelu-1 AB} is the only such object
to be detected, with 3--4 photons from \chandra\ \citep{aob+07}. In each of
these bands, the relationship between rotation and magnetic activity evolves
significantly from what is found for earlier-type dMe stars, showing no
saturation in the radio, the appearance of weakly emitting rapid rotators in
\ha, and an anti-correlated ``super-saturation'' relationship in X-rays
\citep{mb03,bbg+08,bbf+10,rb10,mbr12,cwb14}. Possible physical underpinnings
of these trends include the increasing neutralization of the outer layers of
the (sub)stellar atmosphere \citep{mbs+02} or a shift in the topology of the
large-scale magnetic field \citep{dmp+08,mdp+08,mdp+10}. An understanding of
the dynamo in the lowest-mass stars and BDs should lead to insight into the
magnetic properties of massive exoplanets themselves.

Simultaneous multi-wavelength observations of activity tracers yield insight
into a wide range of topics such as the chromospheric heating mechanism
\citep{bgg+08}, the magnetic field topology \citep{bbg+08}, and role of
variability in biasing empirical relationships between activity tracers
\citep{bbf+10}. Here we present a detailed study of the active, young binary
\nltt\ (= 2MASS\,J13142039$+$1320011) using simultaneous observations in the
radio, optical, \ha, UV, and X-ray bands. \nltt is a unique system in terms of
both its fundamental physical parameters and its magnetic properties. It is a
tight, low-mass binary \citep[$a \apx 2$~AU, $M_\text{tot} \apx
  200$~\mj;][]{sbh+14} resolvable with adaptive optics, so that it promises to
become one of a small sample of benchmark UCD systems with dynamically
measured component masses \citep{bmb+08,dlb+10,kgb+10}. Unlike most other such
systems, \nltt\ is young, with an estimated age of \apx30--200~Myr
\citep{sbh+14}. This youth may be related to another dramatic characteristic:
it is phenomenally magnetically active. It the most radio-luminous UCD system
known \citep{mbi+11}, and, as we demonstrate in this work, it is also one of
the brightest in X-ray and \ha\ emission, with frequent flaring across the
electromagnetic spectrum.

We proceed by reviewing the observed and inferred properties of
\nltt\ (\sect{target}). We then describe the observations (\sect{obs}) and
their analysis (\sect{anal}), which yield a rich multi-wavelength data set
with complex phenomenology that we summarize in \sect{summary}. We discuss the
implications of the data for the system's physical configuration and emission
processes in \sect{disc}. Finally, we summarize our findings and present our
conclusions in \sect{conc}.

\begin{figure*}[tb]
\plotone{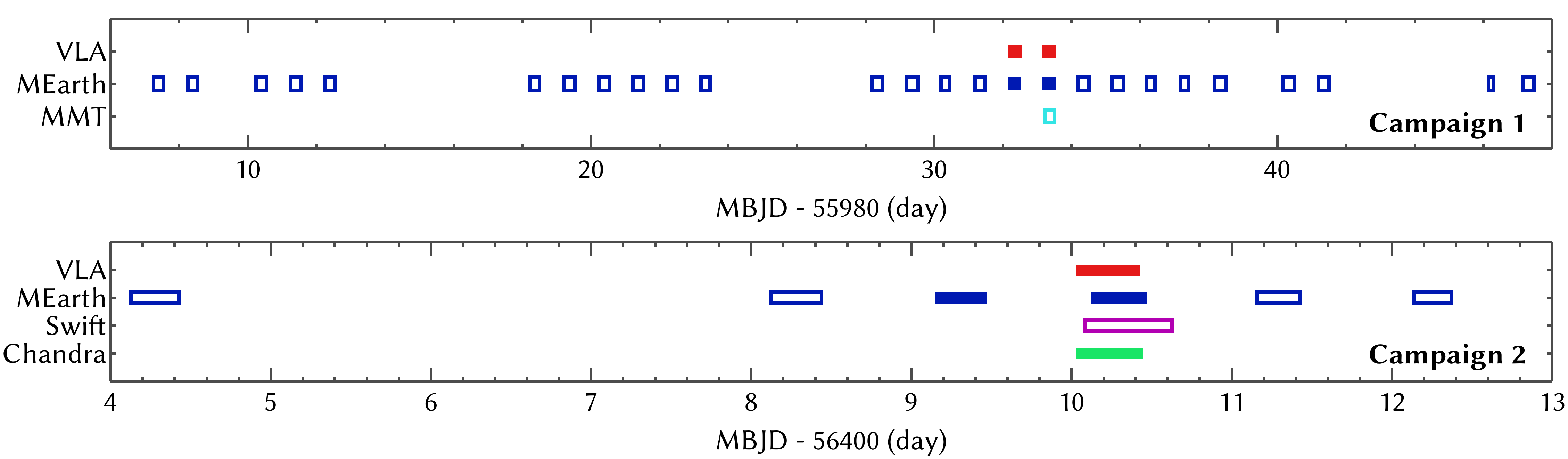}
\caption{Overview of the observing campaigns. Filled blocks indicate
  observations with essentially continuous coverage, while outlined blocks
  indicate observations consisting of distinct snapshots. Note the differing
  horizontal scales; the individual MEarth observations have equal durations
  in both campaigns.}
\label{f.overview}
\end{figure*}

Throughout this work, we use the notation $[x] \equiv \log_{10} x$, with $x$
being measured in cgs units unless specified otherwise. Bolometric
luminosities (\Lb) are measured in units of \ls. Parentheses following numbers
indicate uncertainties in the final digits; for instance, $123.4(56)$ is
shorthand for $123.4 \pm 5.6$. To avoid linguistic contortions we will
sometimes refer to \nltt\ as a single object; depending on the context, such
references should be taken to mean either the blended system, or a single but
unspecified component of the binary.

\section{\nltt}
\label{s.target}

\nltt\ was originally identified as a high-proper-motion object in the New
Luyten Two-Tenths catalogue \citep{thenltt}. It was subsequently recovered by
\citet{thelspmn}, who gave it the identifier \object{LSPM\,J1314$+$1320}.
Further followup assigned a blended spectral type of M7.0e, a tangential
velocity of $23.8(11)$~\kms, and a distance of $16.39(75)$~pc via
trigonometric parallax \citep{ltsr09}. \citet{ltsr09} also measured an
\ha\ equivalent width\footnote{Throughout this work, we report \ewha\ as a
  positive number; some authors, including \citet{sbh+14}, choose to use
  negative values for spectral lines in emission. This work involves no
  discussion of absorption lines so there is no ambiguity.} (\ewha) of
54.1~\AA. Lucky imaging of \nltt\ resolved it into a binary with a separation
of $0.13(2)$~arcsec (\apx2.1~AU; \apx2500~R$_*$) and a companion \apx1~mag
fainter than the primary in the $i'$ band \citep{lhm06}. Although continued
monitoring has refined the binary orbit, showing evolution in position angle
(PA) and separations of \apx0.07~arcsec, the system parameters are still
uncertain \citep{sbh+14}. A recent detailed analysis of the available
astrometry, blended spectroscopy, and resolved photometry by \citet{sbh+14}
has yielded estimates of $\teff = 3200(500)$ and $3100(500)$~K for the two
components, masses of $97^{+41}_{-48}$ and $91^{+41}_{-44}$~\mj, a system
bolometric luminosity $\lb = -2.36(9)$ ($10^{31.2}$~\cgslum), and a young
system age of 30--200~Myr. \nltt\ may be a member of the nearby, young AB~Dor
moving group \citep{sls12}, but such an assignment is still tentative
\citep{sbh+14}.

\citet{mbi+11} detected \nltt\ in the radio as part of a large VLA survey of
UCDs \citep{mbr12}. The emission was bright ($S_\nu \approx 1$~mJy at
4.86~GHz) and broadband ($S_\nu \approx 0.8$~mJy at 22.5~GHz). Furthermore,
\nltt\ was detected at a similar flux density in the Faint Images of the Radio
Sky at Twenty Centimeters survey \citep[FIRST;][]{thefirst}, indicating that
the emission was stable over a \apx10-yr timescale. The radio emission varied
sinusoidally with a period of $3.89(5)$~hr and amplitudes of \apx30\%~(20\%)
at 4.86~(8.46)~GHz. At 4.86~GHz the periodicity was also detected in circular
polarization (CP), with the polarization helicity alternating between left-
and right-handed in phase with the total intensity. \citet{mbi+11} also
obtained broadband optical photometry of \nltt\ from the MEarth survey
\citep{themearth,bic+12} using the Monitor pipeline \citep{iia+07}, finding
periodic modulation with a period of $3.785(2)$~hr and an amplitude of
\apx15~mmag (\apx1.4\%). The optical periodicity is thus marginally shorter
than the radio periodicity, differing by $6(3)$~min. Finally,
\citet{mbi+11} obtained optical spectroscopy and measured $\vsi =
45(5)$~\kms\ and, on two observing sessions separated by \apx100~days,
\ha\ EWs of 9.9 and 14.6~\AA.

\nltt\ has also been observed in the radio using very long baseline
interferometry (VLBI), a technique that can easily resolve the binary and
could potentially resolve the radio emission from each binary component.
\citet{mbi+11} report a detection of one unresolved source with a synthesized
beam of size 2$\times$1~mas, indicating that the emission originates from a
region $\lesssim$50~R$_*$ in size. Additional sources brighter than
\apx0.2~mJy are excluded. Subsequent VLBI observations continue to reveal only
a single radio source (J. Forbrich, 2014, priv. comm.). The system's
astrometric parameters are not sufficiently well-known to identify whether the
primary or the secondary is the radio source. Upcoming observations will
measure these parameters more precisely \citep[cf.][]{sbh+14}, resolving this
question and allowing much more stringent constraints to be placed on the
radio luminosity of the undetected source.

\section{Observations and Data Reduction}
\label{s.obs}

We observed \nltt\ in two simultaneous multiwavelength campaigns in March,
2012 and April, 2013. In \fig{overview} we provide a graphical overview of the
observations. The centerpiece of Campaign 1 (2012) was two consecutive
ten-hour nights of intensive optical and radio observations, with additional
optical spectroscopic observations on the second night. We performed
additional optical monitoring observations over \apx20 days on either side of
the intensive observations. The centerpiece of Campaign 2 (2013) was a
ten-hour session of simultaneous broadband optical, radio, X-ray, and UV
observations, again with additional broadband optical observations surrounding
the period of intensive observing. Below, we describe the observations and
data reduction in more detail.

\begin{figure*}[tb]
\plotone{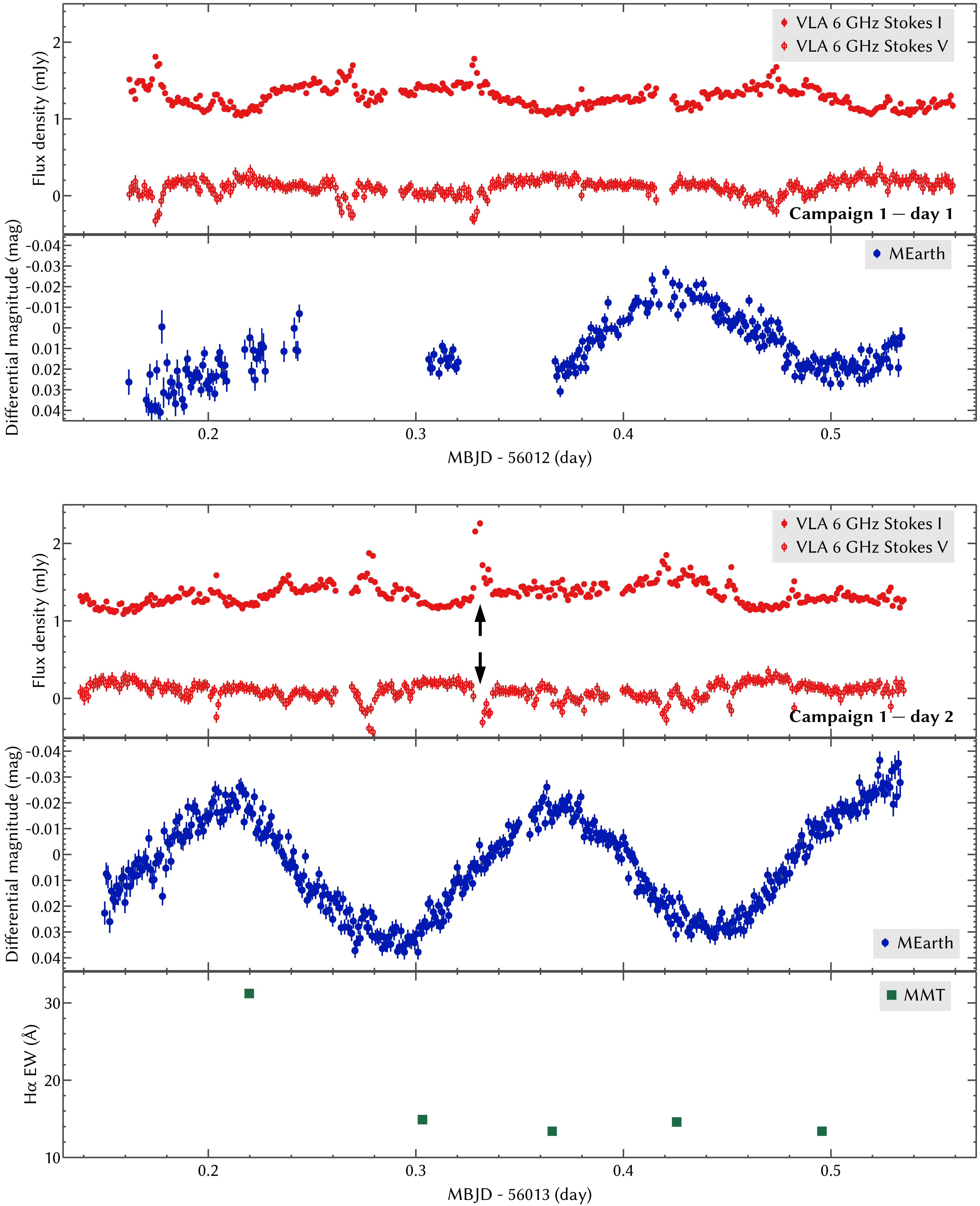}
\caption{Calibrated radio, optical, and \ha\ data from Campaign~1. The upper
  pair of panels show the first day of intensive observations, while the lower
  triptych shows the second. The black arrows in the VLA day~2 panel indicate
  a fully-polarized flare peaking at $\apx$8~mJy that exceeds the plot bounds
  (see \sect{vlaanal}, \fig{vlaflare}). The radio data have been smoothed to a
  cadence of 80~s for legibility. \sect{anal} presents a detailed analysis of
  these observations.}
\label{f.c1}
\end{figure*}

\begin{figure*}[tb]
\plotone{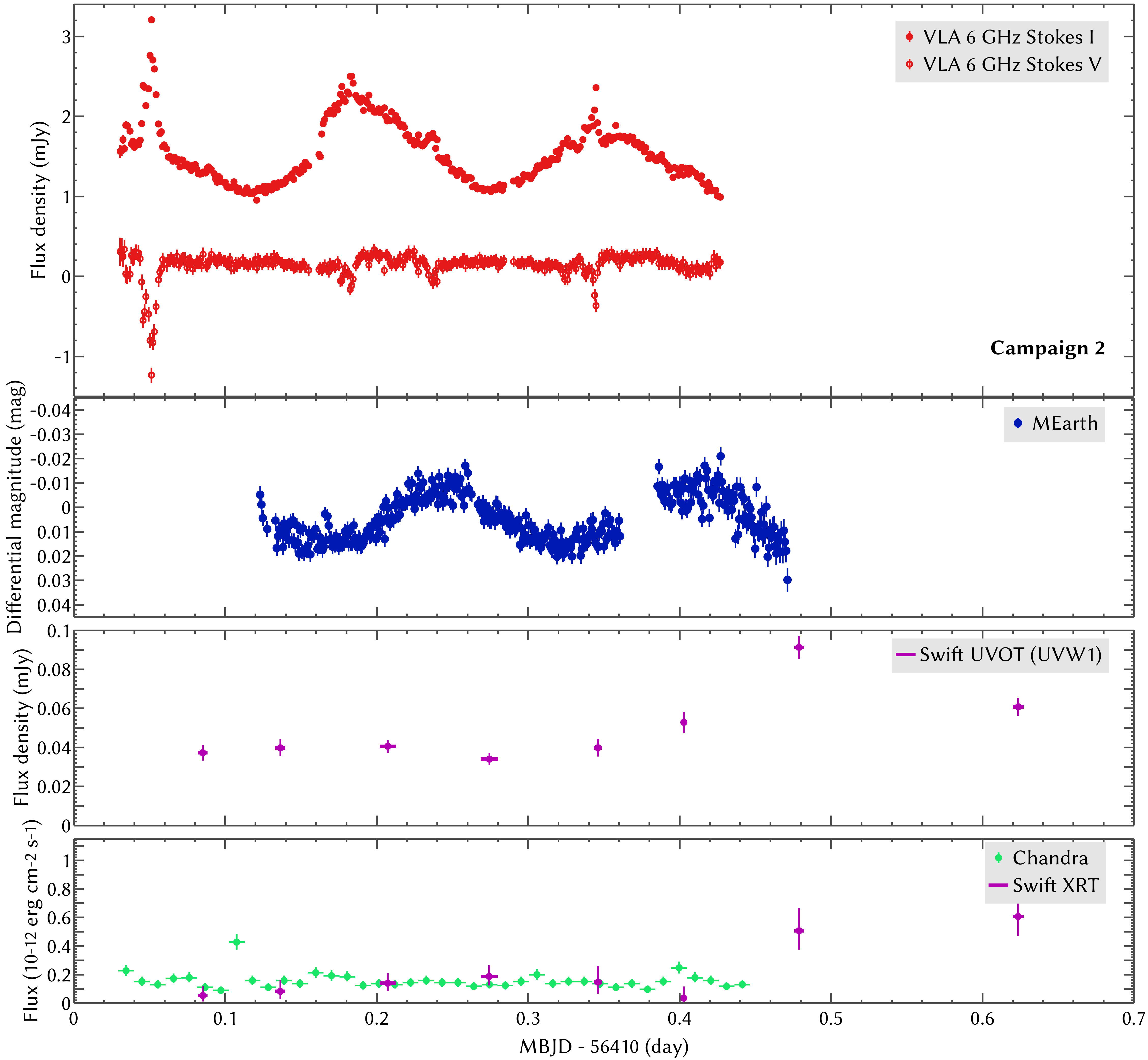}
\caption{Calibrated radio, optical, UV, and X-ray data from the intensive
  segment of Campaign~2. The intensive MEarth observations from the previous
  day are not shown; they are similar to the ones here. X-ray fluxes are in
  the 0.2--2~keV band and represent counts binned at a 15-min cadence. The
  radio data have been smoothed to a cadence of 80~s for legibility.
  \sect{anal} presents a detailed analysis of these observations.}
\label{f.c2}
\end{figure*}

The calibrated photometric data from the intensive observing sessions are
shown in Figures~\ref{f.c1} (Campaign~1) and \ref{f.c2} (Campaign~2). We have
converted all timestamps to modified, barycentric Julian dates (MBJDs) in the
barycentric dynamical time (TDB) timescale, which is the most appropriate
system for long-term timing applications \citep{esg10}. The integration times
of the measurements presented in this work are all well above the \apx5~ms
uncertainties in our conversion routines.

\subsection{MEarth}
\label{s.mearthdata}

We obtained long-term photometric monitoring of \nltt\ with telescopes in the
MEarth array \citep{themearth,bic+12}. The observations of Campaign~1 span
from 2012~March~1 to April~10, covering 27 nights. Those of Campaign~2 span
from 2013~April~22 to April~30, covering 6 nights. The observations were made
with a long-pass RG715 Schott glass filter. This is the same configuration as
used in the observations reported by \citet{mbi+11}, but the instruments and
filters were reconfigured in the interval between the two studies, so the
bandpasses are likely slightly different. All observations were made with an
exposure time of 42~s. The median FWHM in the Campaign~1 observations is
3.1$''$ (4.1~pixels); in Campaign~2, it is 2.6$''$ (3.4~pixels). This is
somewhat smaller than in the observations of \citet{mbi+11} because MEarth
operated with a slight defocus in the 2008--2009 observing season.

Most of the MEarth observations occurred with a cadence of \apx20~min over
the course of each night. However, during two ``intensive'' nights in each
campaign (indicated with filled boxes in \fig{overview}), observations were
essentially continuous, with a cadence of 71~s. These nights were the ones in
which the simultaneous monitoring at other bands occurred. The beginning of
the first intensive night of Campaign~1 was affected by significant cloud
cover, as is discernable in \fig{c1}. The other observations were generally
performed in clear conditions.

We also used previous MEarth observations of \nltt, taken between
2010~February~13 and February~19 and described by \citet{mbi+11}. These data
were reprocessed with a newer version of the MEarth reduction pipeline than
used in that work, but the changes in the outputs are minor. We refer to these
measurements as originating in Campaign~0.

The raw MEarth images are processed automatically using the Monitor project
pipeline \citep{iia+07} with facility-specific improvements as described in
the processing documentation for MEarth Data Release 2
(DR2)\footnote{\url{http://www.cfa.harvard.edu/MEarth/DR2/processing/}}. The
resulting tables of differential photometry are affected by three lingering
systematic effects that can be described with the equation:
\begin{equation}
\begin{split}
m_\text{true} &= m_\text{obs} + \sum_i k_{\text{ZP},i} \delta(S - i) + \\
&k_\text{CM} \text{CM} + k_\text{FWHM} (\text{FWHM} - \text{FWHM}_0).
\end{split}
\end{equation}
Here $m_\text{true}$ is the true differential magnitude, $m_\text{obs}$ is the
observed value reported by the pipeline, the CM term refers to a ``common
mode'' effect due to color-dependent extinction, and the FWHM term refers to
seeing-dependent offsets in the photometry. The $k_{\text{ZP},i}$ sum is a
zero-point term: each MEarth instrument configuration or ``segment'' is
assigned a unique integer identifier (denoted $S$ above) and has its own zero
point. The MEarth DR2 release
notes\footnote{\url{http://www.cfa.harvard.edu/MEarth/DR2/README.txt}} contain
much more detailed information on these effects.

The variables $m_\text{obs}$, $S$, CM, and FWHM are output by the pipeline for
each photometric measurement. In our data, $2 \leq S \leq 5$. The $k$ variables
are calibration terms that must be determined for each source by
simultaneously modeling their values as well as the underlying source
magnitudes $m_\text{true}$. FWHM$_0$ is an arbitrary constant that can aid the
numerical stability of the modeling; it is degenerate with the
$k_{\text{ZP},i}$. The FWHM-dependent term is only significant for crowded
fields, and we found in practice that the parameter $k_\text{FWHM}$ was not
well-constrained in our modeling. We therefore fixed it to zero, leaving five
calibration parameters when modeling the three campaigns.

We discarded individual photometric measurements based on data quality
metrics. Measurements in which the fitted source position was offset from its
expected location by more than 12 pixels in either the $x$ or $y$ direction
(table columns \textsf{Delta\_X}, \textsf{Delta\_Y}) were rejected, as were
those in which the estimated cloud extinction exceeded 1~mag (table column
\textsf{DMag}). The locations of these cutoffs were determined empirically,
and they eliminate 9\% of the measurements. Cuts on other quantities (source
ellipticity, FWHM, etc.) were investigated but were not found to be
beneficial.

\subsection{VLA}
\label{s.vladata}

We monitored \nltt\ in the radio with the Karl G. Jansky Very Large Array
(VLA) on the nights of 2012~March~26, 2012~March~27, and 2013~April~28
(projects VLA/12A-090 and SE0124; PI: Berger). The VLA was in the
C~configuration for first two nights (Campaign~1) and the D~configuration for
the third (Campaign~2). Each observing session lasted 10~hr and consisted of
integrations on the target with periodic visits to a phase calibrator,
\object{J1309$+$1154}, at a 9-min cadence. In all cases, \object{3C286} was
used as a bandpass, flux density, and polarization calibrator, with multiple
visits over the course of the night allowing full polarimetric calibration.
The integration time was 5~s and the C~band receivers were used, with 2048~MHz
of total bandwidth divided into two basebands centered at 5.0 and 7.1~GHz,
each containing 512 spectral channels. In each night, the total integration
time on the target was 496~min, or 83\% of the session.

We additionally obtained multi-band radio photometry of \nltt\ with the VLA on
the night of 2012~March~24 (project VLA/12A-090), just before the intensive
monitoring observations of Campaign~1. The details of these observations and
their outcome will be described in a forthcoming publication.

We calibrated the VLA data using standard procedures in the CASA software
system \citep{thecasa}. Radio-frequency interference was flagged automatically
using the \textsf{aoflagger} tool, which provides post-correlation
\citep{odbb+10} and morphological \citep{ovdgr12} algorithms for identifying
interference. Each observation included four visits to the calibrator
\object{3C\,286}, allowing full polarimetric calibration. The flux density
scale was referenced to \object{3C\,286} using the preliminary, 2010 version
of the scale defined by \citet{pb13}. After calibration, the data sets were
time-averaged to a uniform cadence of 10~s.

To check data quality and develop a model of the radio emission from unrelated
sources in the \nltt\ field, we imaged the data. For each night we made an
image of 2048$\times$2048 pixels, each 1$''$$\times$1$''$. The imaging process
used multi-frequency synthesis \citep{themfs} and CASA's multi-frequency CLEAN
algorithm with 2250 iterations. Because of the wide bandwidth of the data, we
used two spectral Taylor series terms for each CLEAN component, \textit{i.e.}
modeling both the flux and spectral index of each source. The reference
frequency for these images is 6.05~GHz. We detected \nltt\ robustly at all
times.

\begin{figure*}[tb]
\plotone{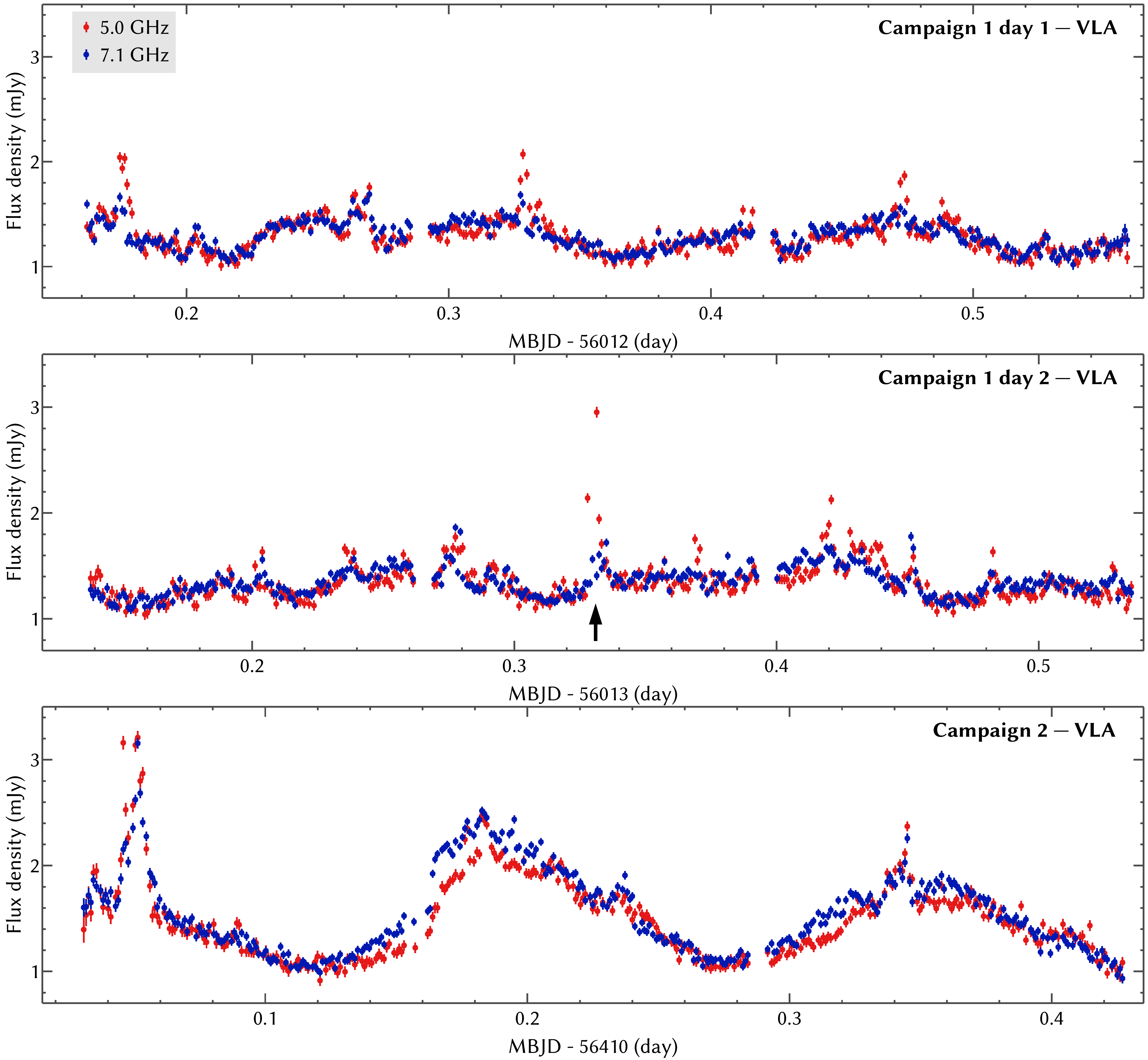}
\caption{VLA \sti\ photometry separated by baseband: 5.0~GHz (red) and 7.1~GHz
  (blue). All panels are on equivalent horizontal and vertical scales,
  emphasizing the change in the radio behavior between the two campaigns. The
  black arrow indicates a fully-polarized flare peaking at $\apx$8~mJy that
  exceeds the plot bounds (see \sect{vlaanal}, \fig{vlaflare}).}
\label{f.vlaspw}
\end{figure*}

We then extracted photometry for \nltt\ from the visibility-domain data
following the procedure described in \citet{wbz13}, using the deep images to
establish a precise source position and model the radio emission in the field.
The \stiv\ light curves are plotted in Figures~\ref{f.c1} and \ref{f.c2}.
Values of $V$ greater (less) than zero denote right (left) circular
polarization (RCP, LCP). We denote the fractional circular polarization $f_c
\equiv V/I$. Stokes parameters are defined such that for any given emission
component, $I \ge |V| \ge 0$ and $|f_c| < 1$. The Stokes parameters for the
superposition of two components are simply sums: $I_{12} = I_1 + I_2$, $V_{12}
= V_1 + V_2$. The radio data are all consistent with zero linear polarization.

\fig{vlaspw} shows the \sti\ photometry of \nltt\ for the two VLA basebands
separately. The common vertical scale of the three panels makes clear the
different variation patterns seen in Campaigns~1 and 2. The flat spectrum
reported by \citet{mbi+11} is evident in the fact that in most cases the flux
densities in the two basebands are virtually identical. The rapid flares,
however, are generally brighter at lower frequencies. Conversely, the rising
periods of the slow variation in Campaign~2 are associated with brighter
emission in the 7.1~GHz baseband.

\subsection{\chandra}

The \chandra\ observations were performed on 2013~April~28 (proposal 14200124;
\chandra\ observation ID 14530; PI: Berger) using the S3 backside-illuminated
chip of the ACIS imager. The total exposure was 35.6~ks, ranging from MBJD
56410.019--56410.457. No grating was used, the data mode was \textsf{VFAINT},
and the exposure mode was ``timed'' (TE).

We analyzed the \chandra\ data in CIAO version 4.6.1 \citep{theciao} using
CalDB version 4.5.9. Following \textsf{VFAINT} reprocessing to eliminate a
substantial fraction of background events, we estimated the residual
background in the data by extracting events in an energy range of 0.3--7~keV
in a large annulus around the astrometrically predicted position of \nltt. The
mean whole-chip background rate is 0.39 s$^{-1}$, which is consistent with
typical nonflaring behavior. We performed a Bayesian blocks analysis
\citep{s98,snjc13} to search for background flares as manifested by
significant changes in the background count rate, using the implementation
described in \citet{wcb14}. No such changes were found.

Based on the astrometric parameters given by \citet{sbh+14}, the predicted
location of \nltt\ at the time of the \chandra\ observation is $\alpha =$
13:14:20.17, $\delta = +$13:19:58.4, with an uncertainty of \apx1$''$. The
\chandra\ data contain an X-ray source of 909 counts (0.3--7~keV) within a
5$''$ aperture at $\alpha =$ 13:14:20.14, $\delta = +$13:19:58.5 (separation
of \apx0.5$''$). This aperture is expected to contain \apx4.2 background
counts. We identify this source with \nltt.

\subsection{\swift}

\nltt\ was observed by \swift\ on 2013~April~28 (target of opportunity ID
4717) with the X-ray Telescope (XRT), UV/Optical Telescope (UVOT), and
$\gamma$-ray Burst Alert Telescope (BAT). No sources were detected with the
BAT. The UVOT had the UVW1 filter (\apx2280--2930~\AA) in place and the XRT
was in photon-counting (PC) mode. A total of 5.0~ks were spent on-source,
divided among 8 visits of durations varying between 384 and 975~s. As can be
seen in \fig{c2}, the total time spanned by the observations was \apx13~hr,
with the final two visits occurring after the end of the simultaneous VLA,
MEarth, and \chandra\ observations.

We analyzed the \swift\ XRT data using \textsf{HEAsoft} version 6.15.1 with
\textsf{CalDB} version 20140120. We calibrated and cleaned the low-level data
with the \textsf{xrtpipeline} task. We then extracted source events from a
region coincident with \nltt\ having a radius of 20~pixels. Over the duration
of the observation, 59 events were detected in the region, with an expectation
of \apx7 of those coming from the background. The bottom panel of \fig{c2}
shows the resulting \swift\ XRT light curve. Here the event rates have been
converted into fluxes using an energy conversion factor (ECF) of $2.14 \times
10^{-11}$~erg~cm$^{-2}$~ct$^{-1}$, which was derived using PIMMS for a 0.7~keV
APEC plasma. This is the temperature obtained in a fit to the \chandra\ data
(\sect{anal}).

\begin{figure*}[tb]
\plotone{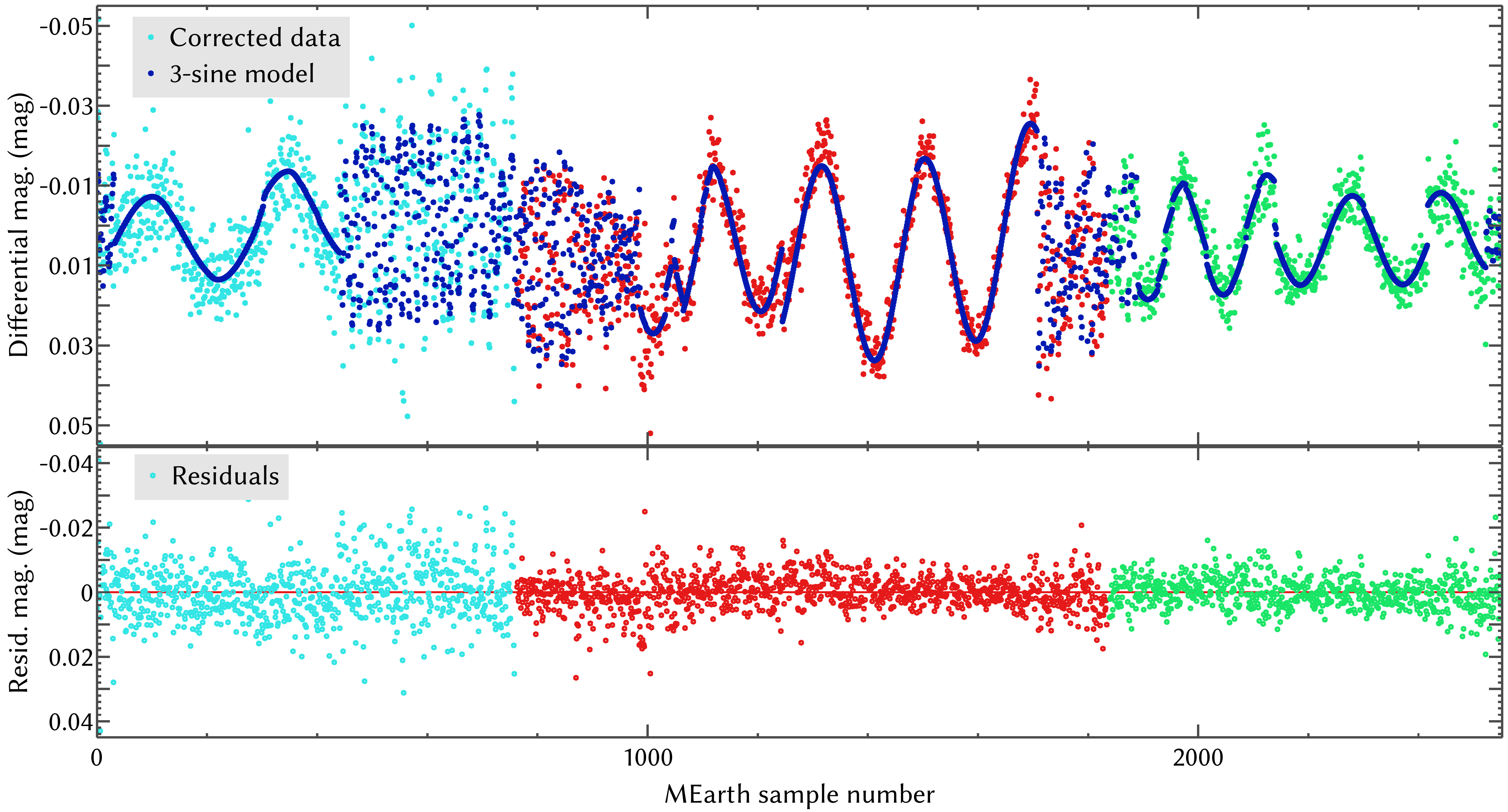}
\caption{Data, model, and residuals for the complete MEarth dataset. For
  clarity, the abscissa is a monotonically-increasing sample number rather
  than, e.g., the observation time; the data span 3.4~yr. Data and residuals
  from Campaigns~0, 1, and 2 are shown in light blue, red, and green,
  respectively. The model is shown in dark blue. The ``intensive'' observing
  runs are apparent as segments of smoothly-varying data. The data shown here
  have been corrected for MEarth photometric systematics as described in
  Sections~\ref{s.mearthdata} and~\ref{s.mearthanal}.}
\label{f.mearth}
\end{figure*}

We analyzed the \swift\ UVOT data using the same software stack as for the
XRT. We used the \textsf{uvotmaghist} task to derive the light curve data
shown in \fig{c2}. The uncertainties shown are sums in quadrature of the
pipeline-reported systematic and statistical uncertainties, which are
dominated by the latter. The UVW1 filter of the UVOT is susceptible to a ``red
leak'' of optical light with wavelengths $\lesssim$4500~\AA\ \citeeg{brm+10}.
While these wavelengths qualify as ``red'' compared to the UV band, they are
still quite blue compared to the photosphere of \nltt. Based on the UVW1$-$B
color of \nltt, the estimates of \citet{brm+10} suggest that the contribution
of the photosphere is $<$0.2~mag, and a convolution of the UVW1 bandpass with
the representative photospheric model described in \sect{sedsummary} yields a
contribution of \apx$2 \times 10^{-17}$~\aafd, \apx10\% of the observed value
in the UVW1 band.

\subsection{MMT}

We obtained optical spectroscopic observations of \nltt\ on 2013~March~27 UT
with the Blue Channel spectrograph mounted on the MMT 6.5-m telescope. Six
300-s exposures were obtained at regular intervals between 05:05 and 11:42 UT,
in conjunction with day 2 of Campaign 1. The observations were obtained with
the 1200 l/mm grating leading to a wavelength coverage of
$5435$--$6750$~\AA\ at a resolution of about 1.5~\AA.

We processed and analyzed the data using standard routines in IRAF, and
measured H$\alpha$ EWs using the task \textsf{splot}. The resulting values are
plotted in \fig{c1}. The errors on each measurement are $0.15$~\AA.

\section{Analysis}
\label{s.anal}

We detected variable emission from \nltt\ in every observed band. We find both
ubiquitous flaring and periodic, non-flaring modulations.

\subsection{MEarth}
\label{s.mearthanal}

The intensive MEarth observations in all three campaigns reveal sinusoidal
variability. While the MEarth data do not show flares, examination of
Figures~\ref{f.c1} and \ref{f.c2} shows that there is variability on the
\apx5~mmag level from one cycle to the next. Here we investigate the
periodicities present in the MEarth data without attempting to model these
low-level variations.

A single sine curve with a period of $3.779816(2)$~hr can reproduce the
phasing of the observations very well over the full 3.4-yr time baseline of
the data. A standard weighted nonlinear least-squares fit to the data yields a
large reduced $\chi^2$ ($\rcs$) of 6.89, however, with periodic structure in
the residuals of the intensive observations. Comparison of Figures~\ref{f.c1}
and \ref{f.c2} suggests that, at a minumum, the oscillation amplitude evolves
from one campaign to another. We performed this fit (and all others in this
section) with a Python implementation of the Levenberg-Marquardt algorithm
based on the classic \textsf{MINPACK} version \citep{m78}.

A sine curve model with a single period but different amplitudes and phases
for each campaign achieves an improved $\rcs = 4.28$ but still shows periodic
residuals. Adding a second sine term yields $\rcs = 2.65$ with much less
marked structure in the residuals. The derived periods in this model are
$3.7859(1)$ and $3.7130(2)$~hr. The former value is consistent with the period
of $3.785$~hr derived by \citet{mbi+11} from the Campaign~0 data alone. A
third sine term yields $\rcs = 2.24$ with the new period being $23.884(9)$~hr.
This is close to the sidereal period ($23.93$~hr), suggesting that this
component represents a systematic effect. The addition of a fourth term yields
a marginal improvement ($\rcs = 2.21$). We use the 3-sine model to guide our
interpretation of the data, treating the first two periodicities as
astrophysical and the third as a systematic. The longer-period astrophysical
term always has the higher amplitude and we refer to it as the ``primary''
component, while the shorter-period term is the ``secondary.'' The 3-sine
model and residuals are shown in \fig{mearth} and the derived parameters are
listed in \tbl{mearthpars}, where the sinusoidal components are given by
\begin{equation}
m_\textrm{true} = A_i \sin \left(\frac{2 \pi (t - T_i)}{P_i}\right).
\label{e.sine}
\end{equation}
Parameter uncertainties are derived from the covariance matrix determined by
the Levenberg-Marquardt minimization. Because this model is defined in terms
of magnitudes, maxima in luminosity correspond to minima in $m_\textrm{true}$.
All plots of MEarth data in this work use $y$ axes such that brighter emission
is closer to the top of the plot.

% TableBuilder table
\begin{deluxetable}{lllr@{}lr@{}l}
%custom preamble

%hardcoded preamble
\tablecolumns{7}
\tablewidth{0em}
\tablecaption{Fitted Parameters from MEarth Modeling\label{t.mearthpars}}
\tablehead{
\colhead{Group} & \colhead{Param.} & \colhead{Units} & \multicolumn{2}{c}{Value} & \multicolumn{2}{c}{Uncert.} \\ \\
\multicolumn{1}{c}{(1)} & \multicolumn{1}{c}{(2)} & \multicolumn{1}{c}{(3)} & \multicolumn{2}{c}{(4)} & \multicolumn{2}{c}{(5)}
}
\startdata
Calibration & $k_{\text{ZP},2}$ & mag & $-0$ & $.0066$ & $0$ & $.0013$ \\
Parameters & $k_{\text{ZP},3}$ & mag & $0$ & $.0020$ & $0$ & $.0013$ \\
 & $k_{\text{ZP},4}$ & mag & $0$ & $.0093$ & $0$ & $.0004$ \\
 & $k_{\text{ZP},5}$ & mag & $0$ & $.0033$ & $0$ & $.0004$ \\
 & $k_\text{CM}$ & --- & $-1$ & $.46$ & $0$ & $.04$ \\
\\
Primary & $P$ & hr & $3$ & $.7859$ & $0$ & $.0001$ \\
Periodicity & $A_{0}$ & mag & $0$ & $.0167$ & $0$ & $.0004$ \\
 & $A_{1}$ & mag & $0$ & $.0164$ & $0$ & $.0002$ \\
 & $A_{2}$ & mag & $0$ & $.0111$ & $0$ & $.0003$ \\
 & $\,T_{0}$ & MBJD & $55247$ & $.6729$ & $0$ & $.0006$ \\
 & $\,T_{1}$ & MBJD & $56010$ & $.8746$ & $0$ & $.0003$ \\
 & $\,T_{2}$ & MBJD & $56409$ & $.8242$ & $0$ & $.0008$ \\
\\
Secondary & $P$ & hr & $3$ & $.7130$ & $0$ & $.0002$ \\
Periodicity & $A_{0}$ & mag & $0$ & $.0067$ & $0$ & $.0004$ \\
 & $A_{1}$ & mag & $0$ & $.0095$ & $0$ & $.0002$ \\
 & $A_{2}$ & mag & $0$ & $.0051$ & $0$ & $.0003$ \\
 & $\,T_{0}$ & MBJD & $55247$ & $.6148$ & $0$ & $.0015$ \\
 & $\,T_{1}$ & MBJD & $56010$ & $.9399$ & $0$ & $.0004$ \\
 & $\,T_{2}$ & MBJD & $56409$ & $.9439$ & $0$ & $.0017$ \\
\\
Systematic & $P$ & hr & $23$ & $.884$ & $0$ & $.009$ \\
Term & $A_{0}$ & mag & $0$ & $.0061$ & $0$ & $.0010$ \\
 & $A_{1}$ & mag & $0$ & $.0097$ & $0$ & $.0004$ \\
 & $A_{2}$ & mag & $0$ & $.0040$ & $0$ & $.0004$ \\
 & $\,T_{0}$ & MBJD & $55247$ & $.9231$ & $0$ & $.0353$ \\
 & $\,T_{1}$ & MBJD & $56011$ & $.0259$ & $0$ & $.0055$ \\
 & $\,T_{2}$ & MBJD & $56410$ & $.0172$ & $0$ & $.0147$
\enddata
\end{deluxetable}
% end TableBuilder table

A more flexible model in which the periods of the sine curves may also vary
between each campaign yields an indistinguishable $\rcs = 2.22$. In this
model, the best-fit periods are all consistent with those obtained above. We
conclude that there is no evidence for \apx year-timescale evolution in the
periodicities present in the data, but cannot rule out changes at the
$\lesssim$0.5\% level.

The phase offsets between the Campaign~1 and 2 signals are $25(25)$~deg for
the primary periodicity and $20(40)$~deg for the secondary. Between Campaign~0
and 1 they are $60(45)$ and $-20(80)$~deg, respectively. A model with constant
phasing between the three epochs achieves an inferior $\rcs = 3.85$, with
large residuals outside of the regions of intensive observations, suggesting
that the least-squares optimizer converged on a solution that worked unusually
well for the densely-sampled points, rather than a solution that fairly
represented the overall system light curve.

\begin{figure}[tb]
\plotone{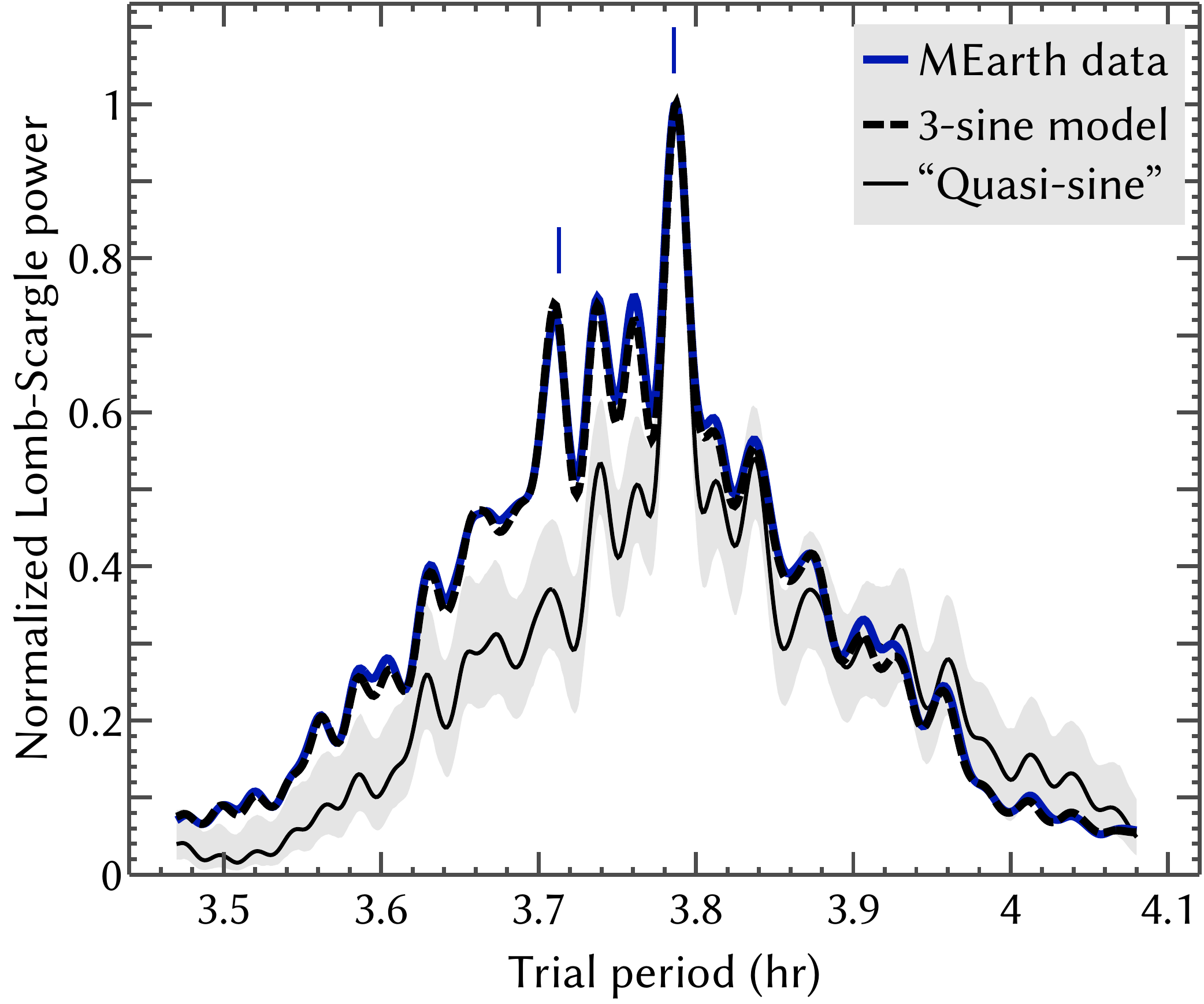}
\caption{Lomb-Scargle periodograms for the Campaign~1 MEarth data. The
  periodogram for the systematics-corrected data is shown in blue with the
  peak normalized to unity. The dashed black line shows the normalized
  periodogram for the 3-sine model, using the same sampling as the data; it is
  nearly identical to the data periodogram. The thin black line shows the
  periodogram for a single-sine fit; its gray envelope shows the 90\% credible
  region for a ``quasi-sine'' model with a single periodicity, but random
  amplitude and phase perturbations that vary with 2--400~hr autocorrelation
  timescales. See \sect{mearthanal} for details. This model does not reproduce
  the observed periodogram structure. The short, vertical blue lines indicate
  the primary and secondary periods determined for the 3-sine model.}
\label{f.c1ls}
\end{figure}

We investigated whether the dual period results might be due to variations
that are not strictly periodic (and hence are broadened in Fourier space)
rather than two discrete components. In particular, we explored an alternative
``quasi-sine'' model in which $A$ and $T$ in Equation~\ref{e.sine} wander
randomly in time (\ie, possibly corresponding to a long-lived photospheric
feature with evolving size and longitude). We realized 512 such models by
drawing the amplitude and phase curves from squared-exponential Gaussian
processes, in which the covariance $C$ between measurements separated by an
interval $\Delta t$ is
\begin{equation}
C(\Delta t) = a \exp \left( \frac{\Delta t}{2 s} \right)^2.
\end{equation}
We drew $a$ uniformly in log-space between $0.01$ and $0.06$ for the amplitude
variation and between $0.01$ and $0.1$~rad for the phase variation. We drew
$s$ uniformly in log-space between $2$ and $400$~hr. The characteristic
excursions were \apx$10a$ in scale. We evaluated these models with the
sampling of Campaign~1, which has the longest time baseline and most
measurements of the three campaigns. \fig{c1ls} compares Lomb-Scargle
periodograms for the Campaign~1 data, the 3-sine model, and the quasi-sine
realizations. The 3-sine model reproduces the observed periodogram well, while
the quasi-sine realizations do not: they cannot recreate the asymmetry of the
data periodogram.

\begin{figure*}[tb]
\plotone{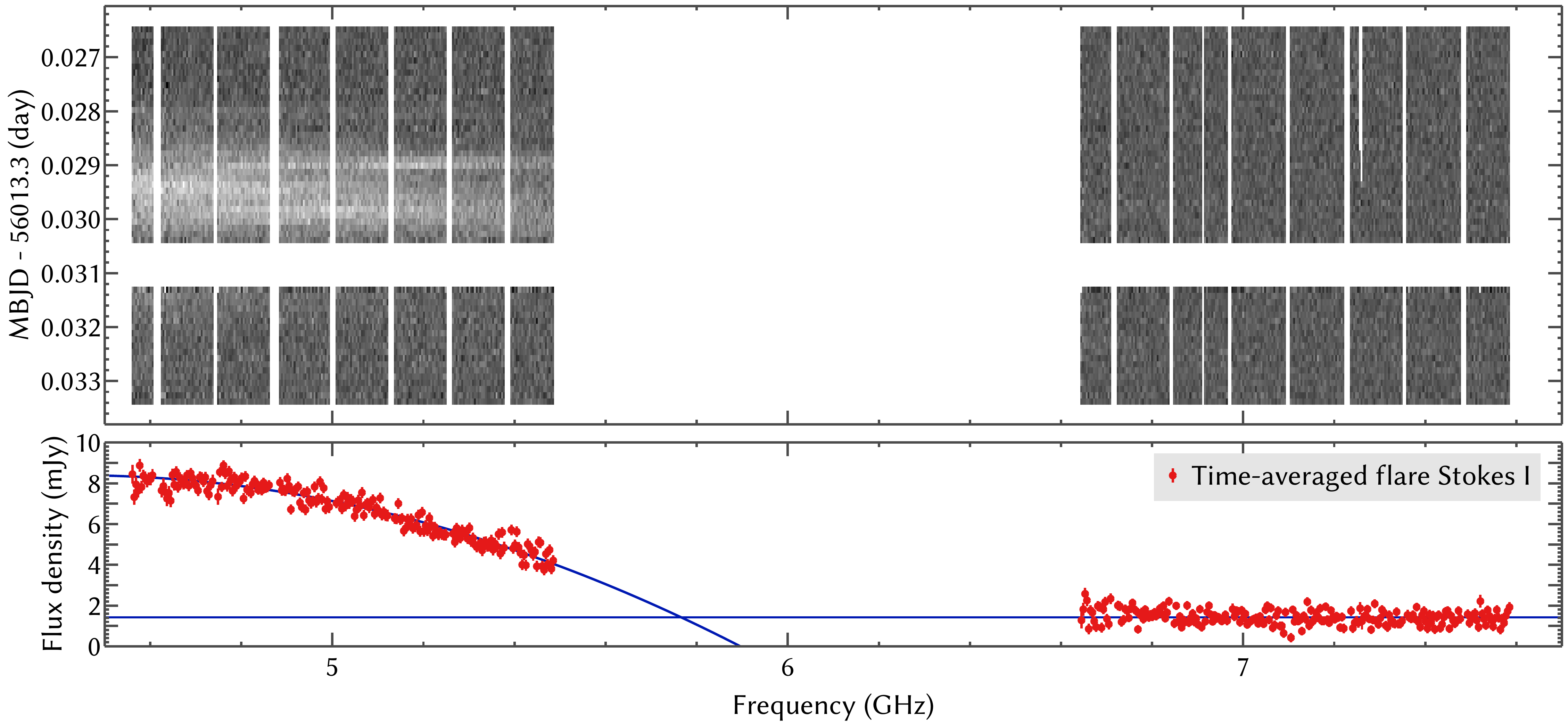}
\caption{Upper panel: \sti\ dynamic spectrum of the bright radio flare seen in
  day~2 of Campaign~1. The measurements are made at a 10~s cadence and the
  grayscale runs linearly from -7.2~mJy (black) to 17.1~mJy (white). The large
  frequency gap in the data is due to the positioning of the VLA basebands;
  all others are due to RFI. The non-flaring variation has not been subtracted
  (\sect{vlaanal}). Lower panel: the flare spectrum averaged between $0.0286 <
  \text{MBJD} - 56013.3 < 0.0304$. The flare is absent from the 7.1~GHz
  frequency window. Also shown are parabolic (constant) fits to the spectrum
  in the lower (upper) frequency windows. These fits suggest that the flare
  cutoff frequency is \apx5.8~GHz. The flare light curve is shown in
  \fig{vlaflare}.}
\label{f.vladynspec}
\end{figure*}

\begin{figure}[tb]
\plotone{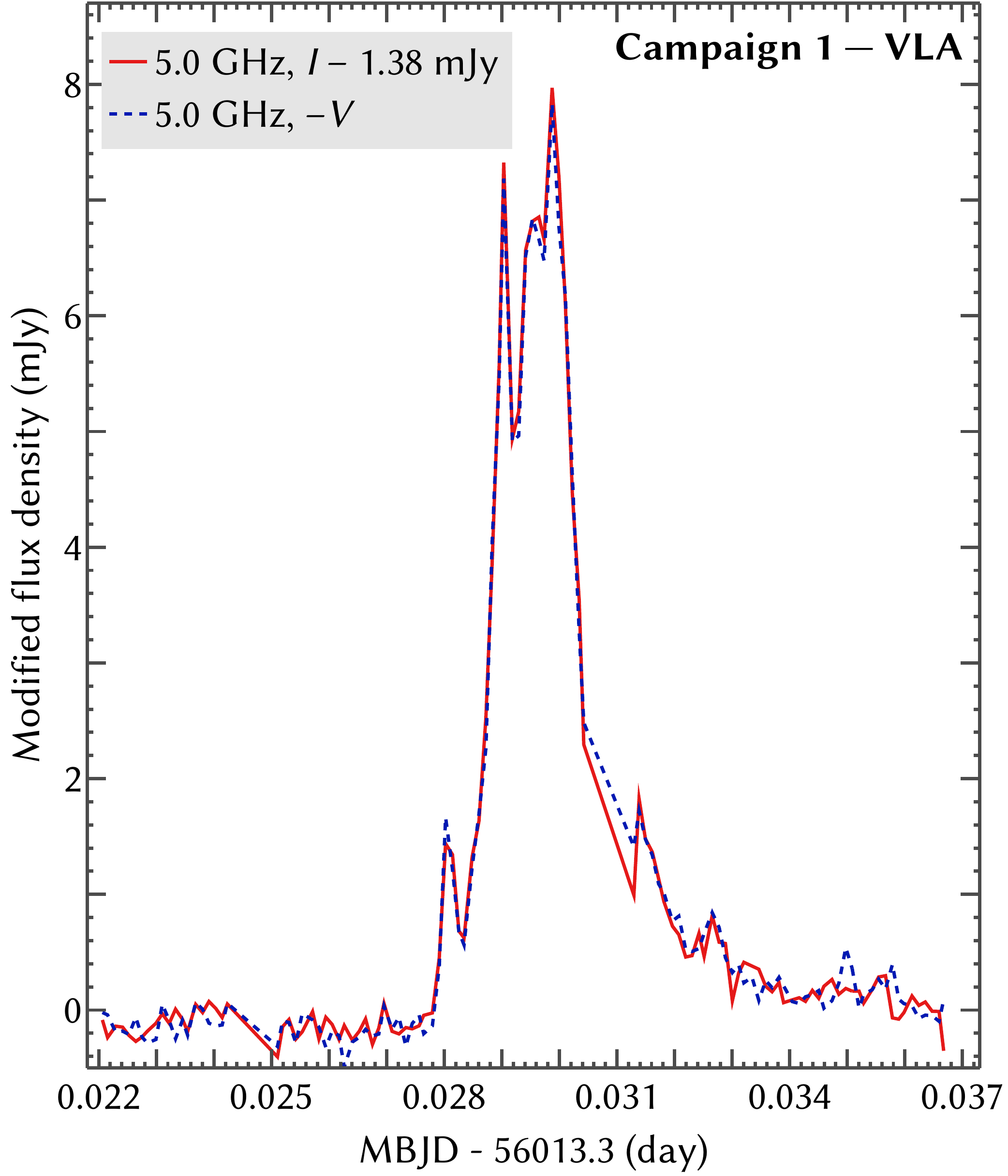}
\caption{Light curve of the radio flare shown in \fig{vladynspec} averaging
  across the lower (5.0~GHz) baseband. Both \stiv\ are shown. Because the
  flare is essentially 100\% LCP, $-V$ is plotted to maintain positivity. The
  nonflaring, unpolarized emission has been removed by subtracting
  $\mathop{\mathrm{median}}(I + V) = 1.38$~mJy from the \sti\ data. Errors are
  \apx0.1~mJy.}
\label{f.vlaflare}
\end{figure}

\begin{figure*}[tb]
\plotone{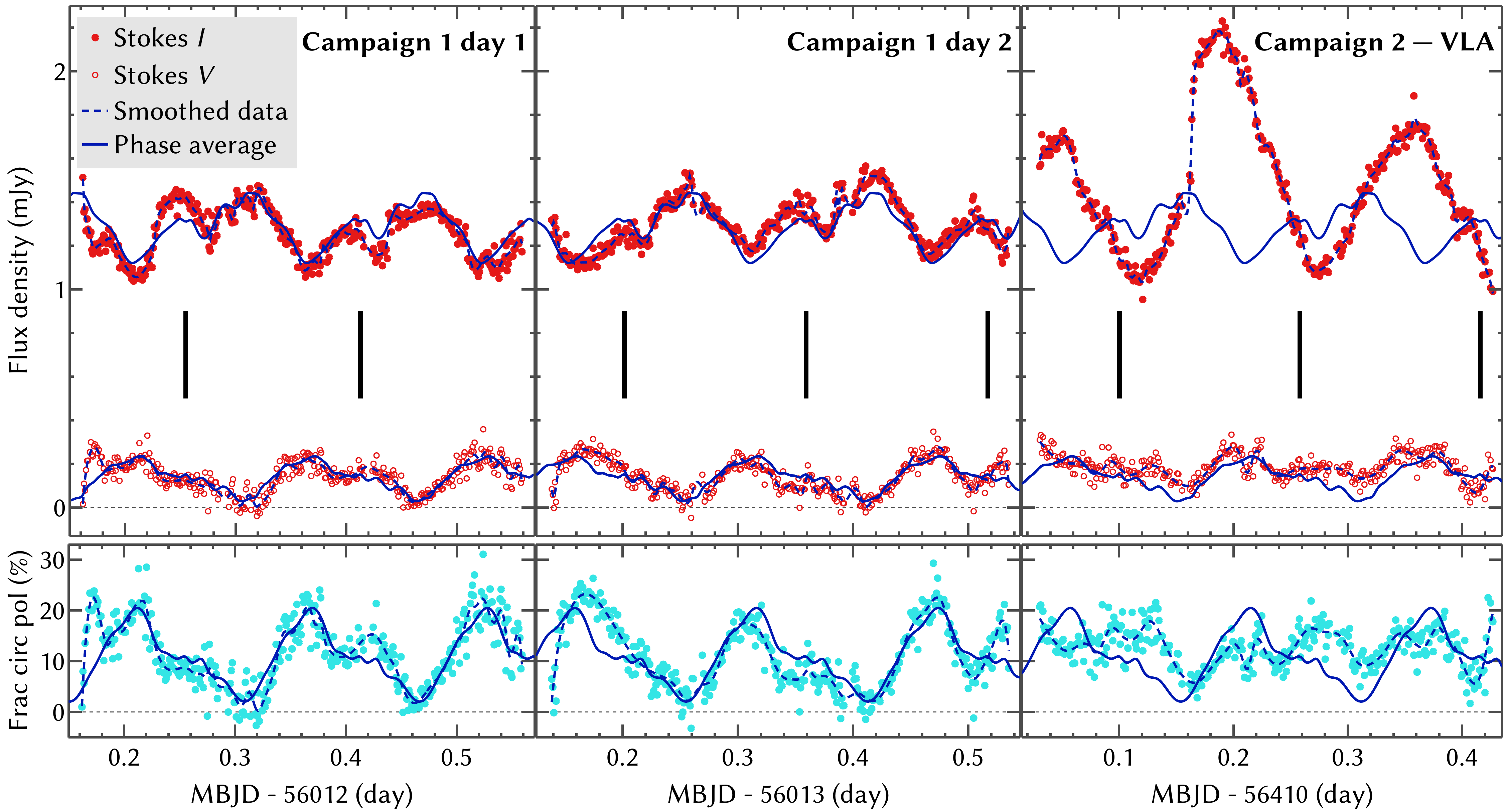}
\caption{VLA light curves with 100\% LCP flares removed. \sti, \stv, and $f_c$
  are shown. Vertical black lines indicate times of maximum flux in the
  primary MEarth periodic component. Dashed blue lines show smoothed spline
  fits to the data. Solid blue lines trace mean profiles of the Campaign~1
  data after phasing to the primary MEarth periodicity (see \sect{vlaanal}).
  While the Campaign~2 \sti\ data disagree with the Campaign~1 mean phase
  profile, the \stv\ data show broadly similar structure.}
\label{f.vlanofl}
\end{figure*}

The sum of two oscillations with frequencies $f_1$ and $f_2$ can be expressed
as the product of two oscillations with frequencies $(f_1 + f_2) / 2$ and
$(f_1 - f_2) / 2$. In this case the corresponding periods are $3.7491(1)$~hr
and $16.07(4)$~d. Detailed investigation of the raw data does not reveal any
evidence that the proximity of the long beat period to an integral number of
days is due to a systematic. This periodicity is not prominent in the MEarth
sampling pattern or any of the systematics parameters such as the CM ``common
mode'' term. Efforts to model the data with a 16~d modulation term do not
perform as well as the 3-sine fit (generally $\rcs \apx 4.5$) and leave a
persistent residual periodicity of \apx3.71~hr. We therefore conclude that the
data require two periods separated by $4.37(1)$ min.

\subsection{VLA}
\label{s.vlaanal}

As previously found by \citet{mbi+11}, the VLA photometry shows clear
variability with a period comparable to that seen in the optical bands. At
most times the emission has moderate RCP, with average values of $V = 0.14$
and $0.18$~mJy outside of flares in Campaigns~1 and 2, respectively. Unlike
the observations of \citet{mbi+11}, however, rapid (\apx5~min), 100\% left
circular polarized (LCP) flares are observed superposed on the steadier
emission in both Campaigns~1 and 2. Previous observations of the radio-active
ultracool dwarf \obj{tvlm} have revealed similar behavior: bright polarized
bursts are sometimes observed \citeeg{hbl+07} and sometimes excluded to high
significance \citeeg{had+06}. One such flare, at MBJD $\apx 56013.33$, reaches
a peak flux density of \apx8~mJy, which would have been clearly detectable in
the data of \citet{mbi+11}. We show this flare's dynamic spectrum in
\fig{vladynspec} and a zoom-in of its light curve in \fig{vlaflare}. The flare
is completely absent from the upper baseband, with a cutoff frequency of
\apx5.8~GHz suggested by extrapolation of the available data.

Such flares are generally interpreted as coherent radio emission
\citep{had+06,had+08,brpb+09,rw12} arising from the electron cyclotron maser
instability \citep[ECMI;][]{theecm,t06}. In the ECMI paradigm, emission cuts
off at approximately the cyclotron frequency $\nu_\text{cyc} = e B / 2 \pi m_e
c \approx 3 (B / 1 \text{ kG})$~GHz. Given the results shown in
\fig{vladynspec}, this suggests $B \apx 2.1$~kG, in line with measurements of
M~dwarfs made through observations of Zeeman broadening of the magnetically
sensitive FeH molecule \citep{rb06,rb07,rb10}. Taking a representative
bandwidth of $\Delta\nu = 5.8$~GHz and flux density of 7~mJy, the flare
luminosities reach $[L_{\text{R},f}/\Lb] \apx -6.1$, where $\lb = -2.36(9)$
\citet{sbh+14}. Using a duration of 3~min (0.002~d), the flare energy output
is $10^{27.4}$~erg.

To better understand the non-flaring emission, we subtracted the rapid LCP
flares from the data as follows. First, we manually identified flare events by
looking for abrupt simultaneous excursions in \stiv. For each flare, we
modeled the underlying \textit{non}-flaring emission in both \stiv\ with cubic
polynomials $I_m(t)$ and $V_m(t)$ fitted to the surrounding flare-free
measurements. For each in-flare measurement, we computed the flare intensity
as the weighted mean of $I-I_m$ and $V_m-V$, then subtracted this quantity to
obtain the non-flare residual. Although not required by our method, the
modeled flare intensity is always consistent with being nonnegative. We
investigated modeling of the flares with fractional circular polarizations
that were high, but not 100\%. The fits with 100\% LCP were qualitatively the
best, and are further justified by the near-perfect agreement between $I$ and
$-V$ seen in \fig{vlaflare}. In Campaign~1, the 100\% LCP flare duty cycle is
$35(5)$\%, while it is $20(5)$\% in Campaign~2. The event rates are
\apx0.9~hr$^{-1}$ and \apx0.5~hr$^{-1}$, respectively.

The non-flaring component of the radio data shows quasi-periodic variations.
We used phase dispersion minimization \citep[PDM;][]{the.pdm} to identify a
period of $3.787(1)$~hr in the Campaign~1 \sti\ data, consistent with the
primary MEarth periodicity, but different from the secondary periodicity at
\apx70$\sigma$. The PDM periodicity of the Campaign~2 \stv\ radio data is
$3.75(5)$~hr; because the observation covered only \apx2.6 rotations, the
precision is insufficient to make informative comparisons against the
Campaign~1 or the MEarth results.

\fig{vlanofl} shows the VLA light curves after removal of the 100\% LCP
flares, along with indicators showing the phasing of the data with regard to
the primary MEarth periodicity. We derived mean phase profiles for the
Campaign~1 data by separately fitting smoothed cubic splines to the \sti,
\stv, and $f_c$ measurements after phasing them to this periodicity.
\fig{vlanofl} shows both these mean profiles and the smoothed un-phased data.
Disagreements between these curves indicate deviations from purely periodic
variation. Disagreements are common even in the Campaign~1 \sti\ light curves:
\nltt's radio emission modulates periodically but varies stochastically as
well. We show the Campaign~1 mean phase profiles alongside the Campaign~2 data
to aid comparison despite the time gap between the observations and the clear
change in nature of the Campaign~2 \sti\ emission.

The non-flaring \stv\ radio emission in both campaigns is similar: it varies
periodically in a sawtoothed pattern with a rapid rise and slow decay. In both
campaigns the midpoint of the \stv\ decay phases with the maximum of the
primary MEarth component, while the phasing with regards to the secondary
MEarth component is not stable. There is a zero-point offset between the
emission in the two campaigns, and the amplitude of the Campaign~2 variation
is somewhat smaller: the smoothed flux density ranges between
\apx$-0.05$--0.36~mJy in Campaign~1 and \apx0.02--0.33~mJy in Campaign~2. The
\stv\ flux density and $f_c$ do not differ significantly between the two VLA
basebands at 5.0 and 7.1~GHz. The range of variation we observe is similar to
that found at 8.5~GHz by \citet{mbi+11}, although those authors were unable to
detect a periodicity in that emission. At 4.9~GHz \citet{mbi+11} found
periodic variation with $-0.3 \lesssim V \lesssim 0.2$~mJy, \ie\ extending to
significantly more net LCP than we observe. This difference may be due to the
presence of low-level LCP flares in their data comparable to the ones we have
removed, but the sensitivity of the pre-upgrade VLA is insufficient to
determine this; we note that these flares are more significant at lower
frequencies (Figures~\ref{f.vlaspw}, \ref{f.vladynspec}). In the non-flaring
data $0 \lesssim f_c \lesssim 25$\%, also consistent with the 8.5~GHz results
of \citet{mbi+11}.

The non-flaring \sti\ radio emission of Campaign~2, on the other hand, differs
substantially from that of Campaign~1. Although the minimal non-flaring
\sti\ flux densities in both campaigns are \apx1.0~mJy, the maximum in
Campaign~2 is \apx2.2~mJy, against \apx1.6~mJy in Campaign~1. While the
Campaign~1 emission shows a sawtoothed, double-humped light curve structure
that mirrors the \stv\ variation, in Campaign~2 the variation is more uniform.
As judged by the locations of the light curve minima, the phasing of the
\sti\ emission relative to MEarth shifts by \apx180\degr\ between the two
campaigns. \citet{mbi+11} measured flux densities somewhat lower than we do,
finding $0.8 \lesssim I \lesssim 1.5$~mJy and $0.8 \lesssim I \lesssim
1.3$~mJy at 4.9 and 8.5~GHz, respectively. The emission maxima (minima) in
that study are decreased (increased) compared to this work because of the
longer averaging interval used (25~m vs. \apx8~m). The modulation amplitudes
are therefore also attenuated. The observed amplitudes are consistent with our
Campaign~1 results but not those of Campaign~2.

Along with the phase shift relative to the optical emission, in Campaign~2 the
rising portions of the non-flaring \sti\ light curve are associated with a
rising spectrum, an effect not seen at any time in the Campaign~1 data
(\fig{vlaspw}). Meanwhile, as discussed, the Campaign~2 \stv\ light curve has
a shape and phasing consistent with that of Campaign~1. These facts suggest
the Campaign~2 radio emission may be the sum of a ``Campaign~1'' term and an
additional component with $f_c = 0$. The phasing and evolution of this
additional component imply that it arises from a distinct region and has a
lifetime of at least a few rotations. As shown in \fig{vlanofl}, however, the
Campaign~2 \sti\ minima fall below what would be expected from Campaign~1.
This could be due to the presence of an absorber, as discussed in
\sect{disc.nonflare}.

The general phenomenology of the Campaign~2 \sti\ emission (temporary increase
in flux density, rising spectrum) is suggestive of optically thick
gyrosynchrotron flares. We applied standard gyrosynchrotron models and
analysis \citep{d85,oha+05} to these periods in Campaign~2, assuming that the
observed emission was the sum of a spectrally flat oscillating component and a
rising-spectrum flare component. We find $B \apx 500$~G, a brightness
temperature $T_B \apx 10^9$~K, and a length scale of \apx1.7~$R_*$ for the
emitting region. These results are in line with previous studies of similar
objects \citep{b02,b06b,obwb02,oha+05,bmzb13} and earlier work on
\nltt\ \citep{mbi+11}. The length scale is compatible with spatially resolved
VLBI observations of M~dwarfs \citep{abg97,bcg98}. We derive a number density
of nonthermal ($E > 10$~keV) electrons $n_e \approx 6000$~\percc, with
substantial uncertainties because the location of gyrosynchrotron spectral
peak is weakly constrained. This value is consistent with the results of
\citet{oha+05} and several orders of magnitude smaller than typical estimates
of overall (thermal and nonthermal) cool star coronal electron densities,
generally found to be $10^8$--$10^{10}$~\percc\ \citep{nsb+02,ngs+04}.

The non-flaring \sti\ spectral luminosity in Campaign~1 ranges between $\slr =
14.52(5)$ and $14.70(5)$, where the uncertainties are due to the source
distance and thus are correlated between the two measurements. In Campaign~2
\slr\ ranges between $14.48(5)$ and $14.86(5)$. Taking $\slr \apx 14.6$ as a
representative value we find $\slrlb \apx -16.6$.

To summarize, the complex radio light curve of \nltt\ appears to combine
emission from at least three separate components: rapid 100\% LCP flares,
periodically modulated emission with moderate RCP, and an additional
unpolarized component seen only in Campaign~2. The complex morphology of the
moderate RCP component implies that it further represents the combined
emission of multiple regions or components. For instance, if the fact that the
total intensity never drops below \apx1.0~mJy is taken as evidence for the
presence of a constant component of that intensity, the periodic modulation in
Campaign~1 would represent the emission of a component with $I \lesssim
0.4$~mJy. Furthermore, the mathematics of Stokes parameters (\sect{vladata}:
$I \ge |V| \ge 0$ within each component) and anti-correlated variability in
$I$ and $V$ then imply that the constant term has $f_c \apx 15$\% while the
modulating term has $f_c \apx -100$\%.

\subsection{Chandra}
\label{s.chandranal}

\begin{figure}[tb]
\plotone{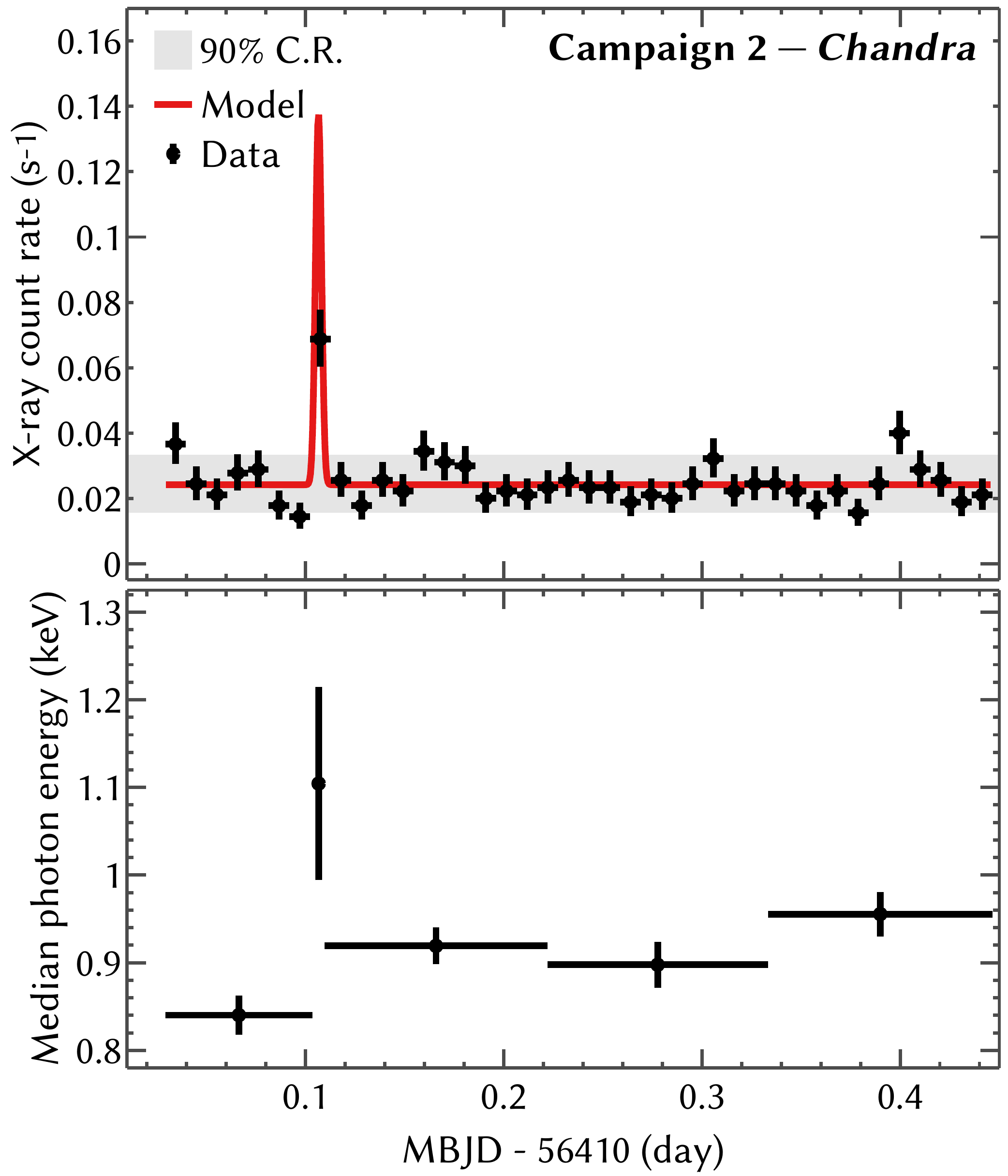}
\caption{\chandra\ observations in Campaign~2. Black points in the upper panel
  show the count rate in uniform 15-min bins. The red line shows an
  analytic model fit to the unbinned data (see \sect{chandranal}). The gray
  band shows the 90\% confidence region for samples of the quasi-quiescent
  emission; most of the apparent variability is not significant. The lower
  panel shows the change in the X-ray spectrum as quantified in the median
  photon energy \citep{hsg04}. The increase in the hardness of the
  quasi-quiescent emission may be a precursor to the long-duration flare
  detected by \swift.}
\label{f.xlc}
\end{figure}

Grouping the X-ray events into forty 15-min bins suggests a largely steady
source with at least one rapid flare (Figures~\ref{f.c2} and \ref{f.xlc}).
\fig{xlc} shows a 90\% confidence region for the binned count rates assuming a
steady source and Poisson statistics. The measurement at MBJD $\apx 56410.11$
is a $7.1\sigma$ outlier. The second-most outlying bin, at MBJD $\apx
56410.40$, contains 36~events and is a 2.9$\sigma$ outlier. In 39~measurements
there is a 7\% chance of obtaining at least one value that large.

\begin{figure}[tb]
\plotone{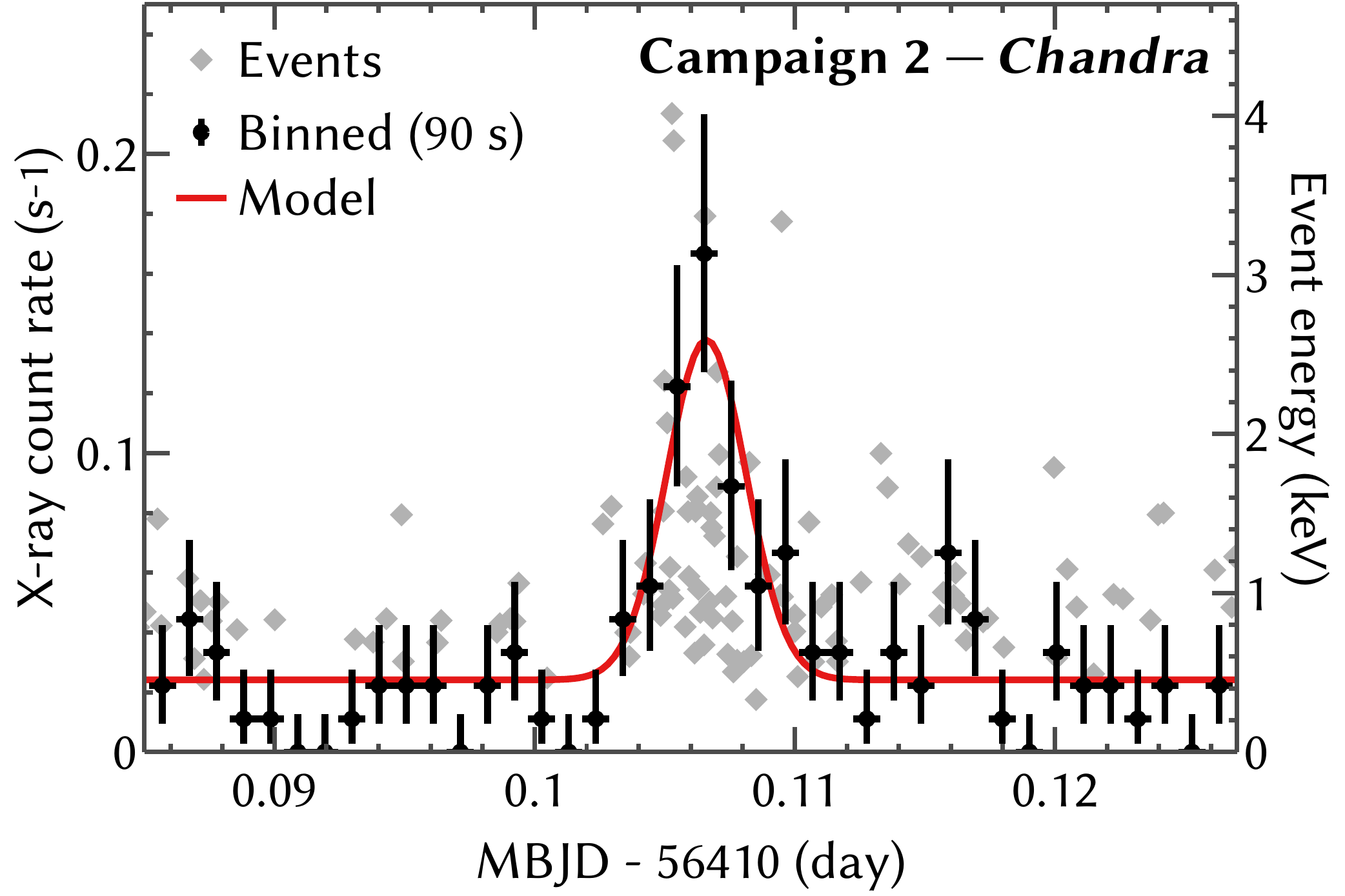}
\caption{Zoom-in of the rapid X-ray flare. Along with the model as in
  \fig{xlc}, individual events are shown with vertical positions indicating
  energy as shown on the \textit{right axis}. Binned and model count rates are
  associated with the \textit{left axis}. The binning interval is 90~s.}
\label{f.xlczoom}
\end{figure}

The red curve in the top panel of \fig{xlc} models the emission of \nltt\ as a
constant (``quasi-quiescent'') term plus a Gaussian flare. We determined the
parameters for this model using a maximum-likelihood technique considering the
arrival time of each observed photon. In particular, using a similar approach
as the Bayesian Blocks technique, we performed a one-dimensional Voronoi
tesselation of the photon arrival times and chose parameters to maximize the
likelihood function of detecting one photon in each bin, evaluating the
instantaneous rate function at each photon's observed arrival time. The fitted
peak flux of the flare (above the quasi-quiescent level) is 0.11 \cps. This is
significantly above the corresponding 15-min bin (\fig{xlc}) because the
modeled flare timescale is $\sigma = 2.2$~min. \fig{xlczoom} zooms in on the
flare, showing a finer (90~s) binning as well as the arrival times and
energies of the individual X-ray events.

The lower panel of \fig{xlc} shows the median photon energy in five time bins.
This quantity provides a more robust quantification of the spectral shape than
a traditional hardness ratio \citep{hsg04}. The second bin encompasses the
$\pm2\sigma$ region of the flare, while the final three bins are uniform in
size. Uncertainties on the medians are calculated using the method of
\citet{mj78}. The higher median photon energy in the flare bin is consistent
with the general finding of elevated temperatures during such events
\citeeg{rs05}. Strikingly, the median energy of the quasi-quiescent emission
increases over the course of the observation. This may be a precursor to the
slowly-evolving flare captured by \swift\ after the end of the
\chandra\ observations (\fig{c2}).

\begin{figure}[tb]
\plotone{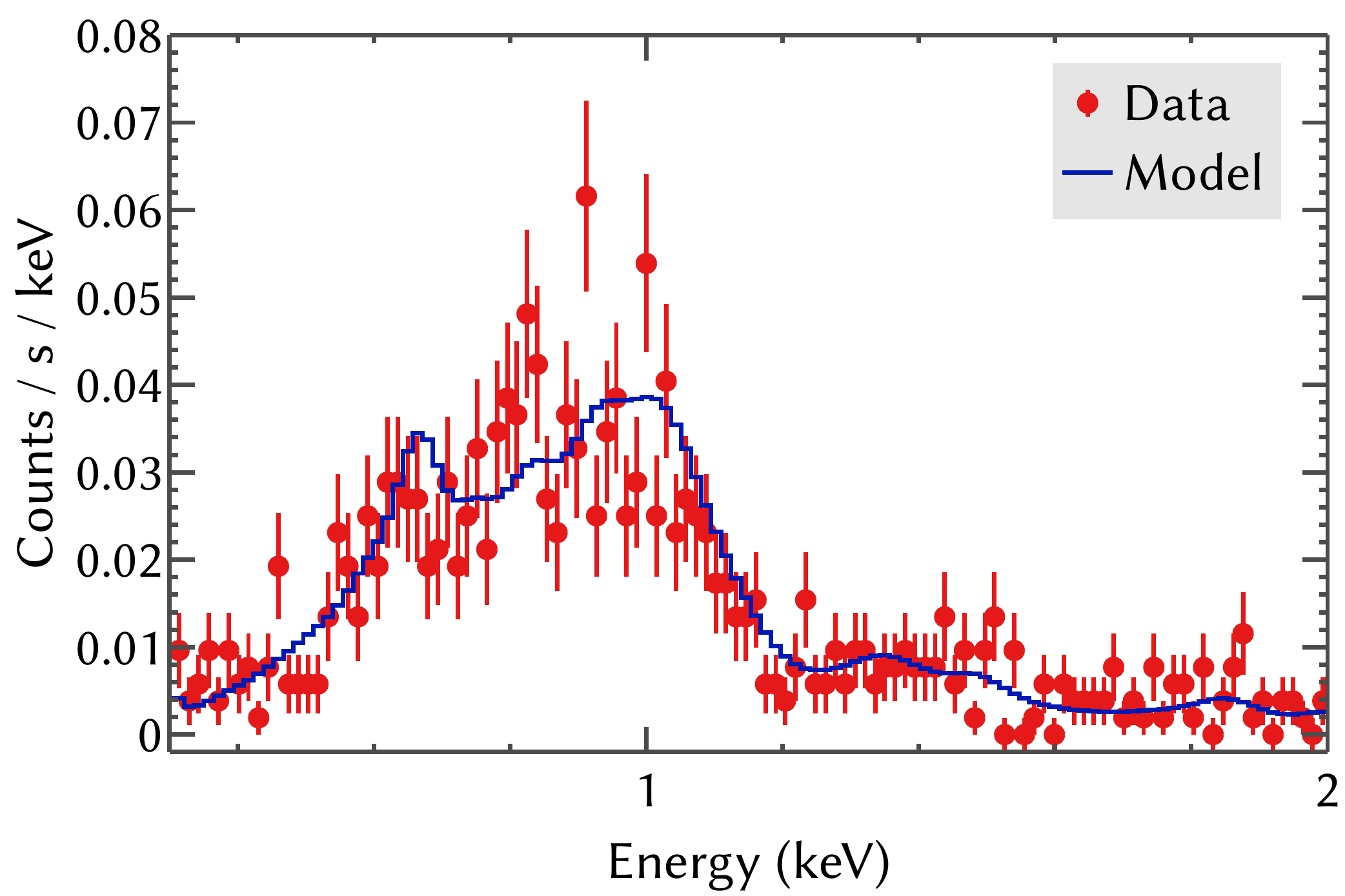}
\caption{\chandra\ X-ray spectrum and a fitted two-temperature solar abundance
  APEC model. The temperatures of the two components are $kT = 0.27(2)$~keV
  and $1.20(6)$~keV.}
\label{f.xspec}
\end{figure}

We modeled the overall X-ray spectrum of \nltt\ with \sherpa\ version~1
\citep{thesherpa}. We used the \sherpa\ implementation of the
\citet{thesimplex} simplex algorithm to optimize the $C$ statistic of
\citet{c79}. We did not group the data in energy or subtract the (negligible)
background. Photons with energies outside of the range 0.3--2.5~keV were
ignored. We used the solar abundances of \citet{l03}. A two-temperature
solar-abundance APEC \citep[Astrophysical Plasma Emission Code;][]{theapec}
model yields a satisfactory fit, achieving a reduced statistic $C_r = 1.27$
with 146 degrees of freedom. We show the data and best-fit model in
\fig{xspec}. The temperatures of the two components are $kT = 0.27(2)$~keV and
$1.20(6)$~keV, consistent with results seen in other active mid-to-late
M~dwarfs \citeeg{rs05,wcb14}. We note that, although there is evidence that
the rapid flare has a hotter spectrum than the quasi-quiescent emission, it
comprises \apx5\% of the total number of observed events, and thus does not
significantly affect the modeling. We have verified this by modeling only the
events from the quasi-quiescent time periods. \tbl{xfluxes} reports the mean
and peak X-ray fluxes based on our spectral modeling. The peak flux is derived
assuming the same energy conversion factor (ECF; ratio of energy flux to count
rate) as the mean emission, and thus may be slightly underestimated if the
flaring spectrum is indeed hotter than the mean emission.

% TableBuilder table
\begin{deluxetable}{lccr@{}l@{\,}l}
%custom preamble

%hardcoded preamble
\tablecolumns{6}
\tablewidth{0em}
\tablecaption{Results of \chandra\ Analysis\label{t.xfluxes}}
\tablehead{
\colhead{State} & \colhead{Integ. Time} & \colhead{Counts} & \multicolumn{3}{c}{[$f_X$]} \\
 & \colhead{(s)} &  & \multicolumn{3}{c}{[\cgsflux]} \\ \\
\multicolumn{1}{c}{(1)} & \multicolumn{1}{c}{(2)} & \multicolumn{1}{c}{(3)} & \multicolumn{3}{c}{(4)}
}
\startdata
Mean emission & $35588$ & $909$ & $-12$ & $.80$  & $\pm\,0.03$ \\
Flare peak & $529$ & $49$ & $-12$ & $.15$ & $\pm\,0.03$
\enddata
\tablecomments{Col. (4) is the X-ray flux in the 0.2--2~keV band. ``Flare
peak'' measures the modeled peak excess above the quasi-quiescent emission.}
\end{deluxetable}
% end TableBuilder table

We investigated alternative models for the X-ray spectrum. In particular,
high-S/N observations of active M~dwarfs often reveal non-solar elemental
abundances, usually in the form of an inverse first ionization potential
(IFIP) effect, in which elements with higher FIPs have elevated abundances
\citeeg{rs05}. No single-temperature model with adjustable elemental
abundances yields a superior fit to the data compared to the two-temperature
solar-abundance model. Two-temperature models with adjustable elemental
abundances show hints of an IFIP effect in \nltt, but the results are not
statistically significant.

We also searched for periodicity in the non-flaring X-ray emission. As may be
judged from \fig{xlc}, any periodicity in the data must be weak, and requires
statistical analysis to uncover. Unfortunately, the small number of observated
rotations (\apx2.6) means that statistical approaches have low sensitivity in
this dataset. Calculations of the Rayleigh test statistic \citep{klw02} showed
that it simply correlated with the candidate rotation period in the parameter
region of interest. The same outcome was found with the Kuiper $V$ statistic
\citep{k60}, a variant of the Kolmogorov-Smirnov test more suited for data on
a circle. The available data do not provide significant evidence for
periodicity in the X-ray emission.

We find $\lxq = 27.71(5)$ and $\lxf = 28.44(5)$, where the $q$ and $f$
subscripts denote quiescent and flaring states. (Here we consider the rapid
X-ray flare observed by \chandra, and not the more poorly-characterized event
in the \swift\ data; both appear to have similar luminosities.) Unlike what is
presented in \tbl{xfluxes}, the flaring luminosity used here includes the
contribution from the quiescent emission, for consistency with other studies.
We find $\lxqlb = -3.5(1)$ and $\lxflb = -2.8(1)$, comparable to the saturated
X-ray emission seen in early-M~dwarfs.

The total energy output of the flare is \apx$10^{30.9}$~erg. No counterpart is
apparent in the radio data, while the MEarth and \swift\ data do not overlap
this event. There is a suggestion of a rapid decrease in optical luminosity
just after the flare in the MEarth photometry, but these points were observed
in partially cloudy conditions and are not reliable enough to be conclusive. A
similar rapid X-ray flare, with a timescale $\tau \lesssim 8$~min, was
observed in the dM4.5e dwarf \object{EV~Lac} \citep{oha+05}, and some X-ray
flares initially evolve on a similarly rapid timescale before shifting to
slower decay \citep[e.g., as seen on \object{LP 412-31};][]{ssml06}. Assuming
$kT = 1.2$~keV and thermal bremsstrahlung emission, the emission measure
($\text{EM} = \int\!\dd V n_e^2$) corresponding to the peak luminosity is
\apx$6 \times 10^{51}$~\percc, in line with previous observations of flare
stars \citeeg{sl02,rs05,ssml06}.

\subsection{\swift\ XRT}

The simultaneous \swift\ and \chandra\ data generally agree, with the
exception of the \swift\ flux measurement at MBJD $\apx 56410.4$, which is
substantially below that of \chandra. In particular, only one source event is
detected in 380~s of observing. Its energy is 1.03~keV. During the precise
\swift\ good-time interval ($0.4005 < \text{MBJD} - 56410 < 0.4049$) the
\chandra\ count rate is consistent with the wider bin shown in \fig{xlc}, as
are the photon energies. We suspect an unidentified instrumental phenomenon
during this \swift\ observation.

% TableBuilder table
\begin{deluxetable*}{lllccccc}
%custom preamble

%hardcoded preamble
\tablecolumns{8}
\tablewidth{0em}
\tablecaption{Summary of emission characteristics.\label{t.summary}}
\tablehead{
\colhead{Mode} & \colhead{Quantity} & \colhead{Units} & \colhead{Radio} & \colhead{Optical} & \colhead{H$\alpha$} & \colhead{UV} & \colhead{X-ray} \\ \\
\multicolumn{1}{c}{(1)} & \multicolumn{1}{c}{(2)} & \multicolumn{1}{c}{(3)} & \multicolumn{1}{c}{(4)} & \multicolumn{1}{c}{(5)} & \multicolumn{1}{c}{(6)} & \multicolumn{1}{c}{(7)} & \multicolumn{1}{c}{(8)}
}
\startdata
Non-flaring & $[L]$ & \cgslum & $25.2$ & $29.5$ & $27.9$ & $27.7$ & $27.7$ \\
 & $[L/\Lb]$ & --- & $-6.0$ & $-1.7$ & $-3.3$ & $-3.6$ & $-3.5$ \\
 & $[F]$ & \cgsflux & $4.0$ & $8.3$ & $6.7$ & $6.5$ & $6.5$ \\
 & Modulation amplitude & --- & 0.15--0.30 & 0.02--0.03 & 0.06 & $\approx$0.08 & $<$0.4 \\
\\Flaring & $L_\text{f}/L_\text{q}$ & --- & 6 & $<$0.05 & 2.3 & 2.7 & 6 \\
 & $[L_\text{f}/\Lb]$ & --- & $-5.3$ & $<$$-3$ & $-3.2$ & $-3.3$ & $-2.8$ \\
 & Duty cycle & --- & 0.20--0.35 & \multicolumn{1}{c}{---} & $\approx$0.2 & $\approx$0.2 & $\approx$0.2 \\
 & Timescale & s & $10^2$--$10^3$ & \multicolumn{1}{c}{---} & $>$$90$ & $10^4$ & $10^4$ \\
 & $[E]$ & erg & $27.4$ & \multicolumn{1}{c}{---} & $>$$30.2$ & $32.1$ & $32.3$
\enddata
\tablecomments{$F$ is the surface flux assuming $R_* = 1.6$~\rj.
    $L_\text{f}/L_\text{q}$ is the ratio of flaring to quiescent luminosity.
    $E$ is a characteristic flare energy release. See \sect{sedsummary} for
    details.}
\end{deluxetable*}
% end TableBuilder table
 % shows up much farther downstream ...

The two \swift\ observations occurring past the end of the VLA and
\chandra\ monitoring suggest a substantial, slowly-evolving (timescale
\apx3~hr) flare. The median photon energies in these bins are $1.0(2)$ and
$1.3(3)$~keV, suggesting a continuation of the hardening trend seen in the
bottom panel of \fig{xlc}. The observed X-ray luminosity in this event is
compatible with that observed in the rapid \chandra\ X-ray flare, but the
overall energy output would be larger by $\approx$2 orders of magnitude if the
apparent timescale is accurate.

\subsection{\swift\ UVOT}

Similar to the XRT data, the UVOT measurements past the end of the
simultaneous \chandra/VLA monitoring suggest a slowly-evolving flare. Assuming
a UVW1 filter bandwidth of 795~\AA\ and pivot wavelength of
2517~\AA\ \citep{blh+11}, the flare luminosity peaks at $\lu \sim 28.0$. It is
tempting to infer a slow rise in UV luminosity in the period $56410.3 \lesssim
\text{MBJD} \lesssim 56410.5$, but we caution that the sparse sampling may
mask more complicated variation. There is a \apx30\% decline in the UV flux
between the first and second samples of the flare light curve (MBJD
\apx\ 56410.48 and 56410.62, respectively). At this time, the X-ray light
curve appears flat, although the measurement uncertainty is nonnegligible.
X-ray and UV variability often track each other fairly closely, with
inter-band delays small compared to the timescales probed here
\citep{obwb02,oba+04,mkhg+05,bbg+08}. The luminosity ratio between the two
bands in this event is consistent with scaling relations found in simultaneous
observations of flare stars \citep{mkhg+05}.

The mean non-flaring luminosity is $\lu = 27.67(4)$, or $\lulb = -3.6(1)$.
Assuming a radius of $1.6(1)$~\rj\ \citep{sbh+14}, the surface flux is
$[F_\text{UVW1}] = 3.55(7)$ (erg~s$^{-1}$~cm$^{-2}$~\AA$^{-1}$), lying between
typical values obtained for main-sequence and classical T~Tauri stars
\citep{jkvl00} in a study performed at somewhat shorter wavelengths (1958~\AA)
and higher \teff\ (3500--4500~K). While the non-flaring UVOT data are
consistent with a constant flux ($\rcs = 0.74$), their variations are
suggestive of an oscillation with a periodicity similar to that of the MEarth
data. Assuming a periodicity fixed to that of the primary MEarth component
(3.7859~hr), we fit a sine curve to the UVOT data. The oscillation amplitude
is $8(7)$\%, the phase relative to the primary MEarth component is
$180(80)$~deg, and $\rcs = 0.43$ (2 degrees of freedom).

\subsection{MMT}

The sparse time sampling of the MMT observations preclude a detailed analysis
of the time variability of \ewha. We interpret the first measurement as a
flare and the subsequent four as quiescent activity. In quiescence, $\ewha =
14.1(7)$~\AA, an unusually high value compared to typical mid-to-late M~dwarfs
\citeeg{lbk10}. This is consistent with the average value of $\ewha =
14.6$~\AA\ reported by \citet{mbi+11}. \citet{ltsr09} and \citet{sbh+14} find
$\ewha = 54.1$ and \apx50~\AA, respectively, in single observations,
suggesting frequent flares of luminosities a factor of \apx2 above what is
contained in our observations.

We derive \Lh\ from \ewha\ using a ``$\chi$ factor'' approach, using the
relation \citep{the.chifactor}
\begin{equation}
\frac{\Lh}{\Lb} = \chi \frac{\ewha}{1\text{ \AA}}.
\end{equation}
\citet{rb08} determine $[\chi]$ as a polynomial function of \teff\ in the
mid-to-late M~dwarf regime. Taking $\teff = 3150(500)$~K, we find $[\chi] =
-4.4^{+0.2}_{-0.7}$. This is compatible with the value found using the
polynomial fit to $\chi$ as a function of $V-I$ color given by \citet{wh08}.
It does not agree well with the value of $\chi$ reported by the latter authors
for objects of spectral type M7 ($\chi \approx -5.3$); however, there is large
scatter to their relation precisely at this spectral type, and it coincides
with an inflection in their $\chi$-vs-SpT data. We therefore prefer the
\citet{rb08} computation. We find in quiescence that $\lhlb =
-3.3^{+0.2}_{-0.7}$ and that $\lh = 27.9^{+0.2}_{-0.7}$. As may be expected
from the large quiescent \ewha, the derived value of \lhlb\ is also unusually
large compared to most M7 dwarfs, which typically have $\lhlb = -4.7$
\citep{whw+04,wmb+11}. \citet{mbi+11} find a lower but compatible value of
$\lhlb = -3.8$ for the quiescent emission from equivalent values of \ewha,
indicating different values for $\chi$ and/or \Lb.

Previous studies suggest that the \ha\ flare timescale is likely
\apx10--100~min \citep{hwhk10,lbk10}. The observed magnitude of the flare can
be assessed as $\max(\ewha)/\min(\ewha) \approx 2.3$, consistent with
observations of flare stars \citep{lbk10}. There is no clear correlation with
any variations in the radio band. While a 100\% LCP flare occurs \apx20~min
before the high \ha\ measurement, the radio events are sufficiently frequent
that there is plausibly happenstance. It is somewhat surprising that the
\ha\ event appears to take place around the maximum of the optical light
curve: in a model in which the optical light curve is modulated by dark spots,
presumed to be associated with enhanced magnetic activity, \ha\ would be
expected to be correlated with optical minima, a trend that has been observed
\citep{fffmc00}.

\begin{figure}[tb]
\plotone{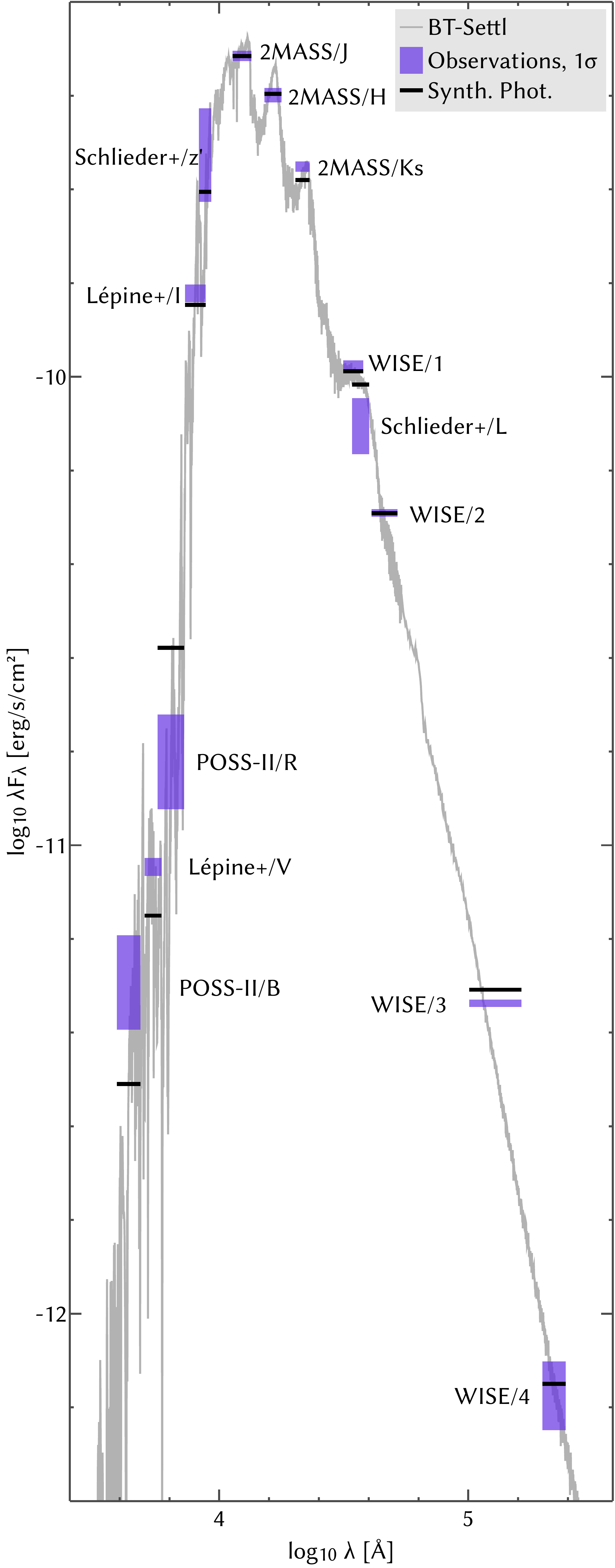}
\caption{Non-simultaneous, blended SED of \nltt\ in the optical/IR (OIR)
  bands. Violet bands show 1$\sigma$ confidence regions for data from
  \citet{ltsr09}, \citet{sbh+14}, 2MASS \citep{the2mass}, WISE
  \citep{thewise}, and the Second Palomar Observatory Sky Survey (POSS-II) as
  digitized in the USNO-B1.0 catalog \citep{theusnob}. Black lines show
  synthetic photometry computed from the BT-Settl model shown (see
  \sect{sedsummary} for details).}
\label{f.oir-sed}
\end{figure}

\begin{figure*}[tb]
\plotone{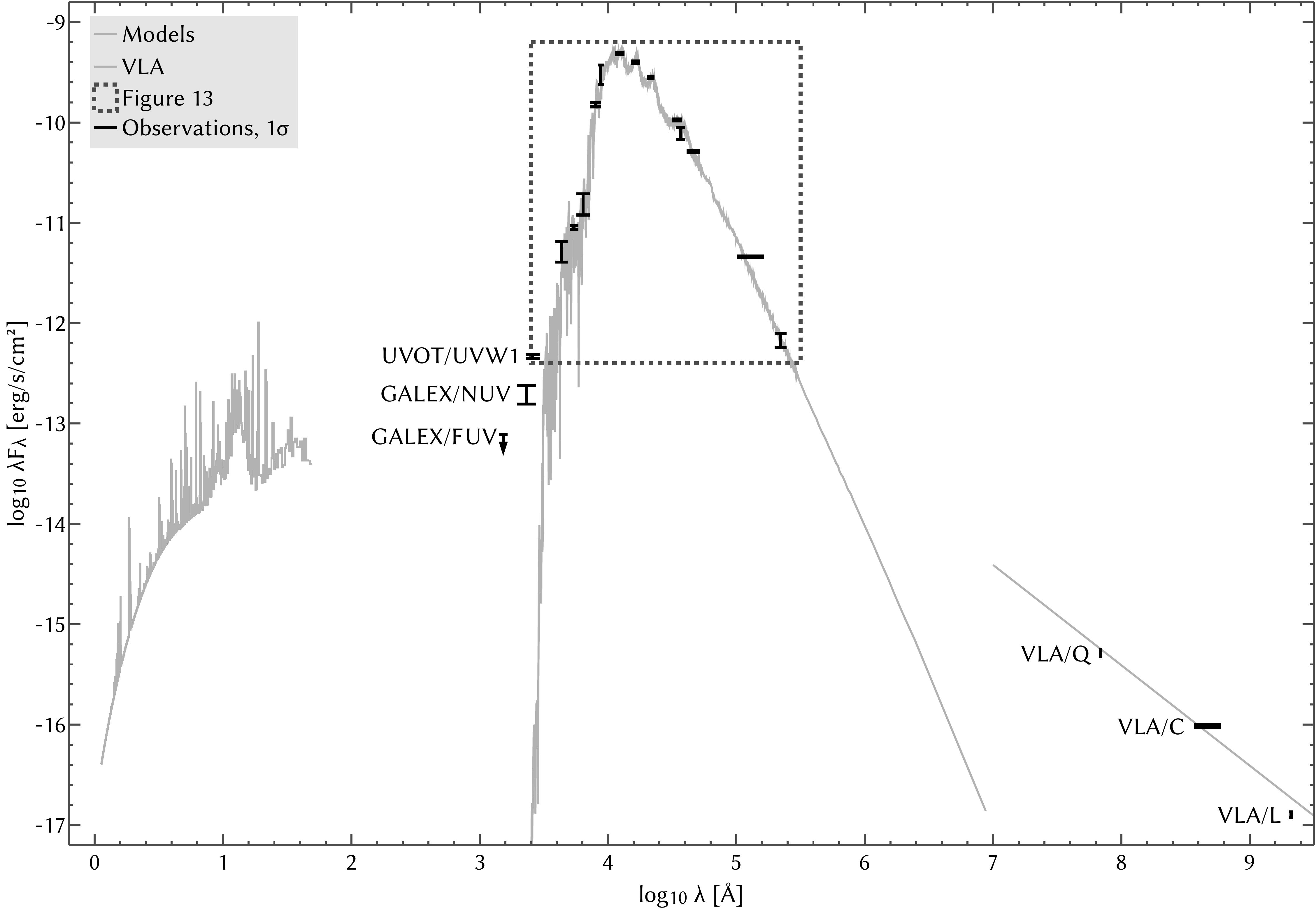}
\caption{The ultra-wideband, non-simultaneous, blended SED of \nltt. The
  dashed box identifies the region shown in \fig{oir-sed}, in which the
  enclosed observational data are identified. The short-wavelength model curve
  shows the best-fitting two-temperature APEC source model fit as described in
  \sect{chandranal}; the observed X-ray data cannot be displayed in flux
  density units because the incident photon energies are only known in a
  probabilistic sense. The middle model curve is the best-fit BT-Settl model
  as described in \sect{sedsummary}. The long-wavelength model curve is a
  constant flux density of 1.3~mJy. Also shown are measurements from GALEX
  \citep{thegalex}, \citet[``VLA/L'']{mbi+11}, and this work (see
  \sect{sedsummary} for details).}
\label{f.full-sed}
\end{figure*}

The non-flaring \ha\ data suggest an oscillation with a periodicity similar to
that of the MEarth data. Here, a fit at the secondary MEarth periodicity gives
a better $\rcs = 4.74$ (1 degree of freedom) compared to that at the primary
periodicity, $\rcs = 6.27$. We find an amplitude of $5.9(7)$\% and a phase
relative to the secondary MEarth component of $120(20)$~deg. The relatively
large values of $\rcs$ and small parameter errors are driven by the small
($0.15$~\AA) statistical error bars on the measurements of \ewha.

\section{Summary of the Phenomenology}
\label{s.summary}

\subsection{Spectral Energy Distribution}
\label{s.sedsummary}

\tbl{summary} summarizes several key parameters regarding the emission of
\nltt\ in the radio, broadband optical, \ha, UV, and X-ray bands. The optical
luminosity is computed for the $R$ band assuming the $k$-corrections of
\citet{br07}. The non-flaring radio parameters are derived for the
\sti\ emission and the flare parameters refer to the rapid 100\% LCP events.
The UV and X-ray flare parameters refer to the slowly-evolving event detected
by \swift. The parameters for the flares outside of the radio band are
uncertain because only portions of single events were observed.

\nltt\ is the brightest UCD in terms of radio flux density \citep[with the
  nearest rival being \object{LSPM J1835+3259} at $525(15)$~\ujy\ in
  quiescence;][]{b06b} and radio spectral luminosity \citep[significantly
  outshining the next most luminous source, \object{2MASS J05181131-3101529}
  at $\slr = 13.9$;][]{mbr12}. It is also one of the brightest UCDs in X-rays,
whether quantified in terms of flux, luminosity, or bolometrically-normalized
luminosity. Only the serendipitously-discovered object
\object{2XMM~J043527.2-144301} may be more X-ray luminous; however, this
object is not well-studied, and in particular has an uncertain
spectrophotometric distance of $67(13)$~pc \citep{ggw11}. In quiescence,
$\Lh \apx \Lu \apx \Lx$. The (non-contemporaneous) quiescent luminosities
between the bands fall on the scaling relations determined by \citet{smm+13}.

Figures~\ref{f.oir-sed} and \ref{f.full-sed} show the SED of the blended
components of \nltt. \fig{oir-sed} shows details in the optical/infrared (OIR)
bands. Included in these plots is a representative BT-Settl model photospheric
spectrum \citep{thebtsettl} with $\teff = 2700$~K, $\log g = 4.5$, and solar
metallicity, computed with the \textsf{CIFIST2011} version of the code, which
uses the solar abundances of \citet{cls+11}. The parameters of this model
match those used by \citet{sbh+14} to model the individual components of the
binary, except for \teff, for which we find that a lower value provides a
better match on the blue end of the OIR SED. The value used here is within
1$\sigma$ of the values found by \citet{sbh+14}, which are $3200(500)$ and
$3100(500)$~K for the A and B components, respectively. The normalization of
this model was set by weighted least-squares optimization of the synthetic
photometry against the measurements shown in \fig{oir-sed}. The reduced
$\chi^2$ of this fit is 5.29; similar fits with other values of \teff, stepped
in 100~K increments, yield inferior results ($\rcs = 18.74$, $10.25$ for
$\teff = 2600$, $2800$~K). We emphasize that this model should not be expected
to precisely match the observations because the data considered here blend the
emission of two distinct objects.

\fig{full-sed} places the OIR photometry in a broader context using our radio
and X-ray observations as well as archival photometry from GALEX
\citep{thegalex}. Magnetic phenomena dominate the photospheric contribution
outside of the comparatively narrow OIR window. The radio component has a
strikingly broad and flat spectrum (in terms of flux density), and our data
show only suggestions of a turnover in the spectrum around \apx40~GHz
(\apx$10^8$~\AA). The radio component is nonetheless energetically
insignificant compared to the X-ray component, as is generally the case in
comparable systems \citep{gb93,wcb14}.

\subsection{Variability}
\label{s.varsummary}

Our data show flares and/or non-flaring modulation in every band. They are
consistent with findings from other detailed multiwavelength studies of cool
flare stars: there is striking variation in the flaring phenomenology, both
within the same band and in multi-band correlations \citep{oba+04,oha+05}. It
is clear that there is no one single observational manifestation of flaring
activity, and the underlying physical phenomena are plausibly as variegated as
their resulting emission.

The only data in which we do \textit{not} observe flares are the broadband
optical MEarth observations. This is not surprising: white-light flares are
expected to be relatively blue, while both the MEarth filter and the
photospheric emission of \nltt\ are red. Using the model spectrum of
\citet{dbk+12}, we find that such flares would need to reach \apx4\% of
\Lb\ to be detectable.

\begin{figure}[tb]
\plotone{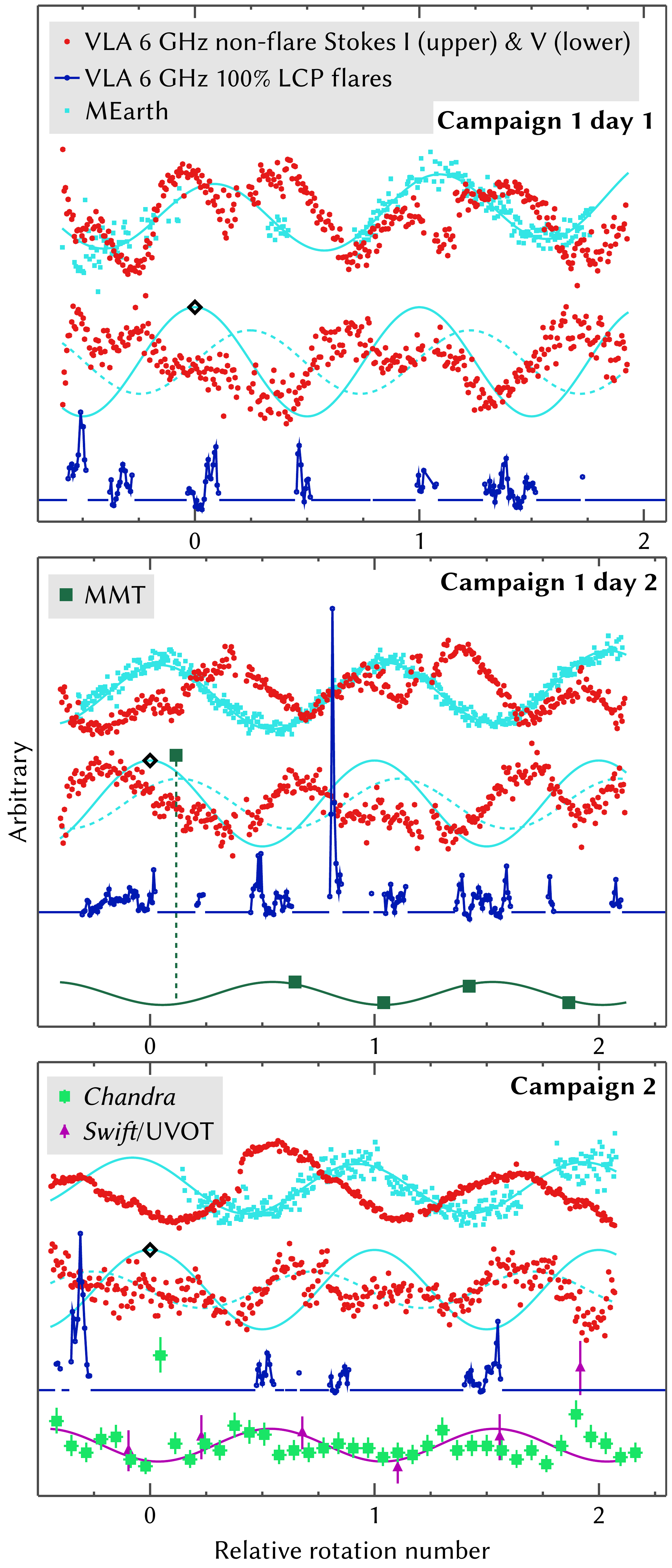}
\caption{Relative phasing of various emission components in the three nights
  of intensive observations. Vertical scalings are chosen to emphasize the
  variation within each band. The horizontal scale for each panel counts
  rotations at a period of 3.7859~hr relative to the black diamonds. Within
  each panel, the upper red and blue light curves trace non-flaring VLA
  \sti\ and MEarth emission, respectively, while the lower curves trace
  non-flaring VLA \stv\ and the MEarth data decomposed into two components
  (see \sect{mearthanal}). Dark blue shows 100\% LCP VLA flares. Dark green in
  the middle panel shows \ewha\ (MMT). In the lower panel, green and purple
  indicate X-ray (\chandra) and UV (\swift) light curves, respectively. See
  \sect{anal} for details on the modeling.}
\label{f.arbscale}
\end{figure}

\fig{arbscale} presents all of the light curves in compact form, with vertical
scalings chosen to emphasize the nature and relative phasing of the
variability in each band. The MEarth data are shown in light blue. The upper
light curves show the data and full emission model, while the lower curves
plot the primary and secondary sine components. The rapid, 100\% LCP radio
flares are shown in dark blue. Double-peaked LCP flares are seen during all 7
observed minima of the primary MEarth component. These flares may also
preferentially occur around the midpoints of the rising portions of both the
primary and secondary MEarth components.

\fig{arbscale} shows the non-flaring radio data in red. The upper (lower)
light curves show \sti\ (\stv) averaged across the two basebands. In
Campaign~1, the \sti\ variations mirror those in \stv\ to a good
approximation. In Campaign~2, the \sti\ variations are of about twice the
amplitude, and the phasing of the light curve minima is shifted by
\apx180\degr\ relative to the primary MEarth component, which we interpret as
being due to the presence of an additional emission and absorption component
(\sect{vlaanal}). The Campaign~2 \sti\ variations are approximately in phase
with the secondary MEarth component.

The middle panel of \fig{arbscale} shows the \ewha\ data in dark green. The
\ha\ flare occurs around the time of a maximum in the MEarth light curve,
although the sparse sampling makes timing analysis difficult. At the time of
the \ha\ flare there is a noticeable dip in the \sti\ radio emission, but no
100\% LCP flaring component. \ewha\ modulates \apx180\degr\ out of phase with
the optical emission.

The lower panel of \fig{arbscale} shows the X-ray data in light green. There
is no evidence for periodic variations in the \chandra\ data. A rapid X-ray
flare occurs around the predicted time of a maximum in the primary MEarth
light curve, although no MEarth data were obtained at that time. The X-ray
flare occurs \apx$1/4$ of a rotation after the second-largest radio flare in
the data set. The UV data are in purple; the last measurement appears to be
associated with the rising phase of the slowly-evolving X-ray/UV flare
(\fig{c2}). The best sinusoidal fit to the UV data is approximately
180\degr\ out of phase with the primary MEarth component.

\section{Discussion}
\label{s.disc}

\subsection{Interpreting the MEarth Periodicities}
\label{s.oneortwo}

We find two similar but distinct periodicities in the MEarth light curves: a
primary component at $3.7859(1)$~hr with amplitude \apx14~mmag and a secondary
at $3.7130(2)$~hr with amplitude \apx7~mmag. Modeling of the individual
campaigns indicates that the periods are stable to levels of $\lesssim0.5$\%
across the 3.4-yr time baseline of the observations. Below we discuss what we
consider to be the most plausible explanations for the data: either there are
multiple spots on one object with differential rotation (DR), or both
components of the binary rotate with very similar periodicities.

\subsubsection{Differential rotation}

The similarity of the periodicities might be taken to suggest the presence of
two spots on one differentially-rotating photosphere. The implied level of DR,
however, is not expected to occur in stars in this mass range. The inferred
absolute shear is $\Delta \Omega = 2\pi/P_1 - 2\pi/P_2 \gtrsim
0.8$~rad~d$^{-1}$ and the relative shear is $\alpha = (P_1 - P_2) / P_1
\gtrsim 0.02$, where both quantities are lower bounds because they are
generally expressed in terms of pole-to-equator variation, and latitudes of
the putative spots are unknown. These values are well within bounds for
\textit{partially} convective rapid rotators \citep[$\teff \gtrsim
  6000$~K;][]{rrb13}, but not fully convective ones. In the coolest stars in
the sample of \citet{rrb13}, $\Delta\Omega \lesssim 0.1$~rad~d$^{-1}$. Zeeman
Doppler imaging (ZDI) observations of mid-M~dwarfs find very low levels of DR
in moderately rapid rotators ($8 \lesssim P \lesssim 100$~hr), with
$\Delta\Omega \lesssim 0.01$~rad~d$^{-1}$ \citep{mdf+08,mdp+08,dmp+08}. In
\object{V374 Peg} (M4), $\alpha \sim 0.0005$ \citep{mdf+08}. These findings
are supported by some magnetohydrodynamic simulations that find that Maxwell
stresses virtually arrest DR in fully-convective objects \citep{b08}.

A single night of spatially resolved photometric monitoring of the binary
could determine if the optical variability is confined to one object. If so,
the inferred presence of DR would be extremely surprising for such a cool
dwarf. We speculate that in this scenario some cool dwarfs may indeed have
high DR, and that this may lead to large radio luminosities; in this case, the
discovery of unusual DR in \nltt\ would be a consequence of its selection in
the radio survey of \citet{mbi+11,mbr12}.

\subsubsection{Similar rotation periods}

Alternatively, it is possible that the two periodicities correspond to
separate signals from the two components of the binary. Although the
similarity of the rotation periods seems implausible, magnetic coupling to the
protostellar disk might synchronize the components' rotation \citep[and
  references therein]{k91,esh+93,s13f}, and subsequent spin-up due to
contraction \citep[which dominates the rotational evolution of a system this
  age;][]{rm12,gb13} would only alter the period ratio by \apx5\%
\citep[assuming the evolutionary models of][]{bcah98}. However, resolved
measurements of \vsi\ in low-mass binaries suggest that \apx50\% of them have
significantly different projected rotational velocities, and it is improbable
that the differences are entirely due to the $\sin i$ term \citep{kgf+12}.

If both components of \nltt\ truly have similar masses, ages, and rotation
rates, one might expect them to have similar levels of magnetic activity.
However, VLBI observations indicate that the system's steady radio emission is
dominated by only one of the components, with the nondetected component being
$\gtrsim$5 times fainter than the detected one \citep{mbi+11}. \nltt\ may
therefore be an excellent laboratory for understanding the apparent bimodality
in the radio and X-ray emission levels of otherwise-similar UCDs reported by
several authors \citep{mbr12,sab+12,wcb14,cwb14}, which may stem from a
bistability in the underlying dynamo \citep{mdp+10,gmd+13}\footnote{Another
  reference point is the binary \object{BL/UV Cet} (dM5.5e+dM5.5e; often
  referred to as UV~Cet~AB), in which the components have indistinguishable
  projected rotational velocities \citep[$\vsi = 31.5(30),
    29.5(30)$~\kms;][]{jpv+05} but the secondary \object{UV~Cet} is far more
  active than the primary \citeeg{ags03}. VLBI observations detected spatially
  extended gyrosynchrotron emission from \object{UV~Cet} but not
  \object{BL~Cet} \citep{bcg98}. Intriguingly, in the same observation
  \object{BL~Cet} was seen to emit rapid polarized pulses, while
  \object{UV~Cet} was not.}. Because \nltt\ is a visual binary and its
components have such similar masses and (possibly) rotation rates, many
potential confounding factors in the effort to understand the radio/X-ray
bimodality could be eliminated.

\subsection{Non-flaring variability}
\label{s.disc.nonflare}

The data presented in this work are consistent with the model proposed by
\citet{mbi+11}, in which the stellar magnetic field is dominated by a dipole
that is misaligned with the rotation axis. While \citet{mbi+11} were unable to
phase-align their optical and radio data, the Campaign~2 results suggesting
anti-phased radio and optical maxima would be consistent with the regions of
maximal field strength hosting cool spots. Extrapolation of our MEarth model
to the VLA data of \citet{mbi+11} using the Campaign~0 phasing suggests that a
phase difference of 100\degr\ between the two modulations, although the
\apx330-day gap between the observations is comparable to the time scale on
which the modulation phase drifts in our data.

The model of a magnetized cool spot is supported by the fact that the
\ha\ modulation, the possible UV modulation, and the Campaign~2 \sti\ radio
emission all peak around the time of optical minimum. Furthermore, we observed
\stv\ radio flares during all 7 optical minima. However, the rapid X-ray and
\ha\ flares occur around optical \textit{maximum}, the \stv\ flares occur at
all phases, and the Campaign~1 \sti\ radio emission modulates approximately in
phase with the optical data. The double-humped shape of the Campaign~1
emission may indeed indicate the presence of two separate emitters. While
magnetic phenomena appear to be enhanced during optical minimum, they clearly
occur at all phases.

A model developed for higher-mass magnetic chemically peculiar (MCP) stars
ascribes radio modulation to the presence of a torus of cold, absorbing plasma
around the magnetic equator \citep{tlu+04,tlu+11}. Such tori are also found
encircling Jupiter and Saturn \citep[e.g.,][and references therein]{k05b} and
thus may plausibly occur in \nltt\ as well. Simulated radio light curves,
spectra, and radio/X-ray luminosity ratios are broadly in line with the
observations of \nltt\ \citep{tlu+04}, and periodic auroral bursts are
predicted that could be consistent with the LCP flares at optical minima
\citep{tlu+11}.

We argue in \sect{vlaanal} that the Campaign~2 emission includes an additional
component that may be due to optically thick gyrosynchrotron flaring,
explaining the apparent phase shift of the radio \sti\ relative to the optical
maxima. However, the decrease of \sti\ in Campaign~2 relative to the expected
Campaign~1 emission (\fig{vlanofl}) requires the presence of an absorber. The
additional emission component may thus be associated with the creation or
expansion of an absorbing plasma torus as suggested by the MCP model. The
component associated with the \stv\ modulation may be confined to the magnetic
poles, where the expanded torus would not alter the observed signal.

Models with large polar spots may seem difficult to reconcile with the
presence of two periodic signals in the MEarth data. However, we note that
magnetic phenomena in the two hemispheres of Saturn are associated with
rotation periods that differ at the 1\% level \citep{glk+09}. We speculate
that one of the components of \nltt\ may host two large polar spots that
similarly rotate at slightly different rates. One caveat is that in Saturn,
this effect may be driven by the Sun \citep{glk+09}; \nltt\ does not have a
similar driver.

\subsection{Flaring conditions}

Although the non-flaring modulation suggests the presence of large-scale
magnetic fields, the rapid flares we observe imply the presence of significant
magnetic energy in small-scale fields as well, as is expected for low-mass
stars in general on both observational \citep{rb09b} and theoretical grounds
\citep{ljm+14}. The frequent (\apx30\% duty cycle) LCP radio flares suggest
the nearly continuous occurrence of magnetic reconnection, especially since
such coherent emission is expected to be strongly beamed. Reconnection
dominated by large-scale (\apx R$_*$) magnetic fields, on the other hand,
would lead to less frequent, larger flares, as observed in some UCDs in X-rays
\citep{ssml06,rps10}; the slow-evolving X-ray/UV flare observed by \swift\ may
be an instance of one of these. Hydrodynamic modeling of flaring loops
suggests a length scale of $10^9\text{ cm} \apx 0.1$~R$_*$ from the \apx150~s
decay timescale of the rapid \chandra\ flare \citep{srj+91}. Frequent
small-scale reconnection events, as suggested by the radio LCP and X-ray
flares, may be related to radio emission above that predicted by the
G\"udel-Benz relation \citep{gb93} both specifically in \nltt\ and more
generally \citep{wcb14}.

Applying the G\"udel-Benz relation to the rapid X-ray flare, the expected
radio luminosity is $\slr \apx 13$, equivalent to a flux density of 0.03~mJy.
Such an enhancement would not be discernible in the radio light curves.
Assuming that \ha\ luminosity represents \apx10\% of the white-light component
of a flare \citep{n89} and that flare soft X-ray luminosity is \apx20\% of
that of the white light \citep{wef+04}, the flaring \ha\ measurement would
correspond to a nearly identical value of $\lx = 28.5$, and have a similarly
insignificant radio component. Combined with the energetic insignificance of
the coherent radio emission (\tbl{summary}), the overall lack of correlation
among rapid flare events at multiple bands is striking, but potentially not
surprising \citeeg{oba+04,oha+05}.

\subsection{Effect of activity on fundamental measurements}

The high levels of activity in \nltt\ have significant implications for its
status as benchmark system for young, low-mass objects. The results of
\citet{sksd12} suggest that mass estimates will not be strongly affected
because \Lb\ is approximately conserved; however, this particular system will
yield dynamical mass measurements regardless. On the other hand, their
relations predict that magnetism may alter \teff\ by \apx$-10$\% and $R_*$ by
\apx+20\%. (Our estimates are \apx2 times larger than those of \citet{sbh+14},
likely due to the use of a different $\chi$ factor, since we use their
\lb\ and equivalent \ewha\ values.) The detection of only a single component
in VLBI \citep{mbi+11} suggests that these effects may be concentrated in only
one component of the binary. However, \citet{sksd12} rely on \ha\ rather than
radio emission as a metric of magnetic activity, and it remains to be
conclusively determined what the most physically relevant tracer truly is.

\section{Summary \& Conclusions}
\label{s.conc}

We have presented a detailed simultaneous study of the magnetic activity of
the young UCD binary \nltt\ from two observational campaigns in the radio,
optical, \ha, UV, and X-ray bands. Some of the key phenomena are:

\begin{itemize}
\item Extreme magnetic activity, with the highest radio luminosity of any
  known UCD, one of the highest X-ray luminosities, and a large
  \ha\ luminosity as well (\tbl{summary}).
\item Periodic modulation of emission in at least the radio and optical bands,
  and plausible UV and \ha\ modulation. The long-term MEarth data set reveals
  two distinct periodicities of $3.7859(1)$ and $3.7130(2)$~hr, with \apx50\%
  evolution in the modulation amplitude and phase from year to year.
\item Significant evolution in the \sti\ radio modulation between the two
  campaigns, with the amplitude increasing by a factor of 2 and the phase
  shifting by \apx180\degr. However, \stv\ is stable and phases consistently
  with the primary MEarth periodicity in both campaigns.
\item A very bright \stv\ flare with a spectral cutoff suggesting a magnetic
  field strength of 2.1~kG in the ECMI interpretation.
\item A typical X-ray spectrum adequately fit by two temperature components of
  $0.27(2)$ and $1.20(6)$~keV. However, the hardness gradually increases
  before a large X-ray/UV flare lasting \apx0.1~d.
\item A general lack of correlation between flares in any band. The exceptions
  are the slow X-ray/UV flare observed by \swift, and the consistent detection
  of \stv\ flares at optical minima.
\end{itemize}

Our analysis leads us to conclude that:

\begin{itemize}
\item The explanation for the two MEarth periodicities is unclear. Although
  they differ by only \apx2\%, even this level of differential rotation is not
  expected in stars this cool. If \nltt\ indeed shows unusual differential
  rotation, it may possibly be a consequence of its selection as an unusually
  magnetically active system.
\item The periodic modulation of the non-flaring emission of \nltt\ may be
  understood in a model with a large-scale magnetic field misaligned with the
  rotation axis. The change in the relative phasing of the radio \sti\ and
  optical modulation between Campaigns~1 and 2 is the most difficult to
  explain in such a context. The stability of the radio \stv\ signal, however,
  suggests that the Campaign~2 radio emission is the sum of that seen in
  Campaign~1 and an additional component that is furthermore associated with
  an absorbing equatorial torus of cold plasma \citep{tlu+04,tlu+11}.
\item The presence of frequent, rapid flares at all rotational phases implies
  the additional presence of significant magnetic energy in small scales that
  is being dissipated in reconnection events nearly continuously. This may be
  related to the excessive radio luminosity of \nltt\ relative to the
  G\"udel-Benz relation \citep{wcb14}.
\item The high levels of magnetic activity in this system may alter \teff\ by
  \apx$-10$\% and $R_*$ by \apx+20\%.
\end{itemize}

Significant progress can be made with currently-available resources. Not only
will spatially resolved astrometric monitoring improve constraints on the
fundamental parameters of the binary components, it will allow much tighter
constraints on the radio emission of the more radio-faint component by
providing knowledge of its position in existing VLBI data sets. Such
monitoring is in progress, as are continued VLBI observations. Spatially
resolved spectroscopy of this system is achievable with state-of-the-art
facilities \citeeg{kgf+12} and would yield both resolved radial velocity
measurements, diagnosing the inclination of the binary orbit, and resolved
\vsi\ measurements, providing key insight into the rotation rates of the two
components. Spatially resolved photometric monitoring would be
resource-intensive but could resolve the $\sin i$ ambiguity in spectroscopic
rotation measurements. With a separation of \apx0.1$''$, \nltt\ is
unfortunately not spatially resolvable by \chandra\ or any planned X-ray
observatories --- unless its orbit turns out to be highly eccentric.

New spatially-resolved observations of \nltt\ will advance the understanding
of habitable exoplanets around low-mass stars because of this system's twofold
importance: it is a benchmark for measurements of both fundamental stellar
properties \textit{and} magnetic activity at the bottom of main sequence. The
former affect the derived properties of exoplanets themselves, and the latter
may strip the atmospheres of close-in planets or otherwise render them
inhospitable to life.

\acknowledgments

We thank Jan Forbrich for sharing preliminary VLBI results and helpful
suggestions. P.~K.~G.~W and E.~B. acknowledge support for this work from the
National Science Foundation through Grant AST-1008361. Z.~K.~B.-T.
acknowledges support from the Torres Fellowship for Exoplanetary Research. The
VLA is operated by the National Radio Astronomy Observatory, a facility of the
National Science Foundation operated under cooperative agreement by Associated
Universities, Inc. This research has made use of the SIMBAD database, operated
at CDS, Strasbourg, France, and NASA's Astrophysics Data System. This research
has made use of the XRT Data Analysis Software (XRTDAS) developed under the
responsibility of the ASI Science Data Center (ASDC), Italy.

This paper makes use of data from the MEarth Project, which is a collaboration
between Harvard University and the Smithsonian Astrophysical Observatory. The
MEarth Project acknowledges funding from the David and Lucile Packard
Fellowship for Science and Engineering and the National Science Foundation
under grants AST-0807690, AST-1109468, and AST-1004488 (Alan T. Waterman
Award), and a grant from the John Templeton Foundation.

This publication makes use of data products from the Wide-field Infrared
Survey Explorer, which is a joint project of the University of California, Los
Angeles, and the Jet Propulsion Laboratory/California Institute of Technology,
funded by the National Aeronautics and Space Administration.

This publication makes use of data products from the Two Micron All Sky
Survey, which is a joint project of the University of Massachusetts and the
Infrared Processing and Analysis Center/California Institute of Technology,
funded by the National Aeronautics and Space Administration and the National
Science Foundation.

Facilities: \facility{\chandra}, \facility{Karl G. Jansky Very Large Array},
\facility{MEarth}, \facility{MMT}, \facility{\swift}.

\bibliographystyle{yahapj}
\bibliography{paper}{}

\begin{thebibliography}{171}
\providecommand\natexlab[1]{#1}
\providecommand\JournalTitle[1]{#1}

\bibitem[{{Alef} {et~al.}(1997){Alef}, {Benz}, \& {G\"{u}del}}]{abg97}
{Alef}, W., {Benz}, A.~O., \& {G\"{u}del}, M. 1997,
  \href{http://adsabs.harvard.edu/abs/1997A\%26A...317..707A}{\JournalTitle{A\&A},
  317, 707}

\bibitem[{{Allard} {et~al.}(2012){Allard}, {Homeier}, \&
  {Freytag}}]{thebtsettl}
{Allard}, F., {Homeier}, D., \& {Freytag}, B. 2012,
  \href{http://dx.doi.org/10.1098/rsta.2011.0269}{\JournalTitle{RSPTA}, 370,
  2765}

\bibitem[{{Audard} {et~al.}(2003){Audard}, {G\"{u}del}, \& {Skinner}}]{ags03}
{Audard}, M., {G\"{u}del}, M., \& {Skinner}, S.~L. 2003,
  \href{http://dx.doi.org/10.1086/374710}{\JournalTitle{ApJ}, 589, 983}

\bibitem[{{Audard} {et~al.}(2007){Audard}, {Osten}, {Brown}, {Briggs},
  {G\"{u}del}, {Hodges-Kluck}, \& {Gizis}}]{aob+07}
{Audard}, M., {Osten}, R.~A., {Brown}, A., {et~al.} 2007,
  \href{http://dx.doi.org/10.1051/0004-6361:20078093}{\JournalTitle{A\&A}, 471,
  L63}

\bibitem[{{Baraffe} {et~al.}(1998){Baraffe}, {Chabrier}, {Allard}, \&
  {Hauschildt}}]{bcah98}
{Baraffe}, I., {Chabrier}, G., {Allard}, F., \& {Hauschildt}, P.~H. 1998,
  \href{http://adsabs.harvard.edu/abs/1998A\%26A...337..403B}{\JournalTitle{A\&A},
  337, 403}

\bibitem[{{Baraffe} {et~al.}(2002){Baraffe}, {Chabrier}, {Allard}, \&
  {Hauschildt}}]{bcah02}
---. 2002,
  \href{http://dx.doi.org/10.1051/0004-6361:20011638}{\JournalTitle{A\&A}, 382,
  563}

\bibitem[{{Barman} {et~al.}(2011){Barman}, {Macintosh}, {Konopacky}, \&
  {Marois}}]{bmkm11}
{Barman}, T.~S., {Macintosh}, B., {Konopacky}, Q.~M., \& {Marois}, C. 2011,
  \href{http://dx.doi.org/10.1088/0004-637X/733/1/65}{\JournalTitle{ApJ}, 733,
  65}

\bibitem[{{Becker} {et~al.}(1995){Becker}, {White}, \& {Helfand}}]{thefirst}
{Becker}, R.~H., {White}, R.~L., \& {Helfand}, D.~J. 1995,
  \href{http://dx.doi.org/10.1086/176166}{\JournalTitle{ApJ}, 450, 559}

\bibitem[{{Benz} {et~al.}(1998){Benz}, {Conway}, \& {G\"{u}del}}]{bcg98}
{Benz}, A.~O., {Conway}, J., \& {G\"{u}del}, M. 1998,
  \href{http://adsabs.harvard.edu/abs/1998A\%26A...331..596B}{\JournalTitle{A\&A},
  331, 596}

\bibitem[{{Benz} \& {G\"{u}del}(1994)}]{bg94}
{Benz}, A.~O., \& {G\"{u}del}, M. 1994,
  \href{http://adsabs.harvard.edu/abs/1994A\%26A...285..621B}{\JournalTitle{A\&A},
  285, 621}

\bibitem[{{Berger}(2002)}]{b02}
{Berger}, E. 2002, \href{http://dx.doi.org/10.1086/340301}{\JournalTitle{ApJ},
  572, 503}

\bibitem[{{Berger}(2006)}]{b06b}
---. 2006, \href{http://dx.doi.org/10.1086/505787}{\JournalTitle{ApJ}, 648,
  629}

\bibitem[{{Berger} {et~al.}(2001){Berger}, {Ball}, {Becker}, {Clarke}, {Frail},
  {Fukuda}, {Hoffman}, {Mellon}, {Momjian}, {Murphy}, {Teng}, {Woodruff},
  {Zauderer}, \& {Zavala}}]{bbb+01}
{Berger}, E., {Ball}, S., {Becker}, K.~M., {et~al.} 2001,
  \href{http://dx.doi.org/10.1038/35066514}{\JournalTitle{Nature}, 410, 338}

\bibitem[{{Berger} {et~al.}(2008{\natexlab{a}}){Berger}, {Basri}, {Gizis},
  {Giampapa}, {Rutledge}, {Liebert}, {Mart\'{\i}n}, {Fleming}, {Johns-Krull},
  {Phan-Bao}, \& {Sherry}}]{bbg+08}
{Berger}, E., {Basri}, G., {Gizis}, J.~E., {et~al.} 2008{\natexlab{a}},
  \href{http://dx.doi.org/10.1086/529131}{\JournalTitle{ApJ}, 676, 1307}

\bibitem[{{Berger} {et~al.}(2008{\natexlab{b}}){Berger}, {Gizis}, {Giampapa},
  {Rutledge}, {Liebert}, {Mart\'{\i}n}, {Basri}, {Fleming}, {Johns-Krull},
  {Phan-Bao}, \& {Sherry}}]{bgg+08}
{Berger}, E., {Gizis}, J.~E., {Giampapa}, M.~S., {et~al.} 2008{\natexlab{b}},
  \href{http://dx.doi.org/10.1086/524769}{\JournalTitle{ApJ}, 673, 1080}

\bibitem[{{Berger} {et~al.}(2009){Berger}, {Rutledge}, {Phan-Bao}, {Basri},
  {Giampapa}, {Gizis}, {Liebert}, {Mart\'{\i}n}, \& {Fleming}}]{brpb+09}
{Berger}, E., {Rutledge}, R.~E., {Phan-Bao}, N., {et~al.} 2009,
  \href{http://dx.doi.org/10.1088/0004-637x/695/1/310}{\JournalTitle{ApJ}, 695,
  310}

\bibitem[{{Berger} {et~al.}(2010){Berger}, {Basri}, {Fleming}, {Giampapa},
  {Gizis}, {Liebert}, {Mart\'{\i}n}, {Phan-Bao}, \& {Rutledge}}]{bbf+10}
{Berger}, E., {Basri}, G., {Fleming}, T.~A., {et~al.} 2010,
  \href{http://dx.doi.org/10.1088/0004-637x/709/1/332}{\JournalTitle{ApJ}, 709,
  332}

\bibitem[{{Berta} {et~al.}(2012){Berta}, {Irwin}, {Charbonneau}, {Burke}, \&
  {Falco}}]{bic+12}
{Berta}, Z.~K., {Irwin}, J., {Charbonneau}, D., {Burke}, C.~J., \& {Falco},
  E.~E. 2012,
  \href{http://dx.doi.org/10.1088/0004-6256/144/5/145}{\JournalTitle{AJ}, 144,
  145}

\bibitem[{{Blanton} \& {Roweis}(2007)}]{br07}
{Blanton}, M.~R., \& {Roweis}, S. 2007,
  \href{http://dx.doi.org/10.1086/510127}{\JournalTitle{AJ}, 133, 734}

\bibitem[{{Bonfils} {et~al.}(2013){Bonfils}, {Delfosse}, {Udry}, {Forveille},
  {Mayor}, {Perrier}, {Bouchy}, {Gillon}, {Lovis}, {Pepe}, {Queloz}, {Santos},
  {S\'{e}gransan}, \& {Bertaux}}]{bdu+13}
{Bonfils}, X., {Delfosse}, X., {Udry}, S., {et~al.} 2013,
  \href{http://dx.doi.org/10.1051/0004-6361/201014704}{\JournalTitle{A\&A},
  549, A109}

\bibitem[{{Bouy} {et~al.}(2008){Bouy}, {Mart\'{\i}n}, {Brandner}, {Forveille},
  {Delfosse}, {Hu\'{e}lamo}, {Basri}, {Girard}, {Zapatero Osorio}, {Stumpf},
  {Ghez}, {Valdivielso}, {Marchis}, {Burgasser}, \& {Cruz}}]{bmb+08}
{Bouy}, H., {Mart\'{\i}n}, E.~L., {Brandner}, W., {et~al.} 2008,
  \href{http://dx.doi.org/10.1051/0004-6361:20078803}{\JournalTitle{A\&A}, 481,
  757}

\bibitem[{{Bowler} {et~al.}(2010){Bowler}, {Liu}, {Dupuy}, \&
  {Cushing}}]{bldc10}
{Bowler}, B.~P., {Liu}, M.~C., {Dupuy}, T.~J., \& {Cushing}, M.~C. 2010,
  \href{http://dx.doi.org/10.1088/0004-637X/723/1/850}{\JournalTitle{ApJ}, 723,
  850}

\bibitem[{{Breeveld} {et~al.}(2011){Breeveld}, {Landsman}, {Holland}, {Roming},
  {Kuin}, \& {Page}}]{blh+11}
{Breeveld}, A.~A., {Landsman}, W., {Holland}, S.~T., {et~al.} 2011,
  \href{http://dx.doi.org/10.1063/1.3621807}{in Gamma Ray Bursts 2010, ed.
  J.~E. {McEnery}, J.~L. {Racusin}, \& N.~{Gehrels}, Vol. 1358} (AIP), 373

\bibitem[{{Brown} {et~al.}(2010{\natexlab{a}}){Brown}, {Browning}, {Brun},
  {Miesch}, \& {Toomre}}]{bbb+10}
{Brown}, B.~P., {Browning}, M.~K., {Brun}, A.~S., {Miesch}, M.~S., \& {Toomre},
  J. 2010{\natexlab{a}},
  \href{http://dx.doi.org/10.1088/0004-637X/711/1/424}{\JournalTitle{ApJ}, 711,
  424}

\bibitem[{{Brown} {et~al.}(2010{\natexlab{b}}){Brown}, {Roming}, {Milne},
  {Bufano}, {Ciardullo}, {Elias-Rosa}, {Filippenko}, {Foley}, {Gehrels},
  {Gronwall}, {Hicken}, {Holland}, {Hoversten}, {Immler}, {Kirshner}, {Li},
  {Mazzali}, {Phillips}, {Pritchard}, {Still}, {Turatto}, \& {Vanden
  Berk}}]{brm+10}
{Brown}, P.~J., {Roming}, P. W.~A., {Milne}, P., {et~al.} 2010{\natexlab{b}},
  \href{http://dx.doi.org/10.1088/0004-637X/721/2/1608}{\JournalTitle{ApJ},
  721, 1608}

\bibitem[{{Browning}(2008)}]{b08}
{Browning}, M.~K. 2008,
  \href{http://dx.doi.org/10.1086/527432}{\JournalTitle{ApJ}, 676, 1262}

\bibitem[{{Burgasser} {et~al.}(2013){Burgasser}, {Melis}, {Zauderer}, \&
  {Berger}}]{bmzb13}
{Burgasser}, A.~J., {Melis}, C., {Zauderer}, B.~A., \& {Berger}, E. 2013,
  \href{http://dx.doi.org/10.1088/2041-8205/762/1/L3}{\JournalTitle{ApJL}, 762,
  L3}

\bibitem[{{Burgasser} \& {Putman}(2005)}]{bp05}
{Burgasser}, A.~J., \& {Putman}, M.~E. 2005,
  \href{http://dx.doi.org/10.1086/429788}{\JournalTitle{ApJ}, 626, 486}

\bibitem[{{Caffau} {et~al.}(2011){Caffau}, {Ludwig}, {Steffen}, {Freytag}, \&
  {Bonifacio}}]{cls+11}
{Caffau}, E., {Ludwig}, H.-G., {Steffen}, M., {Freytag}, B., \& {Bonifacio}, P.
  2011, \href{http://dx.doi.org/10.1007/s11207-010-9541-4}{\JournalTitle{SoPh},
  268, 255}

\bibitem[{{Cash}(1979)}]{c79}
{Cash}, W. 1979, \href{http://dx.doi.org/10.1086/156922}{\JournalTitle{ApJ},
  228, 939}

\bibitem[{{Chabrier} \& {Baraffe}(2000)}]{cb00}
{Chabrier}, G., \& {Baraffe}, I. 2000,
  \href{http://dx.doi.org/10.1146/annurev.astro.38.1.337}{\JournalTitle{ARA\&A},
  38, 337}

\bibitem[{{Chabrier} \& {K\"{u}ker}(2006)}]{ck06}
{Chabrier}, G., \& {K\"{u}ker}, M. 2006,
  \href{http://dx.doi.org/10.1051/0004-6361:20042475}{\JournalTitle{A\&A}, 446,
  1027}

\bibitem[{{Chauvin} {et~al.}(2005){Chauvin}, {Lagrange}, {Zuckerman}, {Dumas},
  {Mouillet}, {Song}, {Beuzit}, {Lowrance}, \& {Bessell}}]{clz05}
{Chauvin}, G., {Lagrange}, A.-M., {Zuckerman}, B., {et~al.} 2005,
  \href{http://dx.doi.org/10.1051/0004-6361:200500111}{\JournalTitle{A\&A},
  438, L29}

\bibitem[{{Cohen} {et~al.}(2014){Cohen}, {Drake}, {Glocer}, {Garraffo},
  {Poppenhaeger}, {Bell}, {Ridley}, \& {Gombosi}}]{cdg+14}
{Cohen}, O., {Drake}, J.~J., {Glocer}, A., {et~al.} 2014,
  \href{http://dx.doi.org/10.1088/0004-637X/790/1/57}{\JournalTitle{ApJ}, 790,
  57}

\bibitem[{{Cohen} {et~al.}(2011){Cohen}, {Kashyap}, {Drake}, {Sokolov},
  {Garraffo}, \& {Gombosi}}]{ckd+11}
{Cohen}, O., {Kashyap}, V.~L., {Drake}, J.~J., {et~al.} 2011,
  \href{http://dx.doi.org/10.1088/0004-637x/733/1/67}{\JournalTitle{ApJ}, 733,
  67}

\bibitem[{{Cook} {et~al.}(2014){Cook}, {Williams}, \& {Berger}}]{cwb14}
{Cook}, B.~A., {Williams}, P. K.~G., \& {Berger}, E. 2014,
  \href{http://dx.doi.org/10.1088/0004-637X/785/1/10}{\JournalTitle{ApJ}, 785,
  10}

\bibitem[{{Davenport} {et~al.}(2012){Davenport}, {Becker}, {Kowalski},
  {Hawley}, {Schmidt}, {Hilton}, {Sesar}, \& {Cutri}}]{dbk+12}
{Davenport}, J. R.~A., {Becker}, A.~C., {Kowalski}, A.~F., {et~al.} 2012,
  \href{http://dx.doi.org/10.1088/0004-637X/748/1/58}{\JournalTitle{ApJ}, 748,
  58}

\bibitem[{{Delorme} {et~al.}(2012){Delorme}, {Gagn\'{e}}, {Malo}, {Reyl\'{e}},
  {Artigau}, {Albert}, {Forveille}, {Delfosse}, {Allard}, \&
  {Homeier}}]{dgm+12}
{Delorme}, P., {Gagn\'{e}}, J., {Malo}, L., {et~al.} 2012,
  \href{http://dx.doi.org/10.1051/0004-6361/201219984}{\JournalTitle{A\&A},
  548, A26}

\bibitem[{{Dobler} {et~al.}(2006){Dobler}, {Stix}, \& {Brandenburg}}]{dsb06}
{Dobler}, W., {Stix}, M., \& {Brandenburg}, A. 2006,
  \href{http://dx.doi.org/10.1086/498634}{\JournalTitle{ApJ}, 638, 336}

\bibitem[{{Donati} {et~al.}(2008){Donati}, {Morin}, {Petit}, {Delfosse},
  {Forveille}, {Auri\`{e}re}, {Cabanac}, {Dintrans}, {Fares}, {Gastine},
  {Jardine}, {Ligni\`{e}res}, {Paletou}, {Ramirez Velez}, \&
  {Th\'{e}ado}}]{dmp+08}
{Donati}, J.~F., {Morin}, J., {Petit}, P., {et~al.} 2008,
  \href{http://dx.doi.org/10.1111/j.1365-2966.2008.13799.x}{\JournalTitle{MNRAS},
  390, 545}

\bibitem[{{Dulk}(1985)}]{d85}
{Dulk}, G.~A. 1985,
  \href{http://dx.doi.org/10.1146/annurev.aa.23.090185.001125}{\JournalTitle{ARA\&A},
  23, 169}

\bibitem[{{Dupuy} \& {Kraus}(2013)}]{dk13}
{Dupuy}, T.~J., \& {Kraus}, A.~L. 2013,
  \href{http://dx.doi.org/10.1126/science.1241917}{\JournalTitle{Sci}, 341,
  1492}

\bibitem[{{Dupuy} {et~al.}(2010){Dupuy}, {Liu}, {Bowler}, {Cushing}, {Helling},
  {Witte}, \& {Hauschildt}}]{dlb+10}
{Dupuy}, T.~J., {Liu}, M.~C., {Bowler}, B.~P., {et~al.} 2010,
  \href{http://dx.doi.org/10.1088/0004-637x/721/2/1725}{\JournalTitle{ApJ},
  721, 1725}

\bibitem[{{Dupuy} {et~al.}(2009){Dupuy}, {Liu}, \& {Ireland}}]{dli09b}
{Dupuy}, T.~J., {Liu}, M.~C., \& {Ireland}, M.~J. 2009,
  \href{http://dx.doi.org/10.1088/0004-637X/692/1/729}{\JournalTitle{ApJ}, 629,
  729}

\bibitem[{{Dupuy} {et~al.}(2014){Dupuy}, {Liu}, \& {Ireland}}]{x.dli14}
---. 2014, \href{http://arxiv.org/abs/1406.1184}{\JournalTitle{ApJ in press}},
  \href{http://arxiv.org/abs/1406.1184}{{\sffamily arxiv:1406.1184}}

\bibitem[{{Durney} {et~al.}(1993){Durney}, {de Young}, \& {Roxburgh}}]{ddyr93}
{Durney}, B.~R., {de Young}, D.~S., \& {Roxburgh}, I.~W. 1993,
  \href{http://dx.doi.org/10.1007/bf00690652}{\JournalTitle{SoPh}, 145, 207}

\bibitem[{{Eastman} {et~al.}(2010){Eastman}, {Siverd}, \& {Gaudi}}]{esg10}
{Eastman}, J., {Siverd}, R., \& {Gaudi}, B.~S. 2010,
  \href{http://dx.doi.org/10.1086/655938}{\JournalTitle{PASP}, 122, 935}

\bibitem[{{Edwards} {et~al.}(1993){Edwards}, {Strom}, {Hartigan}, {Strom},
  {Hillenbrand}, {Herbst}, {Attridge}, {Merrill}, {Probst}, \&
  {Gatley}}]{esh+93}
{Edwards}, S., {Strom}, S.~E., {Hartigan}, P., {et~al.} 1993,
  \href{http://dx.doi.org/10.1086/116646}{\JournalTitle{AJ}, 106, 372}

\bibitem[{{Frasca} {et~al.}(2000){Frasca}, {Freire Ferrero}, {Marilli}, \&
  {Catalano}}]{fffmc00}
{Frasca}, A., {Freire Ferrero}, R., {Marilli}, E., \& {Catalano}, S. 2000,
  \href{http://adsabs.harvard.edu/abs/2000A\%26A...364..179F}{\JournalTitle{A\&A},
  364, 179}

\bibitem[{{Freeman} {et~al.}(2001){Freeman}, {Doe}, \&
  {Siemiginowska}}]{thesherpa}
{Freeman}, P., {Doe}, S., \& {Siemiginowska}, A. 2001,
  \href{http://dx.doi.org/10.1117/12.447161}{\JournalTitle{Proc. SPIE}, 4477,
  76}

\bibitem[{{Fruscione} {et~al.}(2006){Fruscione}, {McDowell}, {Allen},
  {Brickhouse}, {Burke}, {Davis}, {Durham}, {Elvis}, {Galle}, {Harris},
  {Huenemoerder}, {Houck}, {Ishibashi}, {Karovska}, {Nicastro}, {Noble},
  {Nowak}, {Primini}, {Siemiginowska}, {Smith}, \& {Wise}}]{theciao}
{Fruscione}, A., {McDowell}, J.~C., {Allen}, G.~E., {et~al.} 2006,
  \href{http://dx.doi.org/10.1117/12.671760}{\JournalTitle{Proc. SPIE}, 6270,
  62701V}

\bibitem[{{Gallet} \& {Bouvier}(2013)}]{gb13}
{Gallet}, F., \& {Bouvier}, J. 2013,
  \href{http://dx.doi.org/10.1051/0004-6361/201321302}{\JournalTitle{A\&A},
  556, 36}

\bibitem[{{Gastine} {et~al.}(2013){Gastine}, {Morin}, {Duarte}, {Reiners},
  {Christensen}, \& {Wicht}}]{gmd+13}
{Gastine}, T., {Morin}, J., {Duarte}, L., {et~al.} 2013,
  \href{http://dx.doi.org/10.1051/0004-6361/201220317}{\JournalTitle{A\&A},
  549, L5}

\bibitem[{{Gizis} {et~al.}(2000){Gizis}, {Monet}, {Reid}, {Kirkpatrick},
  {Liebert}, \& {Williams}}]{gmr+00}
{Gizis}, J.~E., {Monet}, D.~G., {Reid}, I.~N., {et~al.} 2000,
  \href{http://dx.doi.org/10.1086/301456}{\JournalTitle{AJ}, 120, 1085}

\bibitem[{{G\"{u}del} \& {Benz}(1993)}]{gb93}
{G\"{u}del}, M., \& {Benz}, A.~O. 1993,
  \href{http://dx.doi.org/10.1086/186766}{\JournalTitle{ApJ}, 405, L63}

\bibitem[{{Gupta} {et~al.}(2011){Gupta}, {Galeazzi}, \& {Williams}}]{ggw11}
{Gupta}, A., {Galeazzi}, M., \& {Williams}, B. 2011,
  \href{http://dx.doi.org/10.1088/0004-637X/731/1/63}{\JournalTitle{ApJ}, 731,
  63}

\bibitem[{{Gurnett} {et~al.}(2009){Gurnett}, {Lecacheux}, {Kurth}, {Persoon},
  {Groene}, {Lamy}, {Zarka}, \& {Carbary}}]{glk+09}
{Gurnett}, D.~A., {Lecacheux}, A., {Kurth}, W.~S., {et~al.} 2009,
  \href{http://dx.doi.org/10.1029/2009GL039621}{\JournalTitle{Geophys. Res.
  Lett.}, 36, L16102}

\bibitem[{{Hallinan} {et~al.}(2006){Hallinan}, {Antonova}, {Doyle}, {Bourke},
  {Brisken}, \& {Golden}}]{had+06}
{Hallinan}, G., {Antonova}, A., {Doyle}, J.~G., {et~al.} 2006,
  \href{http://dx.doi.org/10.1086/508678}{\JournalTitle{ApJ}, 653, 690}

\bibitem[{{Hallinan} {et~al.}(2008){Hallinan}, {Antonova}, {Doyle}, {Bourke},
  {Lane}, \& {Golden}}]{had+08}
---. 2008, \href{http://dx.doi.org/10.1086/590360}{\JournalTitle{ApJ}, 684,
  644}

\bibitem[{{Hallinan} {et~al.}(2007){Hallinan}, {Bourke}, {Lane}, {Antonova},
  {Zavala}, {Brisken}, {Boyle}, {Vrba}, {Doyle}, \& {Golden}}]{hbl+07}
{Hallinan}, G., {Bourke}, S., {Lane}, C., {et~al.} 2007,
  \href{http://dx.doi.org/10.1086/519790}{\JournalTitle{ApJL}, 663, L25}

\bibitem[{{Hilton} {et~al.}(2010){Hilton}, {West}, {Hawley}, \&
  {Kowalski}}]{hwhk10}
{Hilton}, E.~J., {West}, A.~A., {Hawley}, S.~L., \& {Kowalski}, A.~F. 2010,
  \href{http://dx.doi.org/10.1088/0004-6256/140/5/1402}{\JournalTitle{AJ}, 140,
  1402}

\bibitem[{{Hong} {et~al.}(2004){Hong}, {Schlegel}, \& {Grindlay}}]{hsg04}
{Hong}, J., {Schlegel}, E.~M., \& {Grindlay}, J.~E. 2004,
  \href{http://dx.doi.org/10.1086/423445}{\JournalTitle{ApJ}, 614, 508}

\bibitem[{{Irwin} \& {Bouvier}(2008)}]{ib08}
{Irwin}, J., \& {Bouvier}, J. 2008,
  \href{http://dx.doi.org/10.1017/s1743921309032025}{\JournalTitle{Proc. IAU},
  258, 363}

\bibitem[{{Irwin} {et~al.}(2007){Irwin}, {Irwin}, {Aigrain}, {Hodgkin}, {Hebb},
  \& {Moraux}}]{iia+07}
{Irwin}, J., {Irwin}, M., {Aigrain}, S., {et~al.} 2007,
  \href{http://dx.doi.org/10.1111/j.1365-2966.2006.11408.x}{\JournalTitle{MNRAS},
  375, 1449}

\bibitem[{{Johns-Krull} {et~al.}(2000){Johns-Krull}, {Valenti}, \&
  {Linsky}}]{jkvl00}
{Johns-Krull}, C.~M., {Valenti}, J.~A., \& {Linsky}, J.~L. 2000,
  \href{http://dx.doi.org/10.1086/309259}{\JournalTitle{ApJ}, 539, 815}

\bibitem[{{Jones} {et~al.}(2005){Jones}, {Pavlenko}, {Viti}, {Barber},
  {Yakovina}, {Pinfield}, \& {Tennyson}}]{jpv+05}
{Jones}, H. R.~A., {Pavlenko}, Y., {Viti}, S., {et~al.} 2005,
  \href{http://dx.doi.org/10.1111/j.1365-2966.2005.08736.x}{\JournalTitle{MNRAS},
  358, 105}

\bibitem[{{Khodachenko} {et~al.}(2007){Khodachenko}, {Ribas}, {Lammer},
  {Grie\ss{}meier}, {Leitner}, {Selsis}, {Eiroa}, {Hanslmeier}, {Biernat},
  {Farrugia}, \& {Rucker}}]{krl+07}
{Khodachenko}, M.~L., {Ribas}, I., {Lammer}, H., {et~al.} 2007,
  \href{http://dx.doi.org/10.1089/ast.2006.0127}{\JournalTitle{AsBio}, 7, 167}

\bibitem[{{Kirkpatrick} {et~al.}(1999){Kirkpatrick}, {Reid}, {Liebert},
  {Cutri}, {Nelson}, {Beichman}, {Dahn}, {Monet}, {Gizis}, \&
  {Skrutskie}}]{krl+99}
{Kirkpatrick}, J.~D., {Reid}, I.~N., {Liebert}, J., {et~al.} 1999,
  \href{http://dx.doi.org/10.1086/307414}{\JournalTitle{ApJ}, 519, 802}

\bibitem[{{Kirkpatrick} {et~al.}(2012){Kirkpatrick}, {Gelino}, {Cushing},
  {Mace}, {Griffith}, {Skrutskie}, {Marsh}, {Wright}, {Eisenhardt}, {McLean},
  {Mainzer}, {Burgasser}, {Tinney}, {Parker}, \& {Salter}}]{kgc+12}
{Kirkpatrick}, J.~D., {Gelino}, C.~R., {Cushing}, M.~C., {et~al.} 2012,
  \href{http://dx.doi.org/10.1088/0004-637X/753/2/156}{\JournalTitle{ApJ}, 753,
  156}

\bibitem[{{Kivelson}(2005)}]{k05b}
{Kivelson}, M.~G. 2005,
  \href{http://dx.doi.org/10.1016/j.asr.2005.05.104}{\JournalTitle{AdSpR}, 36,
  2077}

\bibitem[{{K\"{o}nigl}(1991)}]{k91}
{K\"{o}nigl}, A. 1991,
  \href{http://dx.doi.org/10.1086/185972}{\JournalTitle{ApJL}, 370, 39}

\bibitem[{{Konopacky} {et~al.}(2010){Konopacky}, {Ghez}, {Barman}, {Rice},
  {Bailey}, {White}, {McLean}, \& {Duch\^{e}ne}}]{kgb+10}
{Konopacky}, Q.~M., {Ghez}, A.~M., {Barman}, T.~S., {et~al.} 2010,
  \href{http://dx.doi.org/10.1088/0004-637x/711/2/1087}{\JournalTitle{ApJ},
  711, 1087}

\bibitem[{{Konopacky} {et~al.}(2012){Konopacky}, {Ghez}, {Fabrycky},
  {Macintosh}, {White}, {Barman}, {Rice}, {Hallinan}, \&
  {Duch\^{e}ne}}]{kgf+12}
{Konopacky}, Q.~M., {Ghez}, A.~M., {Fabrycky}, D.~C., {et~al.} 2012,
  \href{http://dx.doi.org/10.1088/0004-637x/750/1/79}{\JournalTitle{ApJ}, 750,
  79}

\bibitem[{{Kopparapu}(2013)}]{k13b}
{Kopparapu}, R.~K. 2013,
  \href{http://dx.doi.org/10.1088/2041-8205/767/1/L8}{\JournalTitle{ApJL}, 767,
  L8}

\bibitem[{{Kruger} {et~al.}(2002){Kruger}, {Loredo}, \& {Wasserman}}]{klw02}
{Kruger}, A.~T., {Loredo}, T.~J., \& {Wasserman}, I. 2002,
  \href{http://dx.doi.org/10.1086/341541}{\JournalTitle{ApJ}, 576, 932}

\bibitem[{{Kuiper}(1960)}]{k60}
{Kuiper}, N.~H. 1960, \JournalTitle{Proceedings of the Koninklijke Nederlandse
  Akademie van Wetenschappen: Series A}, 63, 38

\bibitem[{{Lammer}(2007)}]{l07}
{Lammer}, H. 2007,
  \href{http://dx.doi.org/10.1089/ast.2006.0123}{\JournalTitle{AsBio}, 7, 27}

\bibitem[{{Lang} {et~al.}(2014){Lang}, {Jardine}, {Morin}, {Donati}, {Jeffers},
  {Vidotto}, \& {Fares}}]{ljm+14}
{Lang}, P., {Jardine}, M., {Morin}, J., {et~al.} 2014,
  \href{http://dx.doi.org/10.1093/mnras/stu091}{\JournalTitle{MNRAS}, 439,
  2122}

\bibitem[{{Lanza}(2013)}]{l13b}
{Lanza}, A.~F. 2013,
  \href{http://dx.doi.org/10.1051/0004-6361/201321790}{\JournalTitle{A\&A},
  557, A31}

\bibitem[{{Law} {et~al.}(2006){Law}, {Hodgkin}, \& {Mackay}}]{lhm06}
{Law}, N.~M., {Hodgkin}, S.~T., \& {Mackay}, C.~D. 2006,
  \href{http://dx.doi.org/10.1111/j.1365-2966.2006.10265.x}{\JournalTitle{MNRAS},
  368, 1917}

\bibitem[{{Lecavelier des Etangs}(2007)}]{lde07}
{Lecavelier des Etangs}, A. 2007,
  \href{http://dx.doi.org/10.1051/0004-6361:20065014}{\JournalTitle{A\&A}, 461,
  1185}

\bibitem[{{Lee} {et~al.}(2010){Lee}, {Berger}, \& {Knapp}}]{lbk10}
{Lee}, K.-G., {Berger}, E., \& {Knapp}, G.~R. 2010,
  \href{http://dx.doi.org/10.1088/0004-637x/708/2/1482}{\JournalTitle{ApJ},
  708, 1482}

\bibitem[{{L\'{e}pine} \& {Shara}(2005)}]{thelspmn}
{L\'{e}pine}, S., \& {Shara}, M.~M. 2005,
  \href{http://dx.doi.org/10.1086/427854}{\JournalTitle{AJ}, 129, 1483}

\bibitem[{{L\'{e}pine} {et~al.}(2009){L\'{e}pine}, {Thorstensen}, {Shara}, \&
  {Rich}}]{ltsr09}
{L\'{e}pine}, S., {Thorstensen}, J.~R., {Shara}, M.~M., \& {Rich}, R.~M. 2009,
  \href{http://dx.doi.org/10.1088/0004-6256/137/5/4109}{\JournalTitle{AJ}, 137,
  4109}

\bibitem[{{Linsky} {et~al.}(2014){Linsky}, {Fontenla}, \& {France}}]{lff14}
{Linsky}, J.~L., {Fontenla}, J., \& {France}, K. 2014,
  \href{http://dx.doi.org/10.1088/0004-637X/780/1/61}{\JournalTitle{ApJ}, 780,
  61}

\bibitem[{{Liu} {et~al.}(2013){Liu}, {Magnier}, {Deacon}, {Allers}, {Dupuy},
  {Kotson}, {Aller}, {Burgett}, {Chambers}, {Draper}, {Hodapp}, {Jedicke},
  {Kaiser}, {Kudritzki}, {Metcalfe}, {Morgan}, {Price}, {Tonry}, \&
  {Wainscoat}}]{lmd+13}
{Liu}, M.~C., {Magnier}, E.~A., {Deacon}, N.~R., {et~al.} 2013,
  \href{http://dx.doi.org/10.1088/2041-8205/777/2/L20}{\JournalTitle{ApJL},
  777, L20}

\bibitem[{{Llama} {et~al.}(2013){Llama}, {Vidotto}, {Jardine}, {Wood}, {Fares},
  \& {Gombosi}}]{lvj+13}
{Llama}, J., {Vidotto}, A.~A., {Jardine}, M., {et~al.} 2013,
  \href{http://dx.doi.org/10.1093/mnras/stt1725}{\JournalTitle{MNRAS}, 436,
  2179}

\bibitem[{{Lodders}(2003)}]{l03}
{Lodders}, K. 2003, \href{http://dx.doi.org/10.1086/375492}{\JournalTitle{ApJ},
  591, 1220}

\bibitem[{{L\'{o}pez-Morales}(2007)}]{lm07}
{L\'{o}pez-Morales}, M. 2007,
  \href{http://dx.doi.org/10.1086/513142}{\JournalTitle{ApJ}, 660, 732}

\bibitem[{{Luhman}(2012)}]{l12}
{Luhman}, K.~L. 2012,
  \href{http://dx.doi.org/10.1146/annurev-astro-081811-125528}{\JournalTitle{ARA\&A},
  50, 65}

\bibitem[{{Luhman}(2013)}]{l13}
---. 2013,
  \href{http://dx.doi.org/10.1088/2041-8205/767/1/l1}{\JournalTitle{ApJL}, 767,
  L1}

\bibitem[{{Luhman}(2014)}]{l14}
---. 2014,
  \href{http://dx.doi.org/10.1088/2041-8205/786/2/L18}{\JournalTitle{ApJL},
  786, L18}

\bibitem[{{Luyten}(1979)}]{thenltt}
{Luyten}, W.~J. 1979, New Luyten Catalogue of Stars with Proper Motions Larger
  than Two Tenths of an Arcsecond (Minneapolis, MN: Univ. Minnesota Press)

\bibitem[{{MacDonald} \& {Mullan}(2013)}]{mm13}
{MacDonald}, J., \& {Mullan}, D.~J. 2013,
  \href{http://dx.doi.org/10.1088/0004-637X/765/2/126}{\JournalTitle{ApJ}, 765,
  126}

\bibitem[{{Maritz} \& {Jarrett}(1978)}]{mj78}
{Maritz}, J.~S., \& {Jarrett}, R.~G. 1978,
  \href{http://dx.doi.org/10.1080/01621459.1978.10480027}{\JournalTitle{Journal
  of the American Statistical Association}, 73, 194}

\bibitem[{{Marley} {et~al.}(2013){Marley}, {Ackerman}, {Cuzzi}, \&
  {Kitzmann}}]{mack13}
{Marley}, M.~S., {Ackerman}, A.~S., {Cuzzi}, J.~N., \& {Kitzmann}, D. 2013,
  \href{http://dx.doi.org/10.2458/azu\_uapress\_9780816530595-ch15}{in
  Comparitive Climatology of Terrestrial Planets, ed. S.~J. {Mackwell}, A.~A.
  {Simon-Miller}, J.~W. {Harder}, \& M.~A. {Bullock}} (Tucson, AZ: The
  University of Arizona Press), 367

\bibitem[{{Martin} {et~al.}(2005){Martin}, {Fanson}, {Schiminovich},
  {Morrissey}, {Friedman}, {Barlow}, {Conrow}, {Grange}, {Jelinsky},
  {Milliard}, {Siegmund}, {Bianchi}, {Byun}, {Donas}, {Forster}, {Heckman},
  {Lee}, {Madore}, {Malina}, {Neff}, {Rich}, {Small}, {Surber}, {Szalay},
  {Welsh}, \& {Wyder}}]{thegalex}
{Martin}, D.~C., {Fanson}, J., {Schiminovich}, D., {et~al.} 2005,
  \href{http://dx.doi.org/10.1086/426387}{\JournalTitle{ApJL}, 619, 1}

\bibitem[{{Mart\'{\i}n} {et~al.}(1999){Mart\'{\i}n}, {Delfosse}, {Basri},
  {Goldman}, {Forveille}, \& {Zapatero Osorio}}]{mdb+99}
{Mart\'{\i}n}, E.~L., {Delfosse}, X., {Basri}, G., {et~al.} 1999,
  \href{http://dx.doi.org/10.1086/301107}{\JournalTitle{AJ}, 118, 2466}

\bibitem[{{McIvor} {et~al.}(2006){McIvor}, {Jardine}, \& {Holzwarth}}]{mjh06}
{McIvor}, T., {Jardine}, M., \& {Holzwarth}, V. 2006,
  \href{http://dx.doi.org/10.1111/j.1745-3933.2005.00098.x}{\JournalTitle{MNRAS
  Lett.}, 367, L1}

\bibitem[{{McLean} {et~al.}(2011){McLean}, {Berger}, {Irwin}, {Forbrich}, \&
  {Reiners}}]{mbi+11}
{McLean}, M., {Berger}, E., {Irwin}, J., {Forbrich}, J., \& {Reiners}, A. 2011,
  \href{http://dx.doi.org/10.1088/0004-637x/741/1/27}{\JournalTitle{ApJ}, 741,
  27}

\bibitem[{{McLean} {et~al.}(2012){McLean}, {Berger}, \& {Reiners}}]{mbr12}
{McLean}, M., {Berger}, E., \& {Reiners}, A. 2012,
  \href{http://dx.doi.org/10.1088/0004-637x/746/1/23}{\JournalTitle{ApJ}, 746,
  23}

\bibitem[{{McMullin} {et~al.}(2007){McMullin}, {Waters}, {Schiebel}, {Young},
  \& {Golap}}]{thecasa}
{McMullin}, J.~P., {Waters}, B., {Schiebel}, D., {Young}, W., \& {Golap}, K.
  2007, \href{http://adsabs.harvard.edu/abs/2007ASPC..376..127M}{in
  Astronomical Society of the Pacific Conference Series, Vol. 376, Astronomical
  Data Analysis Software and Systems XVI, ed. R.~A. {Shaw}, F.~{Hill}, \& D.~J.
  {Bell}}, 127

\bibitem[{{Mitra-Kraev} {et~al.}(2005){Mitra-Kraev}, {Harra}, {G\"{u}del},
  {Audard}, {Branduardi-Raymont}, {Kay}, {Mewe}, {Raassen}, \& {van
  Driel-Gesztelyi}}]{mkhg+05}
{Mitra-Kraev}, U., {Harra}, L.~K., {G\"{u}del}, M., {et~al.} 2005,
  \href{http://dx.doi.org/10.1051/0004-6361:20041201}{\JournalTitle{A\&A}, 431,
  679}

\bibitem[{{Mohanty} \& {Basri}(2003)}]{mb03}
{Mohanty}, S., \& {Basri}, G. 2003,
  \href{http://dx.doi.org/10.1086/345097}{\JournalTitle{ApJ}, 583, 451}

\bibitem[{{Mohanty} {et~al.}(2002){Mohanty}, {Basri}, {Shu}, {Allard}, \&
  {Chabrier}}]{mbs+02}
{Mohanty}, S., {Basri}, G., {Shu}, F., {Allard}, F., \& {Chabrier}, G. 2002,
  \href{http://dx.doi.org/10.1086/339911}{\JournalTitle{ApJ}, 571, 469}

\bibitem[{{Monet} {et~al.}(2003){Monet}, {Levine}, {Canzian}, {Ables}, {Bird},
  {Dahn}, {Guetter}, {Harris}, {Henden}, {Leggett}, {Levison}, {Luginbuhl},
  {Martini}, {Monet}, {Munn}, {Pier}, {Rhodes}, {Riepe}, {Sell}, {Stone},
  {Vrba}, {Walker}, {Westerhout}, {Brucato}, {Reid}, {Schoening}, {Hartley},
  {Read}, \& {Tritton}}]{theusnob}
{Monet}, D.~G., {Levine}, S.~E., {Canzian}, B., {et~al.} 2003,
  \href{http://dx.doi.org/10.1086/345888}{\JournalTitle{AJ}, 125, 984}

\bibitem[{{Mor\'{e}}(1978)}]{m78}
{Mor\'{e}}, J.~J. 1978,
  \href{http://dx.doi.org/10.1007/BFb0067700}{\JournalTitle{Lecture Notes in
  Mathematics}, 630, 105}

\bibitem[{{Morin} {et~al.}(2010){Morin}, {Donati}, {Petit}, {Delfosse},
  {Forveille}, \& {Jardine}}]{mdp+10}
{Morin}, J., {Donati}, J.~F., {Petit}, P., {et~al.} 2010,
  \href{http://dx.doi.org/10.1111/j.1365-2966.2010.17101.x}{\JournalTitle{MNRAS},
  407, 2269}

\bibitem[{{Morin} {et~al.}(2008{\natexlab{a}}){Morin}, {Donati}, {Petit},
  {Delfosse}, {Forveille}, {Albert}, {Auri\`{e}re}, {Cabanac}, {Dintrans},
  {Fares}, {Gastine}, {Jardine}, {Ligni\`{e}res}, {Paletou}, {Ramirez Velez},
  \& {Th\'{e}ado}}]{mdp+08}
---. 2008{\natexlab{a}},
  \href{http://dx.doi.org/10.1111/j.1365-2966.2008.13809.x}{\JournalTitle{MNRAS},
  390, 567}

\bibitem[{{Morin} {et~al.}(2008{\natexlab{b}}){Morin}, {Donati}, {Forveille},
  {Delfosse}, {Dobler}, {Petit}, {Jardine}, {Collier Cameron}, {Albert},
  {Manset}, {Dintrans}, {Chabrier}, \& {Valenti}}]{mdf+08}
{Morin}, J., {Donati}, J.-F., {Forveille}, T., {et~al.} 2008{\natexlab{b}},
  \href{http://dx.doi.org/10.1111/j.1365-2966.2007.12709.x}{\JournalTitle{MNRAS},
  384, 77}

\bibitem[{{Neidig}(1989)}]{n89}
{Neidig}, D.~F. 1989,
  \href{http://dx.doi.org/10.1007/BF00161699}{\JournalTitle{SoPh}, 121, 261}

\bibitem[{{Nelder} \& {Mead}(1965)}]{thesimplex}
{Nelder}, J.~A., \& {Mead}, R. 1965,
  \href{http://dx.doi.org/10.1093/comjnl/7.4.308}{\JournalTitle{Comp. J.}, 7,
  308}

\bibitem[{{Ness} {et~al.}(2004){Ness}, {G\"{u}del}, {Schmitt}, {Audard}, \&
  {Telleschi}}]{ngs+04}
{Ness}, J.-U., {G\"{u}del}, M., {Schmitt}, J. H. M.~M., {Audard}, M., \&
  {Telleschi}, A. 2004,
  \href{http://dx.doi.org/10.1051/0004-6361:20040504}{\JournalTitle{A\&A}, 427,
  667}

\bibitem[{{Ness} {et~al.}(2002){Ness}, {Schmitt}, {Burwitz}, {Mewe}, {Raassen},
  {van der Meer}, {Predehl}, \& {Brinkman}}]{nsb+02}
{Ness}, J.-U., {Schmitt}, J. H. M.~M., {Burwitz}, V., {et~al.} 2002,
  \href{http://dx.doi.org/10.1051/0004-6361:20021146}{\JournalTitle{A\&A}, 394,
  911}

\bibitem[{{Nutzman} \& {Charbonneau}(2008)}]{themearth}
{Nutzman}, P., \& {Charbonneau}, D. 2008,
  \href{http://dx.doi.org/10.1086/533420}{\JournalTitle{PASP}, 120, 317}

\bibitem[{{Offringa} {et~al.}(2010){Offringa}, {de Bruyn}, {Biehl}, {Zaroubi},
  {Bernardi}, \& {Pandey}}]{odbb+10}
{Offringa}, A.~R., {de Bruyn}, A.~G., {Biehl}, M., {et~al.} 2010,
  \href{http://dx.doi.org/10.1111/j.1365-2966.2010.16471.x}{\JournalTitle{MNRAS},
  405, 155}

\bibitem[{{Offringa} {et~al.}(2012){Offringa}, {van de Gronde}, \&
  {Roerdink}}]{ovdgr12}
{Offringa}, A.~R., {van de Gronde}, J.~J., \& {Roerdink}, J. B. T.~M. 2012,
  \href{http://dx.doi.org/10.1051/0004-6361/201118497}{\JournalTitle{A\&A},
  539, A95}

\bibitem[{{Ossendrijver}(2003)}]{o03}
{Ossendrijver}, M. 2003,
  \href{http://dx.doi.org/10.1007/s00159-003-0019-3}{\JournalTitle{A\&ARv}, 11,
  287}

\bibitem[{{Osten} {et~al.}(2002){Osten}, {Brown}, {Wood}, \& {Brady}}]{obwb02}
{Osten}, R.~A., {Brown}, A., {Wood}, B.~E., \& {Brady}, P. 2002,
  \href{http://dx.doi.org/10.1086/323666}{\JournalTitle{ApJS}, 138, 99}

\bibitem[{{Osten} {et~al.}(2005){Osten}, {Hawley}, {Allred}, {Johns-Krull}, \&
  {Roark}}]{oha+05}
{Osten}, R.~A., {Hawley}, S.~L., {Allred}, J.~C., {Johns-Krull}, C.~M., \&
  {Roark}, C. 2005, \href{http://dx.doi.org/10.1086/427275}{\JournalTitle{ApJ},
  621, 398}

\bibitem[{{Osten} {et~al.}(2004){Osten}, {Brown}, {Ayres}, {Drake},
  {Franciosini}, {Pallavicini}, {Tagliaferri}, {Stewart}, {Skinner}, \&
  {Linsky}}]{oba+04}
{Osten}, R.~A., {Brown}, A., {Ayres}, T.~R., {et~al.} 2004,
  \href{http://dx.doi.org/10.1086/420770}{\JournalTitle{The Astrophysical
  Journal Supplement Series}, 153, 317}

\bibitem[{{Patience} {et~al.}(2010){Patience}, {King}, {de Rosa}, \&
  {Marois}}]{pkdrm10}
{Patience}, J., {King}, R.~R., {de Rosa}, R.~J., \& {Marois}, C. 2010,
  \href{http://dx.doi.org/10.1051/0004-6361/201014173}{\JournalTitle{A\&A},
  517, A76}

\bibitem[{{Perley} \& {Butler}(2013)}]{pb13}
{Perley}, R.~A., \& {Butler}, B.~J. 2013,
  \href{http://dx.doi.org/10.1088/0067-0049/204/2/19}{\JournalTitle{ApJS}, 204,
  19}

\bibitem[{{Pizzolato} {et~al.}(2003){Pizzolato}, {Maggio}, {Micela},
  {Sciortino}, \& {Ventura}}]{pmm+03}
{Pizzolato}, N., {Maggio}, A., {Micela}, G., {Sciortino}, S., \& {Ventura}, P.
  2003,
  \href{http://dx.doi.org/10.1051/0004-6361:20021560}{\JournalTitle{A\&A}, 397,
  147}

\bibitem[{{Poppenhaeger} {et~al.}(2013){Poppenhaeger}, {Schmitt}, \&
  {Wolk}}]{psw13}
{Poppenhaeger}, K., {Schmitt}, J. H. M.~M., \& {Wolk}, S.~J. 2013,
  \href{http://dx.doi.org/10.1088/0004-637X/773/1/62}{\JournalTitle{ApJ}, 773,
  62}

\bibitem[{{Reid} \& {Gizis}(1997)}]{rg97}
{Reid}, I.~N., \& {Gizis}, J.~E. 1997,
  \href{http://dx.doi.org/10.1086/118436}{\JournalTitle{AJ}, 113, 2246}

\bibitem[{{Reiners} \& {Basri}(2006)}]{rb06}
{Reiners}, A., \& {Basri}, G. 2006,
  \href{http://dx.doi.org/10.1086/503324}{\JournalTitle{ApJ}, 644, 497}

\bibitem[{{Reiners} \& {Basri}(2007)}]{rb07}
---. 2007, \href{http://dx.doi.org/10.1086/510304}{\JournalTitle{ApJ}, 656,
  1121}

\bibitem[{{Reiners} \& {Basri}(2008)}]{rb08}
---. 2008, \href{http://dx.doi.org/10.1086/590073}{\JournalTitle{ApJ}, 684,
  1390}

\bibitem[{{Reiners} \& {Basri}(2009)}]{rb09b}
---. 2009,
  \href{http://dx.doi.org/10.1051/0004-6361:200811450}{\JournalTitle{A\&A},
  496, 787}

\bibitem[{{Reiners} \& {Basri}(2010)}]{rb10}
---. 2010,
  \href{http://dx.doi.org/10.1088/0004-637x/710/2/924}{\JournalTitle{ApJ}, 710,
  924}

\bibitem[{{Reiners} \& {Mohanty}(2012)}]{rm12}
{Reiners}, A., \& {Mohanty}, S. 2012,
  \href{http://dx.doi.org/10.1088/0004-637x/746/1/43}{\JournalTitle{ApJ}, 746,
  43}

\bibitem[{{Reinhold} {et~al.}(2013){Reinhold}, {Reiners}, \& {Basri}}]{rrb13}
{Reinhold}, T., {Reiners}, A., \& {Basri}, G. 2013,
  \href{http://dx.doi.org/10.1051/0004-6361/201321970}{\JournalTitle{A\&A},
  560, A4}

\bibitem[{{Ribas} {et~al.}(2008){Ribas}, {Morales}, {Jordi}, {Baraffe},
  {Chabrier}, \& {Gallardo}}]{rmj+08}
{Ribas}, I., {Morales}, J.~C., {Jordi}, C., {et~al.} 2008,
  \href{http://adsabs.harvard.edu/abs/2008MmSAI..79..562R}{\JournalTitle{MmSAI},
  79, 562}

\bibitem[{{Robrade} {et~al.}(2010){Robrade}, {Poppenhaeger}, \&
  {Schmitt}}]{rps10}
{Robrade}, J., {Poppenhaeger}, K., \& {Schmitt}, J. H. M.~M. 2010,
  \href{http://dx.doi.org/10.1051/0004-6361/200913603}{\JournalTitle{A\&A},
  513, A12}

\bibitem[{{Robrade} \& {Schmitt}(2005)}]{rs05}
{Robrade}, J., \& {Schmitt}, J. H. M.~M. 2005,
  \href{http://dx.doi.org/10.1051/0004-6361:20041941}{\JournalTitle{A\&A}, 435,
  1073}

\bibitem[{{Route} \& {Wolszczan}(2012)}]{rw12}
{Route}, M., \& {Wolszczan}, A. 2012,
  \href{http://dx.doi.org/10.1088/2041-8205/747/2/l22}{\JournalTitle{ApJL},
  747, L22}

\bibitem[{{Sault} \& {Wieringa}(1994)}]{themfs}
{Sault}, R.~J., \& {Wieringa}, M.~H. 1994, \JournalTitle{Astronomy and
  Astrophysics Supplement Series}, 108, 585

\bibitem[{{Scalo} {et~al.}(2007){Scalo}, {Kaltenegger}, {Segura}, {Fridlund},
  {Ribas}, {Kulikov}, {Grenfell}, {Rauer}, {Odert}, {Leitzinger}, {Selsis},
  {Khodachenko}, {Eiroa}, {Kasting}, \& {Lammer}}]{sks+07}
{Scalo}, J., {Kaltenegger}, L., {Segura}, A., {et~al.} 2007,
  \href{http://dx.doi.org/10.1089/ast.2006.0125}{\JournalTitle{AsBio}, 7, 85}

\bibitem[{{Scargle}(1998)}]{s98}
{Scargle}, J.~D. 1998,
  \href{http://dx.doi.org/10.1086/306064}{\JournalTitle{ApJ}, 504, 405}

\bibitem[{{Scargle} {et~al.}(2013){Scargle}, {Norris}, {Jackson}, \&
  {Chiang}}]{snjc13}
{Scargle}, J.~D., {Norris}, J.~P., {Jackson}, B., \& {Chiang}, J. 2013,
  \href{http://dx.doi.org/10.1088/0004-637x/764/2/167}{\JournalTitle{ApJ}, 764,
  167}

\bibitem[{{Schlieder} {et~al.}(2014){Schlieder}, {Bonnefoy}, {Herbst},
  {L\'{e}pine}, {Berger}, {Henning}, {Skemer}, {Chauvin}, {Rice}, {Biller},
  {Girard}, {Lagrange}, {Hinz}, {Defr\`{e}re}, {Bergfors}, {Brandner},
  {Lacour}, {Skrutskie}, \& {Leisenring}}]{sbh+14}
{Schlieder}, J., {Bonnefoy}, M., {Herbst}, T.~M., {et~al.} 2014,
  \href{http://dx.doi.org/10.1088/0004-637x/783/1/27}{\JournalTitle{ApJ}, 783,
  27}

\bibitem[{{Schlieder} {et~al.}(2012){Schlieder}, {L\'{e}pine}, \&
  {Simon}}]{sls12}
{Schlieder}, J.~E., {L\'{e}pine}, S., \& {Simon}, M. 2012,
  \href{http://dx.doi.org/10.1088/0004-6256/143/4/80}{\JournalTitle{AJ}, 143,
  80}

\bibitem[{{Schmitt} \& {Liefke}(2002)}]{sl02}
{Schmitt}, J. H. M.~M., \& {Liefke}, C. 2002,
  \href{http://dx.doi.org/10.1051/0004-6361:20011295}{\JournalTitle{A\&A}, 382,
  L9}

\bibitem[{{Scholz}(2013)}]{s13f}
{Scholz}, A. 2013,
  \href{http://adsabs.harvard.edu/abs/2013MmSAI..84..890S}{\JournalTitle{MmSAI},
  84, 890}

\bibitem[{{Segura} {et~al.}(2005){Segura}, {Kasting}, {Meadows}, {Cohen},
  {Scalo}, {Crisp}, {Butler}, \& {Tinetti}}]{skm+05}
{Segura}, A., {Kasting}, J.~F., {Meadows}, V., {et~al.} 2005,
  \href{http://dx.doi.org/10.1089/ast.2005.5.706}{\JournalTitle{AsBio}, 5, 706}

\bibitem[{{Serio} {et~al.}(1991){Serio}, {Reale}, {Jakimiec}, {Sylwester}, \&
  {Sylwester}}]{srj+91}
{Serio}, S., {Reale}, F., {Jakimiec}, J., {Sylwester}, B., \& {Sylwester}, J.
  1991,
  \href{http://adsabs.harvard.edu/abs/1991A\%26A...241..197S}{\JournalTitle{A\&A},
  241, 197}

\bibitem[{{Shkolnik} \& {Barman}(2014)}]{sb14}
{Shkolnik}, E.~L., \& {Barman}, T.~S. 2014,
  \href{http://arxiv.org/abs/1407.1344}{\JournalTitle{ApJ in press}},
  \href{http://arxiv.org/abs/1407.1344}{{\sffamily arxiv:1407.1344}}

\bibitem[{{Skrutskie} {et~al.}(2006){Skrutskie}, {Cutri}, {Stiening},
  {Weinberg}, {Schneider}, {Carpenter}, {Beichman}, {Capps}, {Chester},
  {Elias}, {Huchra}, {Liebert}, {Lonsdale}, {Monet}, {Price}, {Seitzer},
  {Jarrett}, {Kirkpatrick}, {Gizis}, {Howard}, {Evans}, {Fowler}, {Fullmer},
  {Hurt}, {Light}, {Kopan}, {Marsh}, {McCallon}, {Tam}, {Van Dyk}, \&
  {Wheelock}}]{the2mass}
{Skrutskie}, M.~F., {Cutri}, R.~M., {Stiening}, R., {et~al.} 2006,
  \href{http://dx.doi.org/10.1086/498708}{\JournalTitle{AJ}, 131, 1163}

\bibitem[{{Smith} {et~al.}(2001){Smith}, {Brickhouse}, {Liedahl}, \&
  {Raymond}}]{theapec}
{Smith}, R.~K., {Brickhouse}, N.~S., {Liedahl}, D.~A., \& {Raymond}, J.~C.
  2001, \href{http://dx.doi.org/10.1086/322992}{\JournalTitle{ApJL}, 556, L91}

\bibitem[{{Stassun} {et~al.}(2012){Stassun}, {Kratter}, {Scholz}, \&
  {Dupuy}}]{sksd12}
{Stassun}, K.~G., {Kratter}, K.~M., {Scholz}, A., \& {Dupuy}, T.~J. 2012,
  \href{http://dx.doi.org/10.1088/0004-637X/756/1/47}{\JournalTitle{ApJ}, 756,
  47}

\bibitem[{{Stellingwerf}(1978)}]{the.pdm}
{Stellingwerf}, R.~F. 1978,
  \href{http://dx.doi.org/10.1086/156444}{\JournalTitle{ApJ}, 224, 953}

\bibitem[{{Stelzer} {et~al.}(2013){Stelzer}, {Marino}, {Micela},
  {L\'{o}pez-Santiago}, \& {Liefke}}]{smm+13}
{Stelzer}, B., {Marino}, A., {Micela}, G., {L\'{o}pez-Santiago}, J., \&
  {Liefke}, C. 2013,
  \href{http://dx.doi.org/10.1093/mnras/stt225}{\JournalTitle{MNRAS}, 431,
  2063}

\bibitem[{{Stelzer} {et~al.}(2006{\natexlab{a}}){Stelzer}, {Micela},
  {Flaccomio}, {Neuh\"{a}user}, \& {Jayawardhana}}]{smf+06}
{Stelzer}, B., {Micela}, G., {Flaccomio}, E., {Neuh\"{a}user}, R., \&
  {Jayawardhana}, R. 2006{\natexlab{a}},
  \href{http://dx.doi.org/10.1051/0004-6361:20053677}{\JournalTitle{A\&A}, 448,
  293}

\bibitem[{{Stelzer} {et~al.}(2006{\natexlab{b}}){Stelzer}, {Schmitt}, {Micela},
  \& {Liefke}}]{ssml06}
{Stelzer}, B., {Schmitt}, J. H. M.~M., {Micela}, G., \& {Liefke}, C.
  2006{\natexlab{b}},
  \href{http://dx.doi.org/10.1051/0004-6361:20066488}{\JournalTitle{A\&A}, 460,
  L35}

\bibitem[{{Stelzer} {et~al.}(2012){Stelzer}, {Alcal\'{a}}, {Biazzo},
  {Ercolano}, {Crespo-Chac\'{o}n}, {L\'{o}pez-Santiago},
  {Mart\'{\i}nez-Arn\'{a}iz}, {Schmitt}, {Rigliaco}, {Leone}, \&
  {Cupani}}]{sab+12}
{Stelzer}, B., {Alcal\'{a}}, J., {Biazzo}, K., {et~al.} 2012,
  \href{http://dx.doi.org/10.1051/0004-6361/201118097}{\JournalTitle{A\&A},
  537, A94}

\bibitem[{{Tarter} {et~al.}(2007){Tarter}, {Backus}, {Mancinelli}, {Aurnou},
  {Backman}, {Basri}, {Boss}, {Clarke}, {Deming}, {Doyle}, {Feigelson},
  {Freund}, {Grinspoon}, {Haberle}, {Hauck}, {Heath}, {Henry}, {Hollingsworth},
  {Joshi}, {Kilston}, {Liu}, {Meikle}, {Reid}, {Rothschild}, {Scalo}, {Segura},
  {Tang}, {Tiedje}, {Turnbull}, {Walkowicz}, {Weber}, \& {Young}}]{tbm+07}
{Tarter}, J., {Backus}, P., {Mancinelli}, R., {et~al.} 2007,
  \href{http://dx.doi.org/10.1089/ast.2006.0124}{\JournalTitle{AsBio}, 7, 30}

\bibitem[{{Tian}(2009)}]{t09}
{Tian}, F. 2009,
  \href{http://dx.doi.org/10.1088/0004-637X/703/1/905}{\JournalTitle{ApJ}, 703,
  905}

\bibitem[{{Treumann}(2006)}]{t06}
{Treumann}, R. 2006,
  \href{http://dx.doi.org/10.1007/s00159-006-0001-y}{\JournalTitle{A\&ARv}, 13,
  229}

\bibitem[{{Trigilio} {et~al.}(2011){Trigilio}, {Leto}, {Umana}, {Buemi}, \&
  {Leone}}]{tlu+11}
{Trigilio}, C., {Leto}, P., {Umana}, G., {Buemi}, C.~S., \& {Leone}, F. 2011,
  \href{http://dx.doi.org/10.1088/2041-8205/739/1/l10}{\JournalTitle{ApJL},
  739, L10}

\bibitem[{{Trigilio} {et~al.}(2004){Trigilio}, {Leto}, {Umana}, {Leone}, \&
  {Buemi}}]{tlu+04}
{Trigilio}, C., {Leto}, P., {Umana}, G., {Leone}, F., \& {Buemi}, C.~S. 2004,
  \href{http://dx.doi.org/10.1051/0004-6361:20040060}{\JournalTitle{A\&A}, 418,
  593}

\bibitem[{{Vilhu}(1984)}]{v84}
{Vilhu}, O. 1984,
  \href{http://adsabs.harvard.edu/abs/1984A\%26\%2338\%3BA...133..117V}{\JournalTitle{A\&A},
  133, 117}

\bibitem[{{Walkowicz} {et~al.}(2004){Walkowicz}, {Hawley}, \&
  {West}}]{the.chifactor}
{Walkowicz}, L.~M., {Hawley}, S.~L., \& {West}, A.~A. 2004,
  \href{http://dx.doi.org/10.1086/426792}{\JournalTitle{PASP}, 116, 1105}

\bibitem[{{West} \& {Hawley}(2008)}]{wh08}
{West}, A.~A., \& {Hawley}, S.~L. 2008,
  \href{http://dx.doi.org/10.1086/593024}{\JournalTitle{PASP}, 120, 1161}

\bibitem[{{West} {et~al.}(2004){West}, {Hawley}, {Walkowicz}, {Covey},
  {Silvestri}, {Raymond}, {Harris}, {Munn}, {McGehee}, {Ivezi\'{c}}, \&
  {Brinkmann}}]{whw+04}
{West}, A.~A., {Hawley}, S.~L., {Walkowicz}, L.~M., {et~al.} 2004,
  \href{http://dx.doi.org/10.1086/421364}{\JournalTitle{AJ}, 128, 426}

\bibitem[{{West} {et~al.}(2011){West}, {Morgan}, {Bochanski}, {Andersen},
  {Bell}, {Kowalski}, {Davenport}, {Hawley}, {Schmidt}, {Bernat}, {Hilton},
  {Muirhead}, {Covey}, {Rojas-Ayala}, {Schlawin}, {Gooding}, {Schluns},
  {Dhital}, {Pineda}, \& {Jones}}]{wmb+11}
{West}, A.~A., {Morgan}, D.~P., {Bochanski}, J.~J., {et~al.} 2011,
  \href{http://dx.doi.org/10.1088/0004-6256/141/3/97}{\JournalTitle{AJ}, 141,
  97}

\bibitem[{{Williams} {et~al.}(2013){Williams}, {Berger}, \& {Zauderer}}]{wbz13}
{Williams}, P. K.~G., {Berger}, E., \& {Zauderer}, B.~A. 2013,
  \href{http://dx.doi.org/10.1088/2041-8205/767/2/l30}{\JournalTitle{ApJL},
  767, L30}

\bibitem[{{Williams} {et~al.}(2014){Williams}, {Cook}, \& {Berger}}]{wcb14}
{Williams}, P. K.~G., {Cook}, B.~A., \& {Berger}, E. 2014,
  \href{http://dx.doi.org/10.1088/0004-637X/785/1/9}{\JournalTitle{ApJ}, 785,
  9}

\bibitem[{{Woods} {et~al.}(2004){Woods}, {Eparvier}, {Fontenla}, {Harder},
  {Kopp}, {McClintock}, {Rottman}, {Smiley}, \& {Snow}}]{wef+04}
{Woods}, T.~N., {Eparvier}, F.~G., {Fontenla}, J., {et~al.} 2004,
  \href{http://dx.doi.org/10.1029/2004GL019571}{\JournalTitle{GeoRL}, 31,
  L10802}

\bibitem[{{Wright} {et~al.}(2010){Wright}, {Eisenhardt}, {Mainzer}, {Ressler},
  {Cutri}, {Jarrett}, {Kirkpatrick}, {Padgett}, {McMillan}, {Skrutskie},
  {Stanford}, {Cohen}, {Walker}, {Mather}, {Leisawitz}, {Gautier}, {McLean},
  {Benford}, {Lonsdale}, {Blain}, {Mendez}, {Irace}, {Duval}, {Liu}, {Royer},
  {Heinrichsen}, {Howard}, {Shannon}, {Kendall}, {Walsh}, {Larsen}, {Cardon},
  {Schick}, {Schwalm}, {Abid}, {Fabinsky}, {Naes}, \& {Tsai}}]{thewise}
{Wright}, E.~L., {Eisenhardt}, P. R.~M., {Mainzer}, A.~K., {et~al.} 2010,
  \href{http://dx.doi.org/10.1088/0004-6256/140/6/1868}{\JournalTitle{AJ}, 140,
  1868}

\bibitem[{{Wu} \& {Lee}(1979)}]{theecm}
{Wu}, C.~S., \& {Lee}, L.~C. 1979,
  \href{http://dx.doi.org/10.1086/157120}{\JournalTitle{ApJ}, 230, 621}

\end{thebibliography}

\end{document}